\documentclass[11pt,twoside,a4paper,cmspaper,final,collab]{cms-tdr}
\def\svnVersion{085931e-D}\def\svnDate{2026/04/28}\def\cmsCernNoTag{CERN-EP-2026-101}\def\cmsCernDate{\today}\def\cmsMessage{Submitted to Physical Review D}
\begin{document}\cmsNoteHeader{HIG-24-010}

\newlength\cmsTabSkip\setlength{\cmsTabSkip}{0.3ex}
\newlength\cmsBigSkip\setlength{\cmsBigSkip}{0.6ex}
\providecommand{\cmsTable}[1]{\resizebox{\textwidth}{!}{#1}}
\ifthenelse{\boolean{cms@external}}{
  \providecommand{\cmsLeft}{upper\xspace}
  \providecommand{\cmsRight}{lower\xspace}
  \providecommand{\cmsLLeft}{Upper\xspace}
  \providecommand{\cmsRRight}{Lower\xspace}
  }{
  \providecommand{\cmsLeft}{left\xspace}
  \providecommand{\cmsRight}{right\xspace}
  \providecommand{\cmsLLeft}{Left\xspace}
  \providecommand{\cmsRRight}{Right\xspace}
  }

\newcommand{\PNET}{\textsc{pnet}\xspace}
\newcommand{\GloParT}{\textsc{GloParT}\xspace}
\newcommand{\gjets}{\ensuremath{\gamma+\text{jets}}\xspace}
\newcommand{\Zjets}{\ensuremath{\PZ+\text{jets}}\xspace}
\newcommand{\Vjets}{\ensuremath{\PV+\text{jets}}\xspace}
\newcommand{\gghh}{\ensuremath{\Pg\Pg\PH\PH}\xspace}
\newcommand{\qqhh}{\ensuremath{\PQq\PQq\PH\PH}\xspace}
\newcommand{\zbb}{\ensuremath{\PZ \to \PQb\PAQb}\xspace}
\newcommand{\hbb}{\ensuremath{\PH \to \PQb\PAQb}\xspace}
\newcommand{\xbb}{\ensuremath{\text{X} \to \PQb\PAQb}\xspace}
\newcommand{\zmm}{\ensuremath{\PZ \to \PGm\PGm}\xspace}
\newcommand{\mhh}{\ensuremath{m_{\PH\PH}}\xspace}
\newcommand{\mSD}{\ensuremath{m_\text{SD}}\xspace}
\newcommand{\pTH}{\ensuremath{\pt(\PH)}\xspace}
\newcommand{\detajj}{\ensuremath{\Delta \eta_{\text{jj}}}\xspace}
\newcommand{\mjj}{\ensuremath{m_{\text{jj}}}\xspace}
\newcommand{\pb}{\ensuremath{\mathcal{P}_{\PQb}}\xspace}
\newcommand{\pbb}{\ensuremath{\mathcal{P}_{\text{X}\PQb\PAQb}}\xspace}
\newcommand{\regSRFourb}{\ensuremath{\mathrm{SR_{4\PQb}}}\xspace}
\newcommand{\regCRFourb}{\ensuremath{\mathrm{CR_{4\PQb}}}\xspace}
\newcommand{\regVRFourb}{\ensuremath{\mathrm{VR_{4\PQb}}}\xspace}
\newcommand{\regSRThreeb}{\ensuremath{\mathrm{SR_{3\PQb}}}\xspace}
\newcommand{\regCRThreeb}{\ensuremath{\mathrm{CR_{3\PQb}}}\xspace}
\newcommand{\regSRTwob}{\ensuremath{\mathrm{SR_{2\PQb}}}\xspace}
\newcommand{\regCRTwob}{\ensuremath{\mathrm{CR_{2\PQb}}}\xspace}
\newcommand{\regCRQA}{\ensuremath{\mathrm{CR_{QCD,A}}}\xspace}
\newcommand{\regCRQB}{\ensuremath{\mathrm{CR_{QCD,B}}}\xspace}
\newcommand{\regCRQC}{\ensuremath{\mathrm{CR_{QCD,C}}}\xspace}
\newcommand{\regCRtA}{\ensuremath{\mathrm{CR_{\ttbar,A}}}\xspace}
\newcommand{\regCRtB}{\ensuremath{\mathrm{CR_{\ttbar,B}}}\xspace}
\newcommand{\regCRDaA}{\ensuremath{\mathrm{CR_{data,A}}}\xspace}
\newcommand{\regCRDaB}{\ensuremath{\mathrm{CR_{data,B}}}\xspace}
\newcommand{\kappaVV}{\ensuremath{\kappa_{\text{2V}}}\xspace}
\newcommand{\kappaV}{\ensuremath{\kappa_{\text{V}}}\xspace}
\newcommand{\kappal}{\ensuremath{\kappa_{\lambda}}\xspace}
\newcommand{\muhh}{\ensuremath{\mu_{\PH\PH}}\xspace}
\newcommand{\mugghh}{\ensuremath{\mu_{\Pg\Pg\PH\PH}}\xspace}
\newcommand{\muqqhh}{\ensuremath{\mu_{\PQq\PQq\PH\PH}}\xspace}
\newcommand{\lSM}{\ensuremath{\lambda_{\text{SM}}}\xspace}
\newcommand{\muzz}{\ensuremath{\mu_{\PZ\PZ}}\xspace}
\newcommand{\muzh}{\ensuremath{\mu_{\PZ\PH}}\xspace}
\newcommand{\bbbb}{\ensuremath{\bbbar\bbbar}\xspace}
\newcommand{\bbtautau}{\ensuremath{\bbbar\PGtp\PGtm}\xspace}
\newcommand{\bbgg}{\ensuremath{\bbbar\gamma\gamma}\xspace}
\newcommand{\HHbbbb}{\ensuremath{\PH\PH\to4\PQb}\xspace}
\newcommand{\ZZbbbb}{\ensuremath{\PZ\PZ\to4\PQb}\xspace}
\newcommand{\ZHbbbb}{\ensuremath{\PZ\PH\to4\PQb}\xspace}
\newcommand{\HH}{\ensuremath{\PH\PH}\xspace}
\newcommand{\ZH}{\ensuremath{\PZ\PH}\xspace}
\newcommand{\HZ}{\ensuremath{\PH\PZ}\xspace}
\newcommand{\VV}{\ensuremath{\PV\PV}\xspace}
\newcommand{\ZZ}{\ensuremath{\PZ\PZ}\xspace}
\newcommand{\VH}{\ensuremath{\PV\PH}\xspace}
\newcommand{\FeynNet}{\textsc{FeynNet}\xspace}
\newcommand{\DeepJet}{\textsc{DeepJet}\xspace}
\newcommand{\DeepCSV}{\textsc{DeepCSV}\xspace}
\newcommand{\Prob}{\ensuremath{\mathcal{P}}\xspace}
\newcommand{\mreg}{\ensuremath{m_\text{reg}}\xspace}
\newcommand{\TXbb}{\ensuremath{T_{\text{X}\PQb\PAQb}}\xspace}
\newcommand{\VVH}{\ensuremath{\PV\PV\PH}\xspace}
\newcommand{\VVHH}{\ensuremath{\PV\PV\PH\PH}\xspace}
\renewcommand{\Hone}{\ensuremath{\PH_{1}}\xspace}
\renewcommand{\Htwo}{\ensuremath{\PH_{2}}\xspace}
\newcommand{\CRone}{\ensuremath{\text{CR}_{\text{3T1M}}}\xspace}
\newcommand{\CRtwo}{\ensuremath{\text{CR}_{\text{3T1L}}}\xspace}
\newcommand{\SRone}{\ensuremath{\text{SR}_{\text{3T1M}}}\xspace}
\newcommand{\SRtwo}{\ensuremath{\text{SR}_{\text{3T1L}}}\xspace}
\newcommand{\SRtwob}{\ensuremath{\text{SR}_{2\PQb}}\xspace}
\newcommand{\CRtwob}{\ensuremath{\text{CR}_{2\PQb}}\xspace}
\newcommand{\SRthree}{\ensuremath{\text{SR}_{\text{3MnT}}}\xspace}
\newcommand{\CRthree}{\ensuremath{\text{CR}_{\text{3MnT}}}\xspace}
\newcommand{\DZH}{\ensuremath{\mathcal{D}\text{(\ZH-vs-bkg)}}\xspace}
\newcommand{\DZZ}{\ensuremath{\mathcal{D}\text{(\ZZ-vs-bkg)}}\xspace}
\newcommand{\Dgghh}{\ensuremath{\mathcal{D}\text{(\gghh-vs-bkg)}}\xspace}
\newcommand{\Dqqhh}{\ensuremath{\mathcal{D}\text{(\qqhh-vs-bkg)}}\xspace}
\newcommand{\psig}{\ensuremath{\mathcal{P}_{\text{HH}}}\xspace}
\newcommand{\psigvbf}{\ensuremath{\mathcal{P}_{\text{VBFHH}}}\xspace}
\newcommand{\Nbjet}{\ensuremath{N_{\text{\PQb{}jet}}}\xspace}

\emergencystretch 3em

\cmsNoteHeader{HIG-24-010}
\title{Improved results on Higgs boson pair production in the \texorpdfstring{4\PQb}{4b} final state}

\abstract{Measurements of Higgs boson pair (\HH) production in the four bottom quark (4\PQb) final state are presented using proton-proton ($\Pp\Pp$) collision data at $\sqrt{s}=13.6\TeV$ collected by the CMS experiment at the CERN LHC, corresponding to an integrated luminosity of 62\fbinv. Events in which the Higgs boson decays, \hbb, are separately reconstructed as pairs of small-radius jets (resolved), as well as those where they are reconstructed as single large-radius jets (merged), are studied exclusively. Benefiting from new methods in trigger selection, event selection, and signal extraction, the combination of analyses in the resolved and merged topologies gives an observed (expected) upper limit on the HH signal strength, \muhh, of 4.4~(4.4) at 95\% confidence level (\CL). Compared to previously published LHC results in the 4b final state, the expected limit with an equivalent integrated luminosity is improved by more than a factor of two in the resolved topology and is better in the merged topology as well. An updated analysis of the resolved topology using 138\fbinv of 13\TeV $\Pp\Pp$ collision data yields an observed (expected) 95\% \CL upper limit on \muhh of 10.0\,(5.9), an improvement of about 25\% in the expected limit compared to the published results using the same data. Results in the 4\PQb final state with 13 and 13.6\TeV are combined, resulting in an observed (expected) 95\% \CL upper limit on \muhh of 4.7\,(2.8). The allowed ranges for the Higgs boson trilinear self-coupling and quartic coupling between two Higgs bosons and two vector bosons are also reported. These are the most stringent constraints achieved in the 4b final state to date.}
 
\hypersetup{%
pdfauthor={CMS Collaboration},%
pdftitle={Improved results on Higgs boson pair production in the 4b final state},%
pdfsubject={CMS},%
pdfkeywords={CMS, Higgs self-coupling, Higgs potential, b-jet trigger, data parking, flavour tagging}} 

\maketitle

\section{Introduction}\label{sec:intro}

The discovery of the Higgs boson (\PH)~\cite{PhysRevLett.13.321,Higgs:1964ia,PhysRevLett.13.508,PhysRevLett.13.585,PhysRev.145.1156,PhysRev.155.1554} in 2012 by the ATLAS~\cite{ATLAS:2012yve} and CMS~\cite{CMS:2012qbp} Collaborations at the CERN LHC prompted a broad research program to determine  its properties. To date, all measured properties are found to be consistent with the standard model (SM) predictions~\cite{CMS:2022dwd, ATLAS:2022vkf}. In the SM, after electroweak (EW) symmetry breaking the Higgs field ($\mathcal{H}$) potential can be written as
\begin{equation*}
	V(\mathcal{H}) = \frac{1}{2}m_{\PH}^2\mathcal{H}^2 + \lambda v \mathcal{H}^3 + \frac{1}{4}\lambda \mathcal{H}^4,
\end{equation*}
where $m_{\PH}$ is the Higgs boson mass and the last two terms represent trilinear and quartic \PH self-interactions, with the coupling strengths proportional to $\lambda$. In the SM, $m_{\PH} = \sqrt{2\lambda}v$, where $v$ is the vacuum expectation value of the Higgs field, a parameter tightly constrained by EW precision measurements. The value of $m_{\PH}$ has been precisely measured via direct studies of the Higgs boson at the LHC experiments~\cite{CMS:2024eka,ATLAS:2023oaq}. Altogether, the predicted value of $\lambda$ in the SM is constrained to be about 0.13 and any deviation from this value would be a signature of beyond-the-SM physics with potential cosmological implications~\cite{DiMicco:2019ngk}. Measuring $\lambda$ is thus fundamental in determining the shape of the Higgs potential and one of the key goals of the LHC physics program.

\begin{figure*}[!hbt]
\centering
\includegraphics[width=0.275\textwidth]{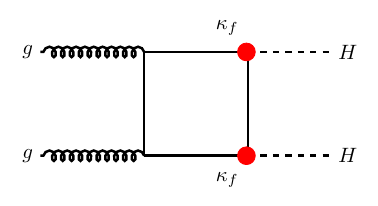}
\includegraphics[width=0.275\textwidth]{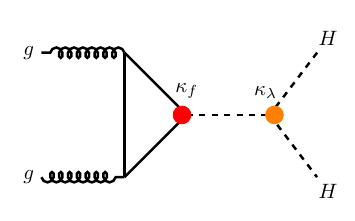}\\
\includegraphics[width=0.275\textwidth]{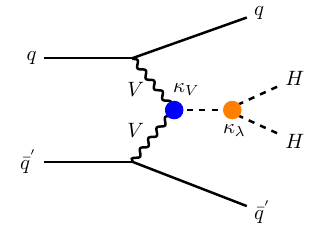} 
\includegraphics[width=0.275\textwidth]{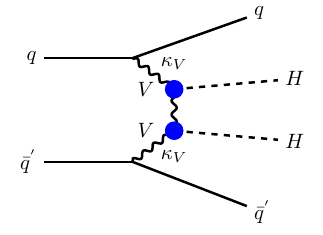} 
\includegraphics[width=0.275\textwidth]{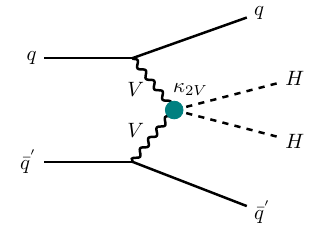}
\caption{Feynman diagrams that contribute to ggF and VBF \HH production at LO with coupling modifiers affecting the Higgs boson coupling strength to fermions ($\kappa_f$), to vector bosons (\kappaV), to two vector boson vertices (\kappaVV), and the Higgs boson self-coupling strength (\kappal).}
\label{fig:HHProduction}
\end{figure*}

At the LHC, $\lambda$ is directly probed through measurements of the Higgs boson pair (\HH) production. The SM \HH production cross section is about three orders of magnitude smaller than that of single \PH production. The primary \HH production mechanism is via gluon-gluon fusion (ggF), followed by vector boson fusion (VBF) production, both depicted at leading order (LO) in Fig.~\ref{fig:HHProduction}. In the SM, the two scattering amplitudes at LO for ggF production are similar in magnitude and interfere destructively.  Deviations of the coupling modifier $\kappal=\lambda/\lSM$ from unity can significantly modify the ggF \HH production cross section and kinematic properties. The VBF \HH production amplitudes depend on three couplings: the \PH trilinear self-coupling, the \PH coupling to two vector bosons (\VVH), and the coupling between two Higgs bosons and two vector bosons (\VVHH), which are parameterized by the coupling modifiers \kappal, \kappaV, and \kappaVV, respectively. While \kappal and \kappaV are primarily constrained by measurements of the ggF \HH and single-\PH processes, respectively, the VBF \HH process provides direct access to \kappaVV and thereby the most stringent constraints on \kappaVV. Deviations of \kappaVV from the SM expectation of unity can result in a significant increase of the VBF \HH cross section for events with Higgs bosons produced at large transverse momentum (\pt)~\cite{Bishara:2016kjn}. It should be noted that the \kappaVV coupling can also be directly accessed by measuring the production rate of a single Higgs boson and two vector bosons via vector boson scattering~\cite{CMS-PAS-HIG-24-003}. 

Using the LHC proton-proton ($\Pp\Pp$) collision data recorded during  2015--2018 (2016--2018) at $\sqrt{s}=13\TeV$, referred to as Run~2, the ATLAS and CMS Collaborations have constrained, at 95\% confidence level (\CL), the \HH signal strength, defined as the observed \HH production cross section divided by the SM prediction ${\muhh = \sigma_{\HH}/\sigma_{\HH}^{\text{SM}}}$, to be ${\muhh<2.9}$~\cite{PhysRevLett.133.101801}~ and $< 3.5$~\cite{CMS:2022dwd,CMS:2025ngq}, respectively. Constraints at 95\% \CL on the coupling modifiers were set by ATLAS to be ${-1.2 < \kappal <7.2}$ and ${0.6 < \kappaVV < 1.5}$~\cite{PhysRevLett.133.101801}, and by CMS to be ${-1.4 < \kappal < 6.4}$ and ${0.64 < \kappaVV < 1.40}$~\cite{CMS:2022dwd,CMS:2025ngq}. The results were achieved by combining several \HH decay modes~\cite{ATLAS:2023gzn,PhysRevD.108.052003,PhysRevD.110.032012,ATLAS:2024lhu,ATLAS:2023elc,CMS:2022hgz,CMS:2024fkb,CMS:2022kdx,CMS:2024rgy,PhysRevLett.133.101801,CMS:2022dwd,ATLAS:2025hhd}, with the dominant contributions resulting from three decay channels with similar sensitivity: \bbbb (referred to as 4\PQb in what follows), \bbtautau, and \bbgg. Of the aforementioned channels, the 4\PQb has the largest branching fraction, constituting about one third of all \HH events expected to be produced at the LHC. 

The primary challenge of the searches in the 4\PQb channel is separating the \HH signal from the overwhelming background and precisely modeling the background contribution in the analysis regions. Multijet events, including heavy-flavor processes, are produced in abundance at the LHC, making it extremely difficult to isolate them from the  \HHbbbb signal. This challenge is tackled with advanced tools and techniques for triggering, heavy-flavor tagging, and machine-learning (ML) algorithms to model backgrounds and extract signal with enhanced sensitivity. 

The ATLAS and CMS Collaborations have reported searches for \HH production in the 4\PQb final state at $\sqrt{s}=13\TeV$ in two distinct event topologies: one in which each \PH candidate is reconstructed from two separable small-radius jets (resolved topology)~\cite{PhysRevD.108.052003, CMS:2022cpr}, and one where each of the \PH  candidates is at high \pt and its decay products merge into a single large-radius jet (merged topology)~\cite{ATLAS:2024lsk, CMS:2022gjd}. Mixed topologies, in which one \PH candidate is reconstructed from small-radius jets and one \PH candidate from a large-radius jet~\cite{CMS:2018vjd}, are not considered in this paper. In the resolved event topology, CMS set observed (expected) upper limits at the 95\% CL on the signal strength \muhh of 3.9~(7.8); constraints were also set on \kappal, with the bounds found to be ${-2.3 < \kappal < 9.4}$~(${-5.0 < \kappal < 12.0}$). In the merged event topology, CMS set observed (expected) upper limits at the 95\% CL on \muhh of 9.9~(5.1); the same analysis also set stringent constraints on \kappaVV, with bounds of ${0.62 < \kappaVV < 1.41}$ (${0.66 < \kappaVV < 1.37}$), and excluding $\kappaVV=0$ with a significance of more than 6 standard deviations.

This paper presents searches for SM \HH production in the 4\PQb final state in both event topologies, and including the ggF and VBF production modes, using the LHC $\Pp\Pp$ collision data produced at $\sqrt{s}=13.6\TeV$ and recorded by the CMS detector during 2022--2023 (referred to herein as Run~3), corresponding to an integrated luminosity of 62\fbinv. For brevity, the analysis of events in the \HHbbbb resolved topology and the corresponding event category is referred to as the ``resolved analysis", while for the merged topology the term ``merged analysis" is used.
Two analysis approaches are presented for each topology, each using different methods for the background estimation, separation of signal from background, and assessment of key systematic uncertainties. The consistency of the results obtained from the two approaches validates the methodologies used and further builds confidence in the improvement in this measurement with respect to previous results, arising from new methods in trigger selection, event reconstruction, and signal extraction. An improved analysis of the \HHbbbb final state in the resolved event topology using Run~2 data corresponding to an integrated luminosity of 138\fbinv is also presented. This analysis uses the techniques developed in the search for \ZZ and \ZH production in the 4\PQb final state by the CMS Collaboration~\cite{CMS:2024tdk}, on the same data, to now target the \HH process, achieving increased sensitivity relative to the previous CMS Run~2 result~\cite{CMS:2022cpr}. These results supersede those reported in Ref.~\cite{CMS:2022cpr}. The merged analysis with Run~2 data described in Ref.~\cite{CMS:2022gjd} is included in the final combination of results. 

The paper is structured as follows. The CMS detector is briefly introduced in Section~\ref{sec:detector}, and the data and simulation samples are described in Section~\ref{sec:dataset}. The improved tools and techniques introduced for the analyses of the Run 3 data, including new triggers and heavy-flavor tagging algorithms, are presented in Section~\ref{sec:objet}. Sections~\ref{sec:resolved_analysis} and~\ref{sec:merged_analysis} describe \HHbbbb analyses in the resolved and merged topologies, respectively. These results all use Run~3 data, except for the updated resolved analysis with Run~2 data described in Section~\ref{sec:results_resolved_run2}. The combination of the Run~3 measurements is presented in Section~\ref{sec:combinations}, while Section~\ref{sec:combination_run23} summarizes the combination of the Run~3 results with the Run~2 resolved analysis described in Section~\ref{sec:results_resolved_run2} and the Run~2 merged analysis~\cite{CMS:2022gjd}. The results presented in Sections~\ref{sec:resolved_analysis} and~\ref{sec:merged_analysis} are in some cases based on partially overlapping data samples, which is addressed in Section~\ref{sec:combinations}. Finally, Section~\ref{sec:summary} summarizes the results and conclusions. Tabulated results from this paper are provided in a corresponding HEPData record~\cite{hepdata}.

\section{The CMS detector}\label{sec:detector}

The CMS apparatus~\cite{CMS:2008xjf,CMS:2023gfb} is a multipurpose, nearly hermetic detector, designed to trigger on~\cite{CMS:2020cmk,CMS:2016ngn,CMS:2024aqx} and identify electrons, muons, photons, and (charged and neutral) hadrons~\cite{CMS:2020uim,CMS:2018rym,CMS:2014pgm}. Its central feature is a superconducting solenoid of 6\unit{m} internal diameter, providing a magnetic field of 3.8\unit{T}. Within the solenoid volume are a silicon pixel and strip tracker, a lead tungstate crystal electromagnetic calorimeter (ECAL), and a brass and scintillator hadron calorimeter, each composed of a barrel and two endcap sections. Forward calorimeters extend the pseudorapidity ($\eta$) coverage provided by the barrel and endcap detectors. Muons are reconstructed using gas-ionization detectors embedded in the steel flux-return yoke outside the solenoid. Events of interest are selected using a two-tiered trigger system. The first level (L1), composed of custom hardware processors, uses information from the calorimeters and muon detectors to select events at a rate of around 100\unit{kHz} within a fixed latency of 4\mus~\cite{CMS:2020cmk}. The second level, known as the high-level trigger (HLT), consists of a farm of processors running a version of the full event reconstruction software optimized for fast processing, and reduces the event rate to a few kHz before data storage~\cite{CMS:2016ngn,CMS:2024aqx}. More detailed descriptions of the CMS detector, together with a definition of the coordinate system used and the relevant kinematic variables, can be found in Refs.~\cite{CMS:2008xjf,CMS:2023gfb}.

\section{Data sets and simulated samples}\label{sec:dataset}

The results presented in this paper are based on ${\sqrt{s}=13.6\TeV}$ $\Pp\Pp$ collision data collected by CMS in 2022 and 2023 during LHC Run~3, corresponding to an integrated luminosity of 62\fbinv, with the exception of an updated resolved analysis, described in Section~\ref{sec:resolved_run2}, which uses $\sqrt{s}=13\TeV$ Run~2 data, corresponding to an integrated luminosity of 138\fbinv. The trigger strategy for the resolved analysis was significantly modified shortly after the start of 2023 data taking, as will be described in Section~\ref{sec:trigger}. The Run~3 data in the resolved analysis are therefore split into two data samples, one before and one after the change in trigger strategy, with integrated luminosities of 41 and 21\fbinv, respectively. 

The dominant background in this measurement stems from SM events composed uniquely of jets produced through the strong interaction, referred to as quantum chromodynamics (QCD) multijet events, which are modeled with a variety of approaches based on control samples in data. The remaining backgrounds, primarily relevant in the merged channel, arise from top quark pair (\ttbar) and diboson (\VV) production, vector boson production in association with jets (\Vjets), and single \PH production (\VH, $\ttbar\PH$). The \HHbbbb signal processes, as well as the minor backgrounds, are modeled with Monte Carlo (MC) simulated events that incorporate detector effects and the same reconstruction algorithms for the physics objects employed for collision data.

The ggF \HHbbbb signal process is simulated at next-to-LO (NLO) precision in perturbative QCD with \POWHEG~2.0~\cite{POWHEG1,POWHEG2,POWHEG3,Bagnaschi:2011tu,Heinrich:2017kxx,Jones:2017giv,Buchalla:2018yce,Heinrich:2019bkc,Heinrich:2020ckp} and corrected as a function of the \HH mass (\mhh) based on Ref.~\cite{Davies:2019dfy}. The VBF \HHbbbb samples are generated at LO precision using \MGvATNLO 2.6.5~\cite{Alwall:2014hca}. A set of \HHbbbb signal samples corresponding to different combinations of the \kappaV, \kappaVV, and \kappal coupling modifiers is also generated. Signal predictions for \kappal and \kappaVV values not explicitly generated are constructed via linear combinations of the simulated signal samples by applying event weights, as described in Ref.~\cite{CMS:2022cpr}. The ggF \HHbbbb samples are normalized to the next-to-NLO (NNLO) cross sections~\cite{PhysRevD.58.115012,Shao:2013bz,PhysRevLett.111.201801,deFlorian:2015moa,PhysRevLett.117.012001,Grazzini:2018bsd,Baglio:2018lrj,PhysRevD.103.056002}. The VBF \HHbbbb sample with SM couplings is normalized to the next-to-NNLO (N$^3$LO) cross section~\cite{Dreyer:2018qbw}, and the SM N$^3$LO/LO relative correction is applied to the VBF predictions with non-SM \kappal and \kappaVV values. For brevity, the ggF \HHbbbb and VBF \HHbbbb processes will be referred to as ``\gghh'' and ``\qqhh'', respectively.

The \ttbar background, which is the most significant after QCD multijet, is simulated at NLO precision in QCD using \POWHEG~v2.0~\cite{POWHEG1,POWHEG2,POWHEG3,Frixione:2007nw,POWHEG-TT} and normalized to the cross section calculated at NNLO precision using {{\textsc{Top++} v2.0}}~\cite{Czakon:2011xx}. The \ttbar differential cross section as a function of the top quark \pt is corrected to NNLO QCD and NLO EW precision~\cite{Czakon:2017wor}. Samples of single \PW and \PZ boson production in association with jets are generated at LO using \MGvATNLO, including up to four partons in the final state. Other minor backgrounds, such as \VV production, single top quark, and single \PH production are generated at NLO using \POWHEG, with the exception of $\ttbar\PH$ and $\Pg\Pg \to \ZH$, which are generated at LO.

The increase in center-of-mass energy from ${\sqrt{s}=13\TeV}$ in Run~2 to 13.6\TeV in Run~3 raised the \HH production cross section by about 10\%~\cite{Grazzini:2018bsd}. The cross sections of the major background processes, \ttbar and QCD multijet, increased by roughly the same factor.

Parton showering, hadronization, and the underlying event are modeled by \PYTHIA8.306~\cite{Bierlich:2022pfr} using the CP5 tune~\cite{CMS:2019csb}. The NNPDF~3.0~\cite{Ball:2014uwa} and 3.1~\cite{Ball:2017} parton distribution functions (PDFs) are used in the generation of all simulated samples. The \GEANTfour~\cite{Agostinelli:2002hh} package is used to model the response of the CMS detector, and simulated minimum bias $\Pp\Pp$ interactions are mixed with the hard parton interactions in simulated events to model additional $\Pp\Pp$ interactions within the same or nearby bunch crossings (pileup). All simulated events used in this result are weighted to match the pileup distributions measured in data.

\section{Object and event reconstruction}\label{sec:objet}

The particle-flow (PF) algorithm~\cite{CMS:2017yfk} aims to reconstruct and identify each individual particle (PF candidate) in an event, with an optimized combination of information from the various elements of the CMS detector. The PF candidates are clustered into jets using the anti-\kt algorithm~\cite{Cacciari:2008gp} implemented in \FASTJET~\cite{Cacciari:2011ma} with a distance parameter of 0.4 (``AK4'' jets) or 0.8 (``AK8'' jets). 

Pileup interactions can contribute to the jet momentum by adding tracks and energy depositions in the calorimeters. To mitigate this effect, the pileup-per-particle identification algorithm~\cite{Bertolini:2014bba,CMS:2020ebo} is used in Run~3 data to assign a weight to each particle prior to jet clustering based on the likelihood of the particle originating from the primary vertex. For AK4 jets in Run 2 data, the pileup effect was instead mitigated by removing charged particles matched to pileup vertices and an offset correction was applied to correct for remaining neutral-particle contributions~\cite{CMS:2020ebo}.

Jet momentum is determined as the vectorial sum of all particle momenta in the jet, and is found from simulation to be, on average, within 5--10\% of the true momentum over the entire \pt spectrum and detector acceptance. Jet energy corrections are derived from simulation studies so that the average measured energy of jets becomes identical to that of particle-level jets. In situ measurements of the momentum balance in dijet, \gjets, and \Zjets events are used to determine any residual differences between the jet energy scale (JES) in data and in simulation, and appropriate corrections are made~\cite{CMS:2016lmd}. Additional selection criteria are applied to each jet to remove jets potentially mismeasured due to instrumental effects or reconstruction failures~\cite{CMS:2020ebo}. The missing transverse momentum \ptvecmiss is the negative vector \pt sum of all the reconstructed particles in an event, and its magnitude is denoted as \ptmiss~\cite{CMS:2019ctu}.

Muon candidates, within the geometrical acceptance of the muon detectors (${\abs{\eta}< 2.4}$), are reconstructed by combining the information from the silicon tracker and the muon chambers and are required to satisfy a set of quality criteria applied to the fitted track~\cite{CMS:2018rym}. Electron candidates within ${\abs{\eta}< 2.5}$ are reconstructed using an algorithm that associates fitted tracks in the silicon tracker with electromagnetic energy clusters in the ECAL~\cite{CMS:2020uim}. To reduce the misidentification rate, these candidates are required to satisfy identification criteria based on the properties of associated tracks and energy deposits in the calorimeters.

Identified muons and electrons are required to be isolated from hadronic activity in the event. The isolation is defined by summing the \pt of all the PF candidates in a cone of radius $\Delta R=\sqrt{\smash[b]{(\Delta\eta)^2+(\Delta\phi)^2}} = 0.4~(0.3)$ around the muon (electron) track and is corrected for the contribution of neutral particles from pileup interactions~\cite{CMS:2018rym,CMS:2020uim}. Events containing an isolated muon (electron) with ${\pt>10~(15)\GeV}$ are rejected in both the resolved and merged analyses, except for in certain control regions described in this section and in Section~\ref{sec:merged_analysis}.

The AK4 jets arising from the fragmentation of \PQb quarks (\PQb jets) are identified with a ``\PQb tagging'' algorithm~\cite{CMS:2017wtu}. Two \PQb tagging algorithms are used for Run 3 data, one tailored for use in the HLT (Section ~\ref{sec:trigger}) and one for the offline reconstruction (Section ~\ref{sec:ak4btag}). The offline algorithm also provides a correction to the reconstructed \PQb jet \pt, as described in Section~\ref{sec:ak4_btag_pt_regression}. Similar ``\bbbar tagging'' algorithms are used to identify AK8 jets produced by heavy resonance decays to \bbbar (\xbb) and to estimate the resonance mass, as described in Section~\ref{sec:glopart}.

\subsection{Trigger overview}\label{sec:trigger}

A major challenge in the search for \HHbbbb is maintaining high signal efficiency while rejecting the large QCD multijet background in the trigger decision, which is based on a low-latency and limited-precision event reconstruction. The enormous cross section of the QCD multijet background processes warrants strict trigger selection criteria to maintain a manageable trigger output rate, significantly impacting the \HHbbbb signal efficiency. It is also critical to collect large samples of QCD multijet events in control regions (CRs) with similar kinematic features to that of the \HHbbbb signal regions (SRs), providing flexibility in the design and validation of the background estimate based on control samples in data for the offline analysis.

\subsubsection{The L1 selection}
Event candidates are initially selected by the L1 trigger based on the overall L1 jet activity (\HT), defined as the scalar transverse energy (\et) sum of all L1 jets with ${\et>30\GeV}$ and ${\abs{\eta}<2.5}$~\cite{CMS:2016ngn}. The jet \et response is calibrated in situ and contributions from pileup interactions are subtracted on average. In 2022 (2023) the L1 \HT trigger required ${\HT>360~(280)\GeV}$. To recover the efficiency for signal events with lower-\et jets, a trigger requiring ${\HT>240\GeV}$ and a reconstructed muon with ${\pt>6\GeV}$ was also used, targeting the semileptonic decays of \PQb hadrons that produce muons in the final state and are reconstructed with high efficiency by the CMS detector. The combined efficiency of these triggers was measured as a function of \HT in \ttbar-enriched CRs selected from muon triggers. Auxiliary triggers, complementing those already described, are included for the resolved analysis that require the presence of at least four jets and ${\HT > 320\GeV}$, while for the merged analysis an additional trigger requires the presence of a single high-\et jet, with a threshold of 180\GeV. The absolute L1 trigger efficiency in simulation, relative to the inclusive yield without any selections applied, for the SM \gghh signal in 2022 (2023) was approximately 58 (68)\%.

\subsubsection{The HLT selection}

The most important improvements to the trigger strategy for Run~3 concerned the HLT. One of the most critical steps in the HLT decision is the \PQb tagging, since \PQb jets are expected for the \HHbbbb signal but are a small fraction of the QCD multijet background. The \PQb tagging performance in the trigger system was a major limiting factor of the Run 2 trigger designed for the \HHbbbb analysis in the resolved topology~\cite{CMS:2022cpr}, which required at least four AK4 jets in the event including at least three identified as \PQb jets. The \Pb tagging criteria applied at HLT in Run~2 had a relatively low efficiency and high background rate, significantly limiting the \HHbbbb signal efficiency and restricting the data sample available for the background prediction. The trigger acceptance for high-\pt \hbb events with the merged topology was also limited in Run 2, requiring high jet \pt thresholds that rejected a significant fraction of merged \HHbbbb signal events.

An innovative algorithm was developed for the LHC Run 3, \textsc{pnet@hlt}, customized to perform \PQb tagging at the HLT and to overcome the limitations of the Run~2 trigger criteria. The algorithm is based on \textsc{ParticleNet} (\PNET), a graph neural network (GNN) introduced in Ref.~\cite{Qu:2019gqs}. There are two versions of the \textsc{pnet@hlt} algorithm-- one to identify AK4 \PQb jets and another one to identify AK8 \xbb jets. Both \textsc{pnet@hlt} versions are based on a training sample of jets with simulated HLT reconstruction conditions and were designed to satisfy the strict inference time constraints of the trigger system. 

\begin{figure*}[!htb]
  \centering
    \includegraphics[width=0.45\textwidth]{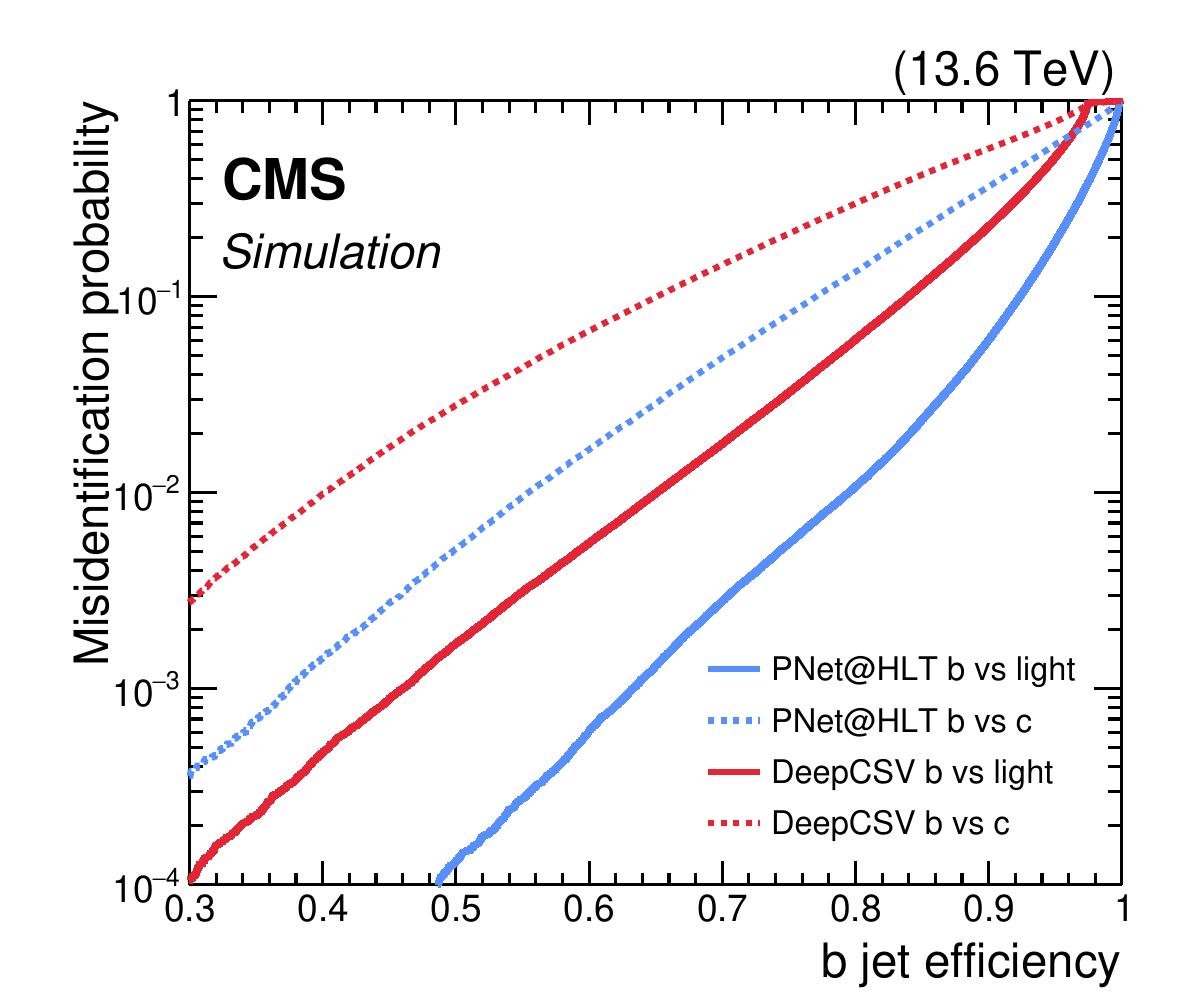}
    \includegraphics[width=0.45\textwidth]{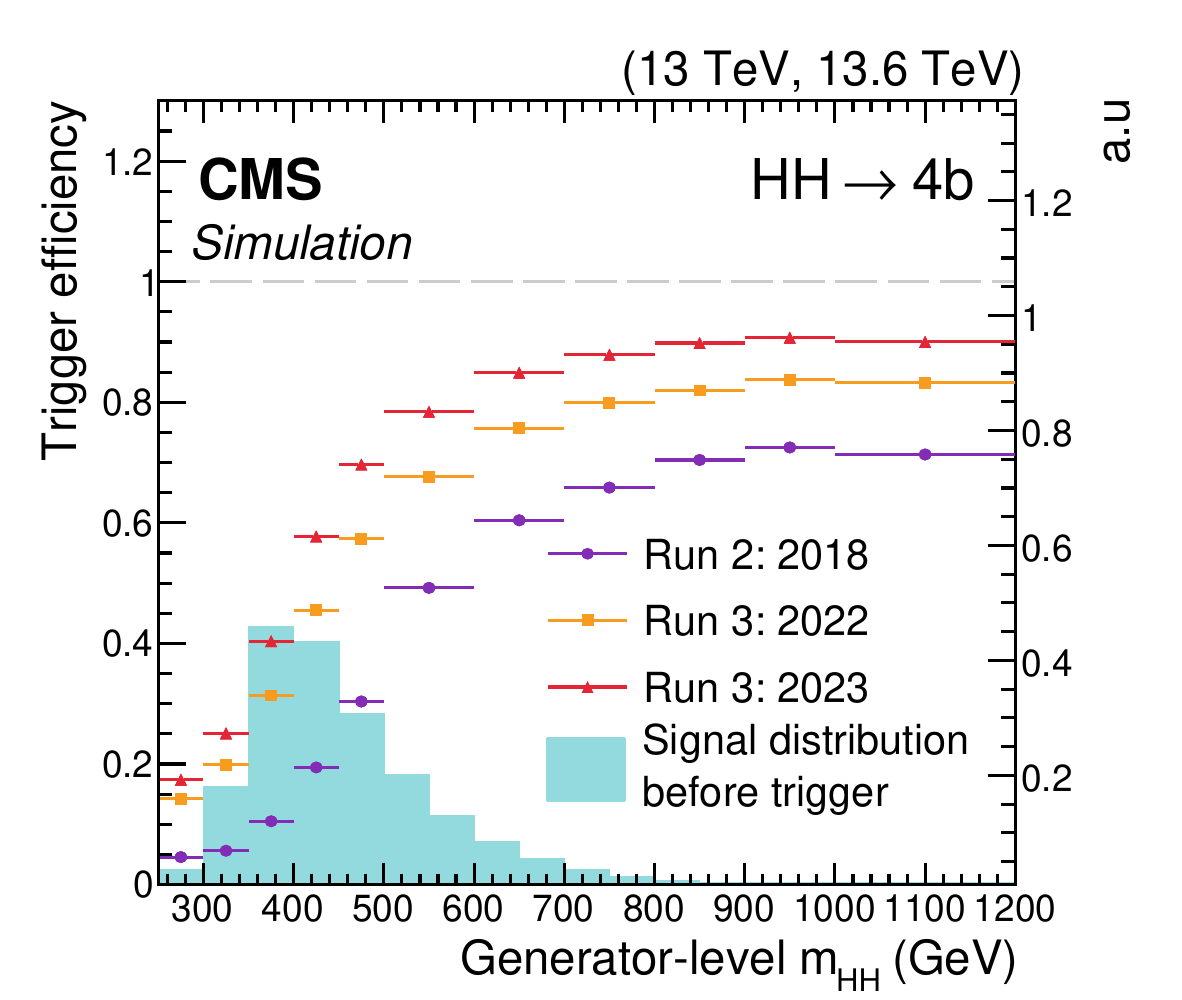}
	  \caption{Left: the \PQb tagging performance of the Run 3 \textsc{pnet@hlt} algorithm (in blue) compared to the best performing algorithm deployed at HLT in Run 2 (\DeepCSV, in red), as evaluated from \ttbar simulation on trigger-level AK4 jets with ${\pt>30 \GeV}$ and ${\abs{\eta}<2.5}$. Right: efficiencies of the triggers targeting the resolved \HHbbbb topology as functions of the generator-level \mhh in simulated SM \HHbbbb signal events in which the generator-level jets from the b quarks produced by \hbb decays have $\pt>25\GeV$ and ${\abs{\eta}<2.5}$. The teal histogram shows the expected distribution for the SM \gghh signal at $\sqrt{s}=13.6\TeV$ prior to any trigger selections, scaled by an arbitrary constant for visibility. 
    }              
    \label{fig:trigger_resolved}
          
\end{figure*}

As shown in Fig.~\ref{fig:trigger_resolved} (left), the AK4 jet \textsc{pnet@hlt} algorithm significantly outperforms previous methods, increasing the background rejection for a given \PQb tagging efficiency by about a factor of 5 with respect to the best performing tagger employed in the HLT during Run 2~\cite{CMS-DP-2023-021}. The AK8 \textsc{pnet@hlt} \hbb algorithm, shown in Fig.~\ref{fig:trigger_merged} (left), is the first tagger that achieves the performance necessary to design a single highly efficient trigger for merged \hbb decays, without relying primarily on other more inclusive hadronic triggers. These major advancements enabled a new design of the selections employed in Run 3 triggers, more than doubling the absolute \HHbbbb signal efficiency~\cite{CMS-DP-23-050}. In addition, the more inclusive event selection made possible by these developments has also enabled a variety of new methods in the background prediction and its validation in the offline analysis.

\subsubsection{Strategy for the resolved topology}\label{sec:trigger_resolved}

At the HLT, events in the resolved topology are required to contain at least four AK4 jets with ${\abs{\eta}<2.5}$, with at least two of them identified as \PQb jets by means of a selection on their \textsc{pnet@hlt} \PQb jet output score, \pb. In 2022, this trigger operated at a data output rate of approximately 60\unit{Hz}, with jet \pt thresholds at 70, 50, 40, and 35\GeV on the four \pt-leading AK4 jets, respectively. In 2023, a larger allocated rate, up to 150\unit{Hz}, was made possible by the development of a dedicated data parking stream, denoted ``ParkingHH''~\cite{CMS:2023gfb,CMS-DP-23-050,CMS:2024zhe}. The ParkingHH trigger was re-optimized with looser jet \pt thresholds, set at 30\GeV for all four leading AK4 jets, as well as a looser selection on the \pb score. With the ParkingHH strategy, the absolute trigger acceptance for the SM \HHbbbb signal increased by about 22\% relative to the 2022 data-taking conditions, at a cost of more than twice the background rate. Figure~\ref{fig:trigger_resolved} (right) shows the efficiencies of these triggers as functions of the generator-level \mhh for the SM \gghh production, for events in which the generator-level jets produced by \hbb decays have ${\pt>25\GeV}$ and ${\abs{\eta}<2.5}$. The trigger efficiency for ${\mhh>1\TeV}$ decreases because of limitations in the track reconstruction for jets with large \pt, where the jet constituents are highly collimated. For the SM \gghh signal, the absolute trigger efficiency for the 2022 (2023) version of the triggers was 37 (45)\%, an increase of more than a factor of two compared to that achieved in Run 2 (18\%).

Corrections to the trigger efficiency, as predicted in simulated events, were measured independently for each trigger selection, following a factorized approach. The efficiencies of jet \pt requirements were measured, as a function of the offline jet \pt, using a ``tag-and-probe`` method~\cite{CMS:2011aa} in CRs enriched in QCD multijet or \ttbar events. The efficiency of the \PNET \PQb tagging selection was assessed, as a function of the offline average \pb score of the two jets with highest \pb, in a high-purity sample of \PQb quark jets in a ${\ttbar \to \Pe\mu \PQb \PAQb}$ enriched CR~\cite{CMS-DP-2023-089}. The ratio between the efficiencies measured in data and simulation defines scale factors (SFs) that are used to correct the simulated \HHbbbb prediction. These SFs are typically consistent with unity, within uncertainties.

\subsubsection{Strategy for the merged topology}\label{sec:trigger_merged}

\begin{figure*}[!htb]
  \centering
    \includegraphics[width=0.45\textwidth]{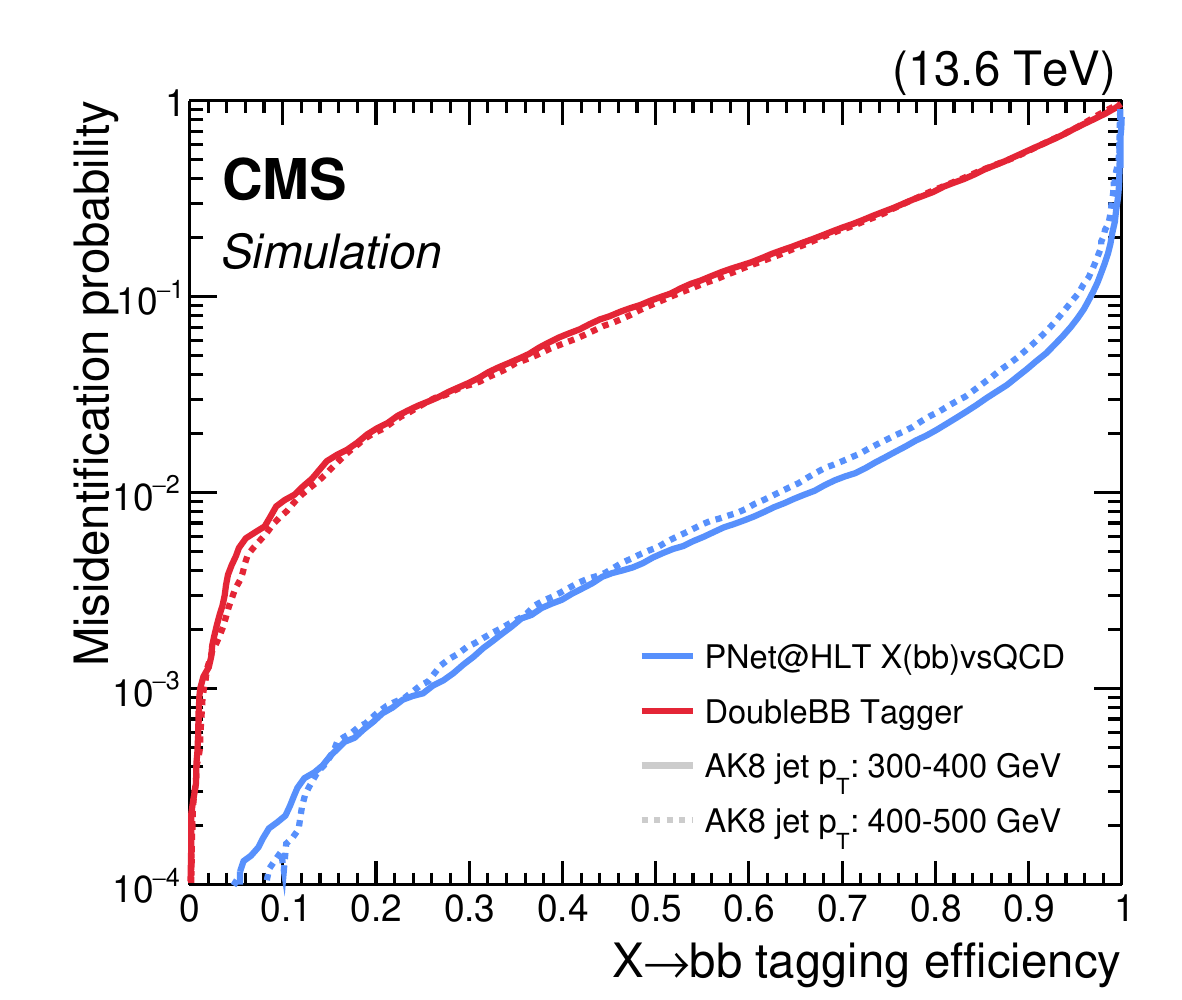}
    \includegraphics[width=0.45\textwidth]{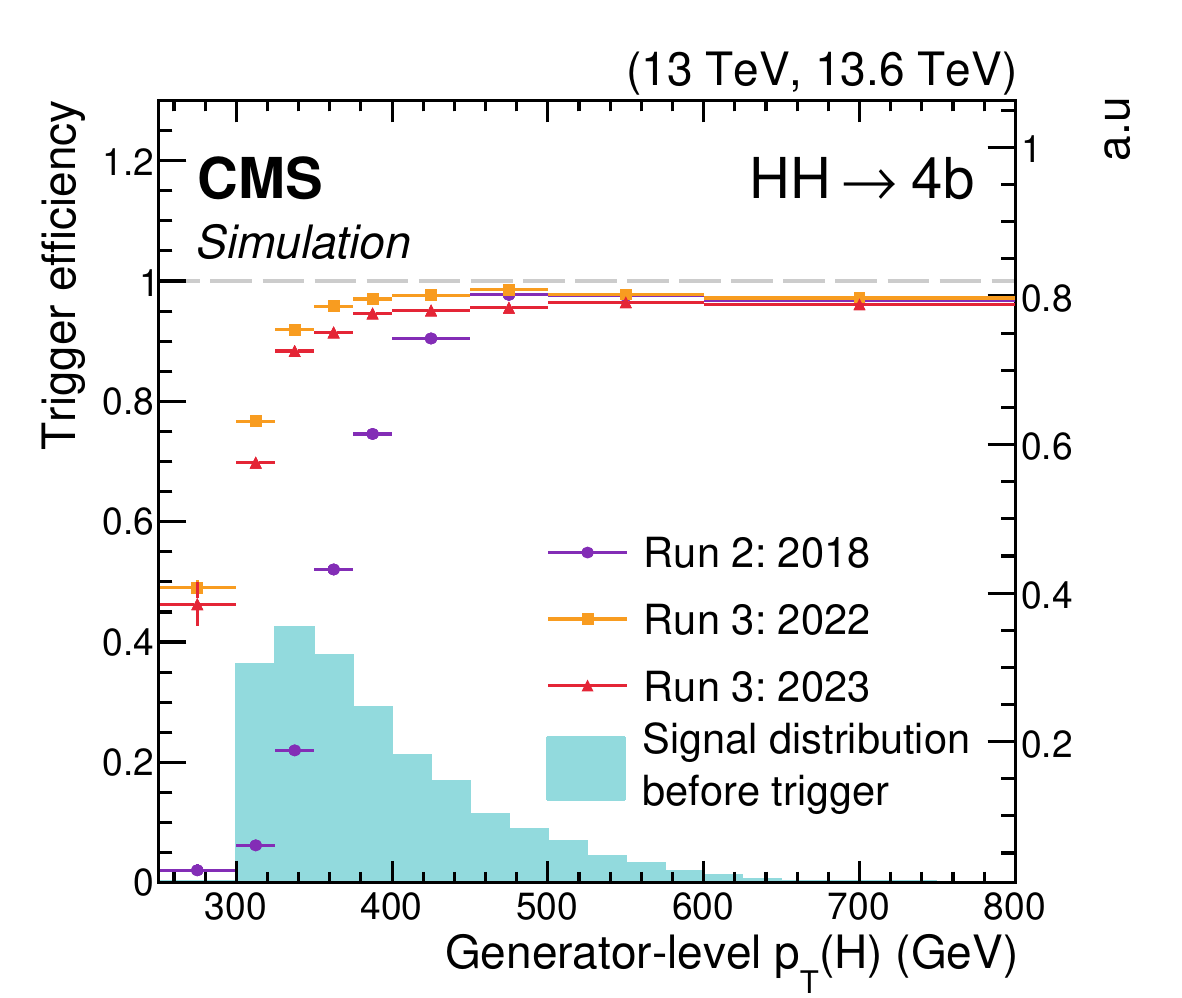}
    \caption{Left: the $\PQb\PAQb$ tagging performance of the \textsc{pnet@hlt} algorithm (in blue) compared to the best performing Run 2 algorithm (\textsc{DoubleBB}, in red), as evaluated on AK8 jets in the HLT from simulated \HHbbbb and QCD multijet events with ${\pt>300 \GeV}$ and ${\abs{\eta}<2.5}$. Right: efficiency of the logical or of the trigger paths developed for the merged topology, as a function of the generator-level leading \PH candidate \pt in simulated SM \HHbbbb events in which ${\Delta R(\PQb,\PAQb)<0.8}$. The teal histogram shows the expected distribution for the SM \gghh signal at $\sqrt{s}=13.6\TeV$ prior to any selections, scaled by an arbitrary constant for visibility.}
    \label{fig:trigger_merged}

\end{figure*}

The \HHbbbb merged analysis targets events in which both the \hbb decays are reconstructed as AK8 jets, a significant fraction of the \HHbbbb signal events once the $\pt(\PH)$ exceed about 200\GeV. In this regime, the sets of trigger-level event selection criteria, referred to as ``trigger paths'', designed for the resolved \HHbbbb topology become inefficient. Therefore, a set of triggers was specifically developed to collect events with at least one merged \hbb candidate. Two trigger paths were developed for the HLT. The first one uses the AK8 version of the \textsc{pnet@hlt} algorithm to identify AK8 jets produced by the decay of a heavy particle to $\PQb\PAQb$ (\xbb), referred to as ``$\PQb\PAQb$ tagging'', following a strategy similar to that used for offline event reconstruction in Refs.~\cite{CMS:2022psv,BTV-22-001}. In 2022, the trigger required at least one AK8 jet with ${\pt>250\GeV}$, $\abs{\eta}<2.5$, a soft-drop~\cite{Larkoski:2014wba} invariant mass ${\mSD>40\GeV}$, and a selection on the \textsc{pnet@hlt} $\PQb\PAQb$ tagging score (\pbb). The \textsc{pnet@hlt} training was improved for 2023 data taking and the trigger was updated for 2023 by lowering the \pt threshold from 250 to 230\GeV, as well as relaxing the requirement on the \pbb score. In addition, to recover a roughly 5\% signal inefficiency in the \pbb requirement, a second trigger path requiring at least one AK8 jet with ${\pt>420\GeV}$, ${\abs{\eta}<2.5}$, and ${\mSD>40\GeV}$ was included in both 2022 and 2023. Figure~\ref{fig:trigger_merged} (right) shows the efficiency of the 2022 (orange) and 2023 (red) merged-topology triggers compared to the Run 2 (purple). The Run 3 triggers are nearly fully efficient for the merged \HHbbbb signal, in which ${\Delta R(\PQb,\PAQb)<0.8}$ for both \PH decays, with an absolute efficiency for the SM merged \HHbbbb process above 90\%, a significant improvement with respect to Run 2. The Run~3 efficiency gain is concentrated in the region with leading $\pt(\PH)<500 \GeV$, which constitutes the majority of the merged \HHbbbb signal events and was accessible during Run~2 only through a mix of partially efficient trigger paths. The trigger efficiency for $\pt(\PH)>500\GeV$ degrades slightly because of inefficiencies in the soft-drop mass reconstruction for large jet \pt.  

The Run 2 merged-topology trigger selection~\cite{CMS:2022gjd} was a combination of several trigger paths requiring large \HT and high-\pt AK8 jets, with \pt thresholds in the range of 330--400\GeV. The excellent performance during Run~3 of the HLT $\PQb\PAQb$ tagging algorithm enabled a major simplification in the trigger strategy used for the \HHbbbb merged topology, lowering significantly the online jet \pt threshold and collecting the bulk of the merged \HHbbbb signal with a single trigger path. The Run 2 analysis furthermore relied on the precise modeling of trigger efficiency turn-on curves, whereas in Run 3 the efficiency turn-on has little impact since the trigger efficiency is typically high for the targeted events.

Corrections to the predicted trigger efficiency from simulated samples are measured for each trigger requirement independently. Because of the challenge in isolating \zbb events from the large hadronic backgrounds, jets from ${\PW\to\mathrm{\PQq\PQq'}}$ decays in \ttbar events are used as a proxy of the merged \hbb signal jets. Such events are collected by a reference trigger requiring one muon with ${\pt>55\GeV}$. It is additionally required that the muon and the closest \PQb-tagged jet recoil against an AK8 jet with ${\pt>200\GeV}$ and ${\mSD>30\GeV}$. This selection provides a pure sample of AK8 jets containing the decay products of high \pt ${\PW\to\mathrm{\PQq\PQq'}}$ decays and, for jets with high \pbb score, merged top decays in which the jet contains both the \PQb quark and the \PW boson decay products. The efficiency of the \pt and \mSD requirements is measured as a function of the offline AK8 jet \pt and \mSD in both data and simulation. The efficiency of the \pbb selection is measured as a function of the offline \pbb score for AK8 jets with ${\pt>300\GeV}$ and ${\mSD>50\GeV}$. The ratio between the efficiencies measured in data and simulation defines a SF that is used to correct the simulated \HHbbbb prediction in the offline analysis. This procedure assumes that the relative simulation-to-data efficiency corrections as measured for ${\PW\to\mathrm{\PQq\PQq'}}$ jets can be applied to the simulated \hbb signal.

\subsection{Small-radius jet flavor identification}\label{sec:ak4btag}

The offline AK4 \PQb jets are identified using the \PNET~\cite{Qu:2019gqs, CMS-DP-2024-024} algorithm as developed for \PQb tagging. Three \PQb tagging working points (WPs) are used, the loose (L), medium (M), and tight (T), corresponding to misidentification rates for light-flavor quark and gluon jets of about 10\%, 1\% and 0.1\%, and \PQb tagging efficiencies of approximately 92\%, 83\%, and 70\%, respectively. For the analysis using Run 2 data, discussed in Section~\ref{sec:resolved_run2}, jets are tagged with the \DeepJet~\cite{Bols:2020bkb, CMS-DP-2023-005} \PQb tagging algorithm. For a fixed misidentification rate, \PNET improves the absolute \PQb tagging efficiency with respect to the Run 2 \DeepJet algorithm by 3--7\%, depending on the WP and jet \pt. The \PQb tagging efficiency in simulated events is corrected to match the one observed in data with correction factors applied to the simulation~\cite{CMS-DP-2024-025, CMS-DP-2023-005}. These correction factors are derived in samples enriched in \ttbar, \Zjets, and QCD multijet events with muons~\cite{CMS-DP-2024-025, CMS-DP-2023-005} and as function of the \pt, $\eta$, and flavor of the initial quark or gluon. The selection of \PQb jets for the different analysis regions is further discussed in Section ~\ref{sec:resolved_analysis}.

\subsection{Small-radius jet \texorpdfstring{\pt}{pt} reconstruction}\label{sec:ak4_btag_pt_regression}

It is also important to infer the \PQb quark momentum from the reconstructed AK4 jet as precisely as possible, a difficult task because of the presence of neutrinos from semileptonic \PQb hadron decays and limitations in the reconstruction of the jet constituent particles. The CMS Run 2 search employed a deep neural network (DNN) based algorithm to improve the \PQb jet \pt estimate~\cite{CMS:2019uxx}, correcting for such effects. The regression aimed to infer the generator-level jet \pt, defined as the \pt of the corresponding jet clustered from the constituent particles predicted at generator level, including neutrinos. This network was trained on a sample exclusively containing \PQb-tagged jets produced in \ttbar events, considerably improving the \PQb jet \pt scale and resolution over the jet reconstruction baseline.

In this paper, we introduce a novel approach to jet \pt reconstruction where a single algorithm simultaneously performs the jet flavor identification and estimates a jet \pt correction. This is achieved by training the \PNET algorithm described in the context of \PQb tagging in Section~\ref{sec:ak4btag} to minimize a loss function defined as the sum of two separate terms: one loss term for learning the differences in properties among \PQb jets and other jet classes, and another loss term for learning the relative correction necessary to adjust the reconstructed jet \pt to match the generator-level \pt. The \PQb tagging performance achieved by this network is summarized in Section~\ref{sec:ak4btag}. 

This new approach, referred to as the ``jet \pt regression'', differs significantly from prior methods in that it is not trained on any specific jet flavor sample (\PQb, c, etc.), but rather learns the jet flavor for each jet and tailors the \pt correction for jets of that flavor type. This makes the \pt regression a general method that can be applied to jets of any flavor and prior to any particular \PQb tagging selection, in contrast to previous approaches. The regression is used to correct the jet \pt prior to the application of JES or jet energy resolution (JER) corrections, making it possible to deploy the regression universally for all jets reconstructed in CMS across data-taking periods. The new approach improves the jet \pt resolution by 10--25\%, depending upon the jet flavor and \pt, and reduces the flavor dependence of the jet energy corrections, an important source of uncertainty.  

\begin{figure}[!htb]
  \centering
    \includegraphics[width=0.45\textwidth]{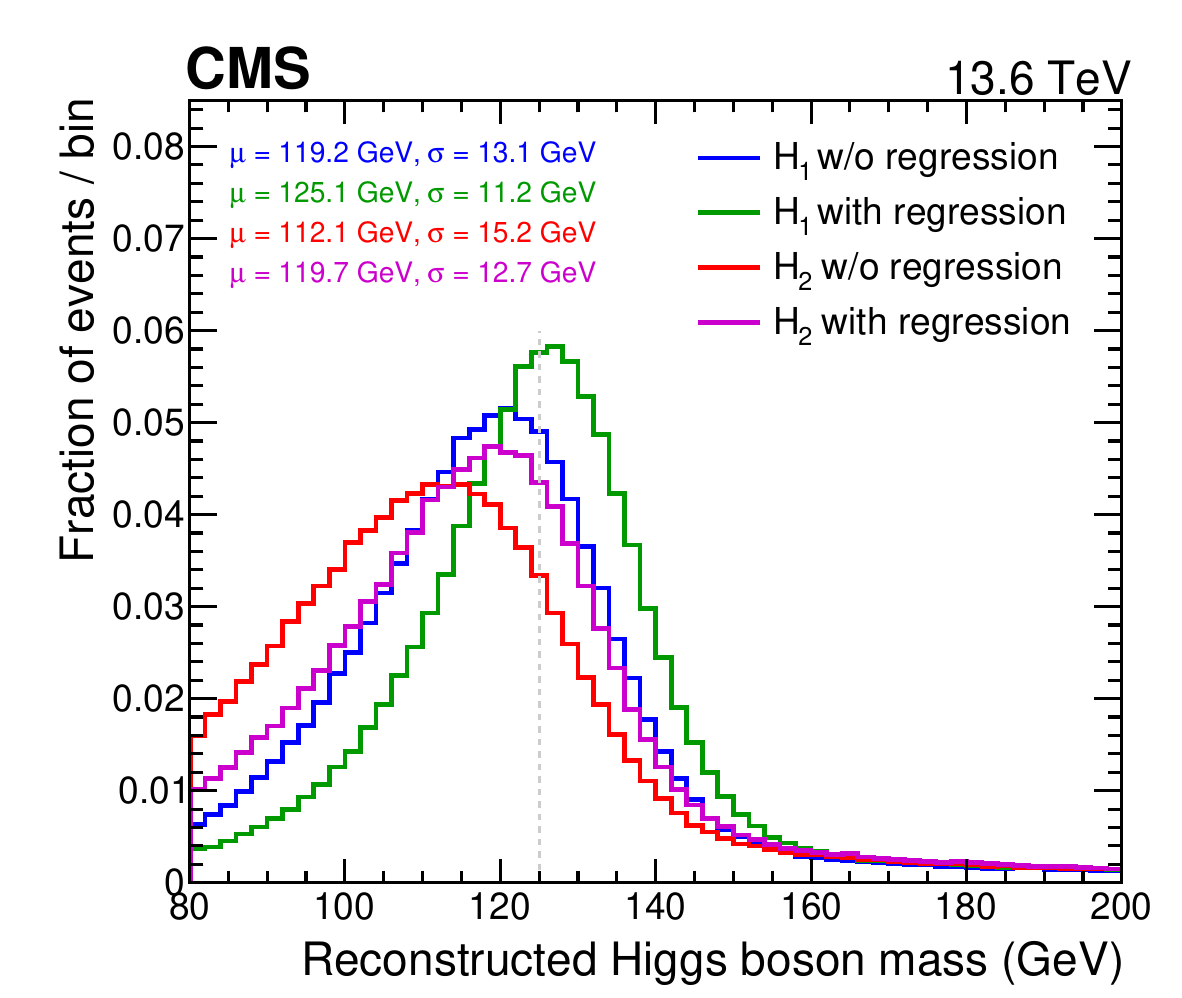}
    \includegraphics[width=0.45\textwidth]{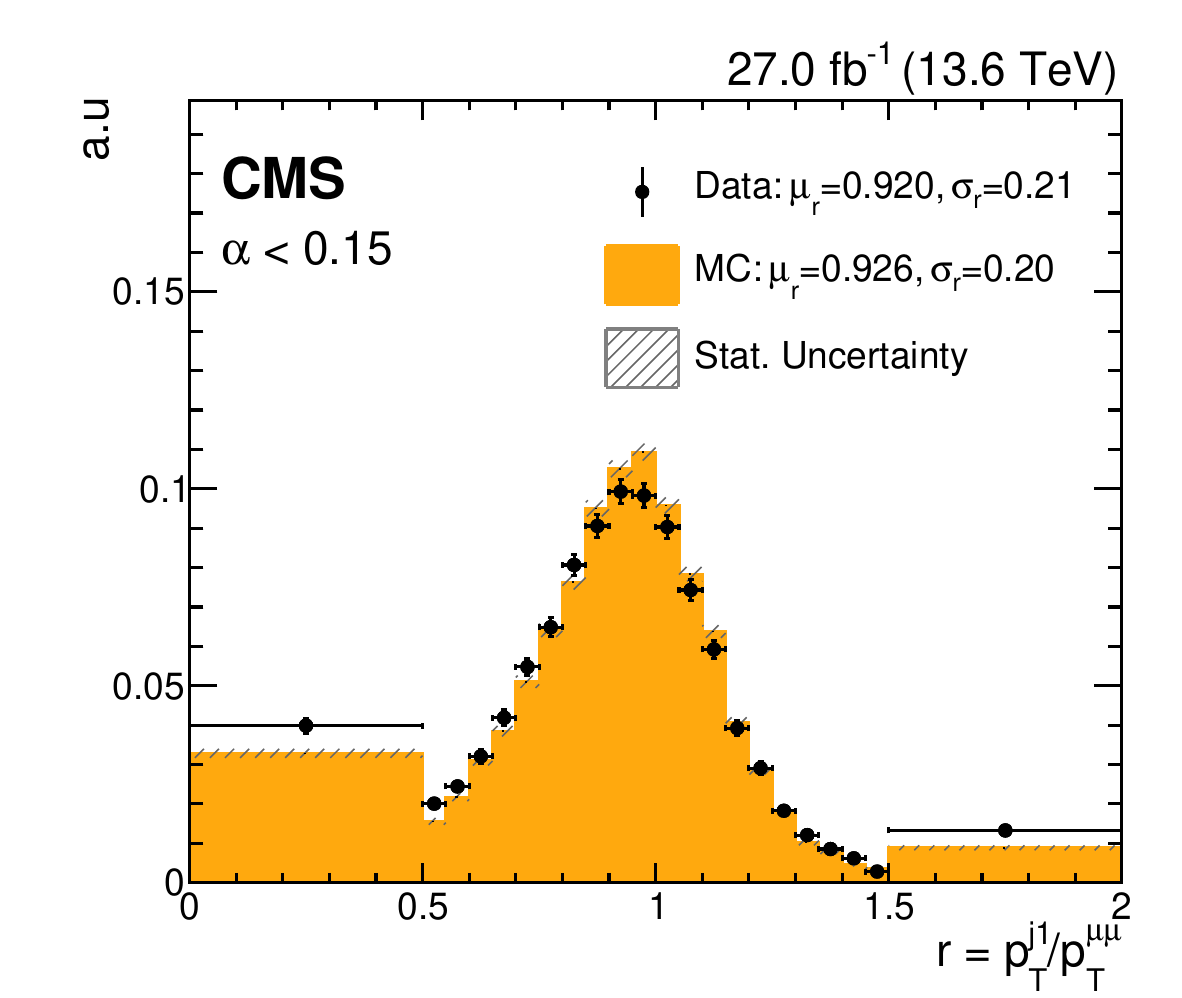}
    \caption{\cmsLLeft: invariant mass distributions for the leading ($m_{\Hone}$)  and subleading ($m_{\Htwo}$) \pt \PH candidates in SM \HHbbbb events obtained before and after the application of the \PNET jet \pt regression. \cmsRRight: distribution of the \pt balance, $r=\pt(\mathrm{j_{1}})/\pt(\PGm\PGm)$, in 2023 data and simulated events for the selected $\PZ(\PGm\PGm)+\PQb\text{ jet}$ region with $\alpha=\pt(\mathrm{j_{2}})/\pt(\PGm\PGm)<0.15$, obtained after applying the \PNET jet \pt regression. The $\mu$ values quoted in the legend correspond to the mean of  the distributions.}
    \label{fig:ak4_jet_regression}

\end{figure}

The invariant mass distributions of the reconstructed \pt-leading (\Hone) and subleading (\Htwo) \PH candidates in simulated SM \HHbbbb events are shown in Fig.~\ref{fig:ak4_jet_regression} (\cmsLeft), with and without application of the regression. The means ($\mu$) and widths ($\sigma$) of the distributions are obtained from a parametric fit with a Double-sided Crystal Ball function. The regression improves the \hbb mass resolution for each \PH candidate in this analysis by about 20\% and better aligns the reconstructed mass scale to that at the generator level. This is about a 10\% relative improvement in performance compared to the previous \PQb jet \pt regression method employed by CMS~\cite{CMS:2019uxx}.

\subsubsection{Validation in data}

Dedicated corrections to the jet \pt response after applying the regression are derived following the general CMS jet calibration procedures described in the introduction of this section. However, these corrections are derived with jet samples dominated by jets originating from light-flavor quark and gluon jets. It is therefore important to further validate these corrections and the effect of the jet \pt regression on a data sample with a high purity of \PQb jets.

The corrections are validated in events where one \PQb jet, passing the \PNET medium WP with ${\pt(\mathrm{j_{1}})>30\GeV}$ and ${\abs{\eta}<2.5}$, recoils against a Z boson that decays into a $\mu^{+}\mu^{-}$ pair with ${\pt(\PGm\PGm)>100\GeV}$. Since the \pt of muons is measured with high precision by the CMS detector, the ratio of the reconstructed \PQb jet \pt to that of the \PZ boson, $r=\pt(\mathrm{j_{1}})/\pt(\PGm\PGm)$, allows for a validation of both the \PQb jet \pt scale and resolution. The mean value of $r$ ($\mu_{r}$) is sensitive to the jet \pt scale while its spread ($\sigma_{r}$), defined as the half-difference between the 84\% and 16\% quantiles of the distribution, is correlated to the \pt resolution. Events are categorized according to a measure of the additional jet activity, ${\alpha=\pt(\mathrm{j_{2}})/\pt(\PGm\PGm)}$, where $\pt(\mathrm{j_{2}})$ refers to the maximum \pt among additional AK4 jets in the event with ${\pt>15\GeV}$ and ${\abs{\eta}<4.7}$, if any ($\alpha=0$ otherwise). The smaller the value of $\alpha$, the more similar the \PQb jet \pt is expected to be to the $\pt(\PGm\PGm)$ and therefore the higher the correlation of $\mu_{r}$ and $\sigma_{r}$ with the true \PQb jet \pt scale and resolution, respectively.

The distribution of $r$ for $\PZ(\PGm\PGm)+\PQb\text{ jet}$ selected events in 2023 data with $\alpha<0.15$ is shown in Fig.~\ref{fig:ak4_jet_regression} (\cmsRight). The compatibility between data and simulation in $\mu_{r}$ was assessed independently for all data-taking periods, with the largest nonclosure observed to be 1--2\%. An additional 2\% uncertainty in the post-regression \pt scale in simulation is assigned to cover for any such nonclosure. The \pt resolution in data and simulation is assessed by measuring $\sigma_{r}/\mu_{r}$ for progressively tighter selections on $\alpha$, ranging from $\alpha<0.3$ to $\alpha<0.15$. The \pt resolution in data is observed to be wider than that of simulation by 4 to 8\%, depending on the data-taking period. These values are used to assess an additional uncertainty in the post-regression jet \pt resolution in simulation.

\subsection{Large-radius jet flavor identification and mass reconstruction}\label{sec:glopart}

To select \hbb candidates from offline AK8 jets and reconstruct their mass, the global particle transformer (\GloParT) algorithm~\cite{Qu:2022mxj,CMS:2026uph} is used.
The \GloParT algorithm generalizes the jet identification and mass regression tasks performed by its predecessor, \PNET~\cite{Qu:2019gqs,CMS-DP-2020-002,CMS-DP-2021-017}, by expanding the number of training classes to include a variety of jets from \PH and top quark decays with a wide range of \pt and mass values.

The training of \GloParT adopts the \PNET~\cite{CMS:2022psv} jet mass decorrelation approach, where the jets are selectively sampled so that the training sample is approximately uniformly distributed in jet \pt and mass. The \GloParT algorithm is also trained to regress a correction to the jet mass based on a generator-level definition. For final states originating from \PH and top quark like resonances, the generator-level mass corresponds to the invariant mass of the visible event-generator-level particles that initiate the jet.
For QCD jets, this mass corresponds to the \mSD of the corresponding jet clustered using event-generator-level particles excluding neutrinos~\cite{CMS-DP-2021-017}. 

Three types of inputs are considered: neutral and charged PF candidates, ``lost tracks'' not associated with a charged PF candidate, and secondary vertices. The \GloParT algorithm uses the state-of-the-art particle transformer architecture~\cite{Qu:2022mxj}, which processes input particles and secondary vertices as a permutation-invariant point cloud, updating each object's features through a series of particle attention blocks that consider pairwise features between the inputs~\cite{CMS:2026uph}.

The outputs for each jet correspond to the probabilities for the jet to originate from each of the classes as well as additional mass regression outputs.
To evaluate the performance of the tagger, a single discriminant differentiating between \hbb jets and QCD jets is defined as
\begin{equation*}\label{eq:thww}
    \TXbb = \frac{\Prob(\xbb)}{\Prob(\xbb) + \Prob(\text{QCD})},
\end{equation*}
where $\Prob(\xbb)$ and $\Prob(\text{QCD})$ are the predicted probabilities of the respective categories.

\begin{figure}[!htb]
    \centering
    \includegraphics[width=0.425\textwidth]{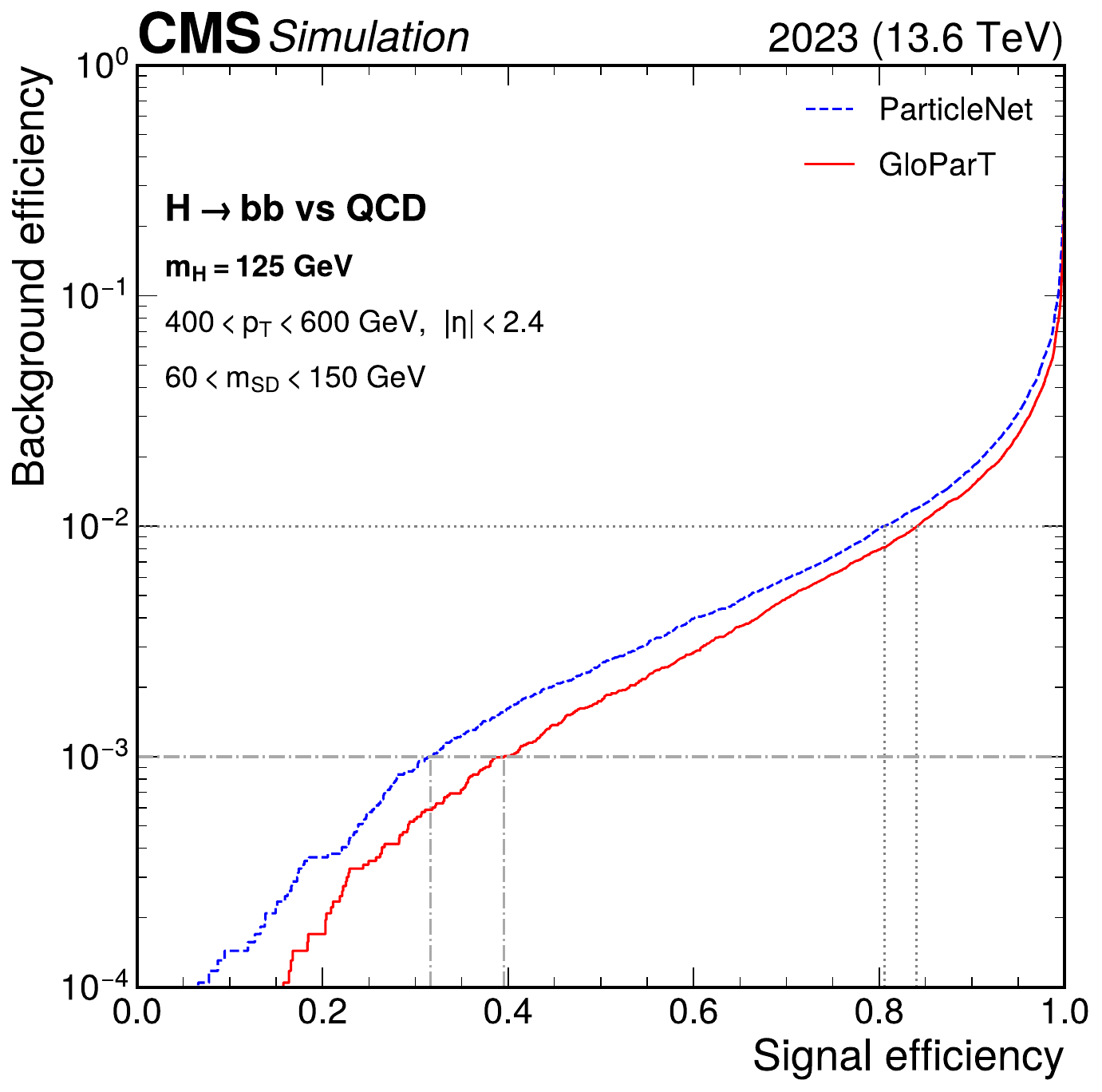}
    \includegraphics[width=0.425\textwidth]{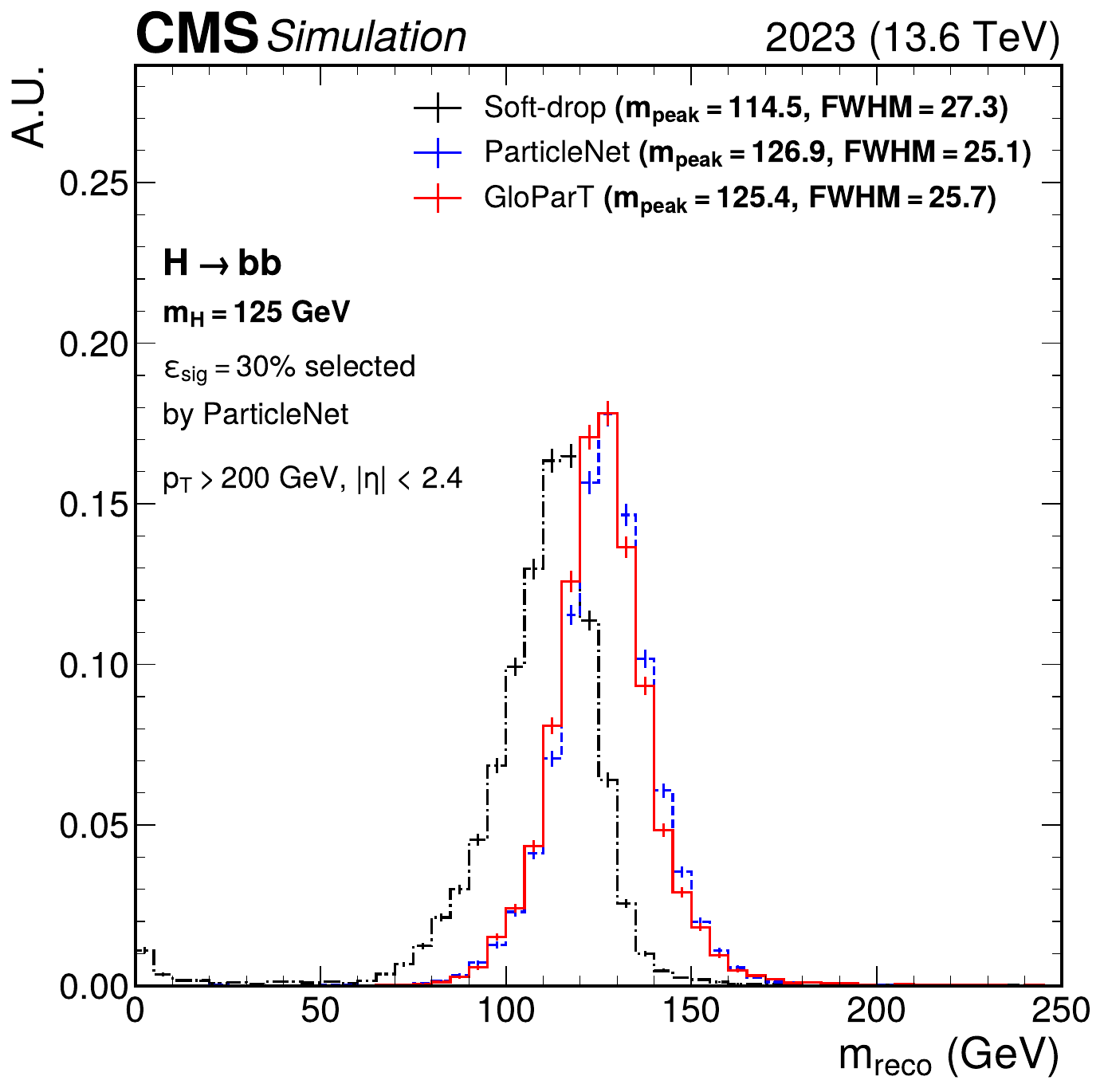}
    \caption{\cmsLLeft: the receiver operating characteristic curve for \GloParT and \PNET for discriminating \hbb from QCD jets with ${400 < \pt < 600\GeV}$, ${\abs{\eta} < 2.4}$, and ${60 < \mSD < 150\GeV}$. The dotted and dash-dotted gray lines denote the signal efficiency at background efficiencies of 1 and 0.1\%, respectively. \cmsRRight: the mass regression performance for \GloParT and \PNET for jets with ${\pt > 200\GeV}$, ${\abs{\eta} < 2.4}$, and satisfying a \PNET selection corresponding to 30\% signal efficiency. The conditions correspond to those during data collection in 2023.}
    \label{fig:glopart}
\end{figure}

For \hbb jets in simulation, \GloParT provides about a 25 (5)\% relative signal efficiency improvement compared to \PNET in the classification against QCD jets at a fixed 0.1 (1)\% background efficiency for ${400 < \pt < 600\GeV}$, ${\abs{\eta} < 2.4}$, and ${60 < \mSD < 150\GeV}$, as shown in Fig.~\ref{fig:glopart} (\cmsLeft). The mass distribution peaks at 125\GeV for \hbb jets and is shown in Fig.~\ref{fig:glopart} (\cmsRight). The \GloParT algorithm has a slightly improved mass regression performance compared to \PNET, and improves the mass resolution, quantified as the full width at half maximum, by about 9\% compared to the \mSD, as predicted by simulated MC signal and background events. 

\subsubsection{Calibration with \texorpdfstring{\zbb}{Zbb} data}\label{sec:zbb}

A dedicated measurement of $\PQb\PAQb$ tagging efficiency SFs is performed using \zbb jets as a proxy for \hbb jets. A sample of \zbb events is selected via a combination of jet-based triggers, without imposing any requirements on the \TXbb scores of AK8 jets. The trigger path with highest efficiency requires one AK8 jet with ${\pt>420\GeV}$ and ${\mSD>40\GeV}$. Offline \zbb candidate events are required to have at least two AK8 jets with leading (subleading) ${\pt>450}~(200)\GeV$ and ${\Delta \phi>\pi/2}$. This selection ensures that the selected events correspond to the trigger efficiency plateau. The AK8 jet with the largest \TXbb score is chosen as the \zbb candidate. The expected \zbb decays are taken from \Zjets events simulated at LO with NLO QCD and EW corrections from Ref.~\cite{Lindert:2017olm}. Events are then further divided into three exclusive categories corresponding to the \GloParT purity selections previously described. For each category, a $\PQb\PAQb$ tagging efficiency SF is defined as:
\begin{equation*}
        \mathrm{SF}=\frac{\varepsilon_{\text{data}}}{\varepsilon_{\mathrm{MC}}}=\frac{N^{\text{data}}_{\text{pass}}}{N^{\mathrm{MC}}_{\text{pass}}} \frac{N^{\mathrm{MC}}_{\text{tot}}}{N^{\text{data}}_{\text{tot}}}=\frac{\mu_{\text{pass}}}{\mu_{\text{tot}}},
\end{equation*}
where $\mu_{\text{tot}}$ and $\mu_{\text{pass}}$ are defined based on the number of \zbb decays measured without and with the corresponding tagging requirements, respectively. 

Due to the overwhelming QCD multijet background before any $\PQb\PAQb$ tagging requirement, as well as contamination from \PW and \PZ boson decays to lighter quarks, the measurement of the \zbb signal strength in the inclusive selected sample ($\mu_{\text{tot}}$) is very difficult. The $\mu_{\text{tot}}$ can instead be measured using $\PZ(\PGm\PGm)+\text{jets}$ events as a proxy~\cite{ATL-PHYS-PUB-2021-035}. To mimic the region enriched with \zbb decays, \zmm candidates are selected by requiring two muons with leading (subleading) ${\pt > 50}~(30)\GeV$, an invariant mass ${75<m_{\PGm\PGm}<105\GeV}$, opposite charge, and ${\pt(\PGm\PGm)>450\GeV}$. In addition, at least one AK8 jet not overlapping with the two muons, $\Delta R(\mu,\text{j})>0.8$, must be present in the event with ${\pt>200\GeV}$. Efficiency SFs are  extracted from a simultaneous binned likelihood fit to the observed $m_{\PGm\PGm}$ distribution in the \zmm region and the \GloParT regressed jet mass distributions for the \zbb selected events in three mutually exclusive categories--medium purity (MP), high purity (HP), and very high purity (VHP)--each containing approximately 10--15\% of the inclusive merged \zbb signal.

The distributions for signal (\zbb, \zmm) and background processes are modeled via parametric functions. In the \zmm event category, both the \zmm signal and the total background are parameterized with the convolution of a Breit--Wigner and a double-sided Gaussian function summed with a second-order polynomial, which allows for a more accurate description of the tails. The same parameterization is used for the \zbb signal in the various purity-based categories. In each \zbb category, the background is dominated by QCD multijet production, for which the \GloParT regressed jet mass distribution is described using Chebyshev polynomials. The order of the polynomial is chosen, independently for each tagging category, via a Fisher F-test~\cite{10.1111/j.2397-2335.1922.tb00768.x}. 

Predictions for the \zbb and \zmm rates are affected by several experimental and theoretical uncertainties related to the renormalization ($\mu_{\text{R}}$) and factorization ($\mu_{\text{F}}$) scales, and PDF variations, NLO EW corrections, integrated luminosity, pileup, trigger efficiency corrections, JES, JER, and muon identification. Differences between data and simulation in the scale and resolution of the regressed mass, as well as the \pt scale and resolution of the muons, are accounted for by allowing the parameters defining the \zbb and \zmm peak positions and widths to adjust in the fit. In addition, in the \zbb categories both the normalization and the coefficients of the Chebyshev polynomials are floated in the fit, constrained by data in the mass sidebands.

\begin{figure*}[!htb]
  \centering
    \includegraphics[width=0.32\textwidth]{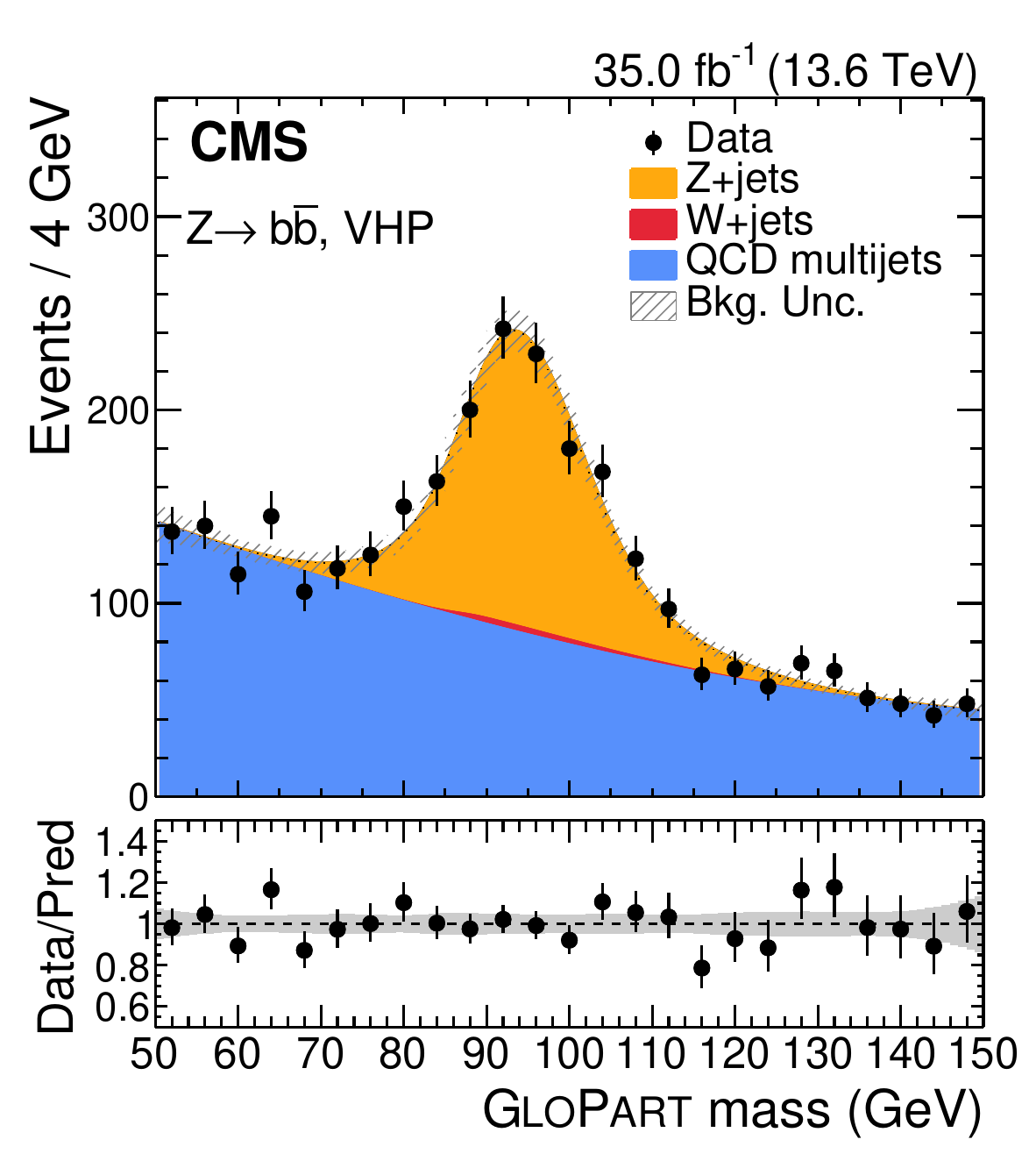}
    \includegraphics[width=0.32\textwidth]{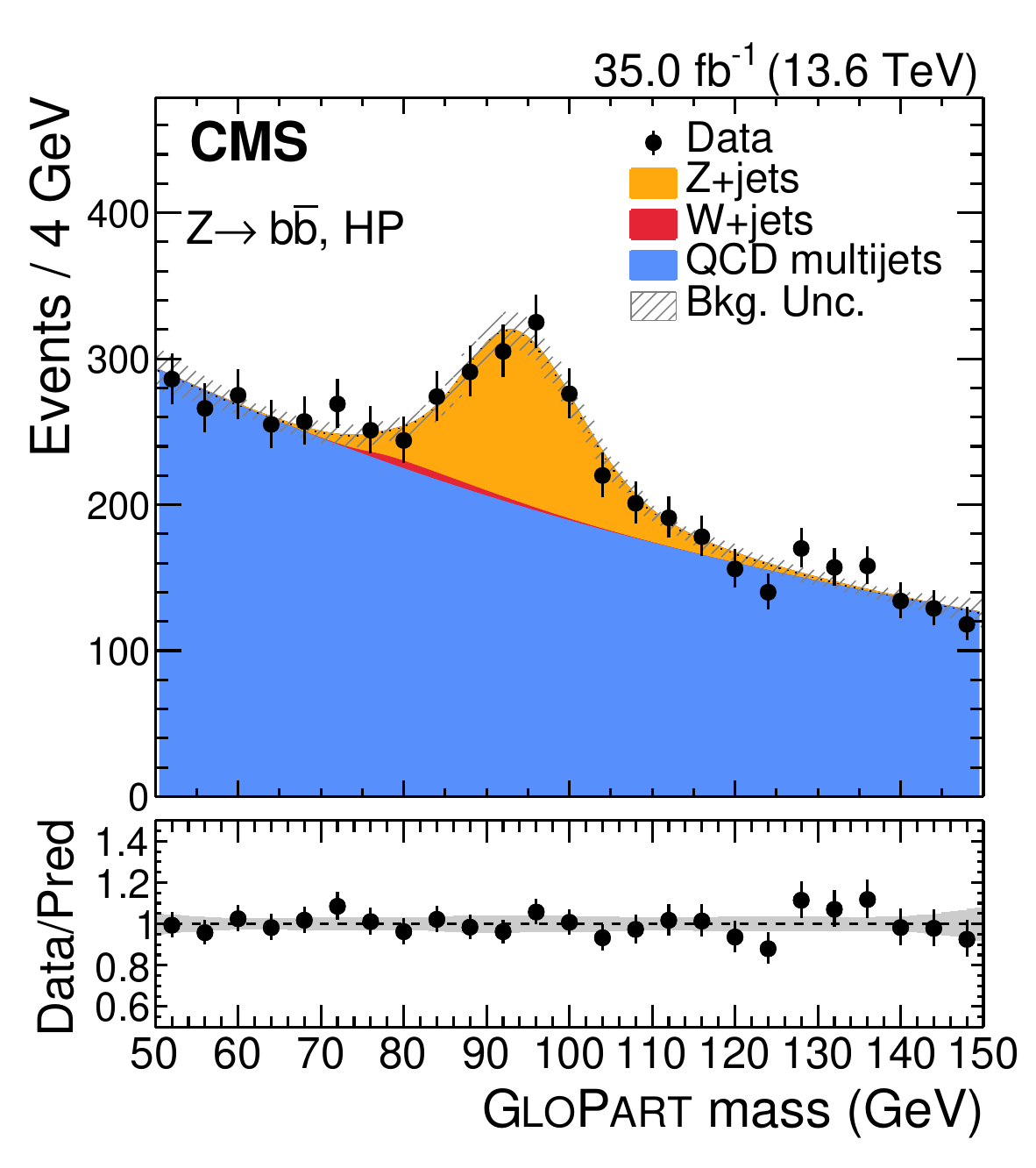}
    \includegraphics[width=0.32\textwidth]{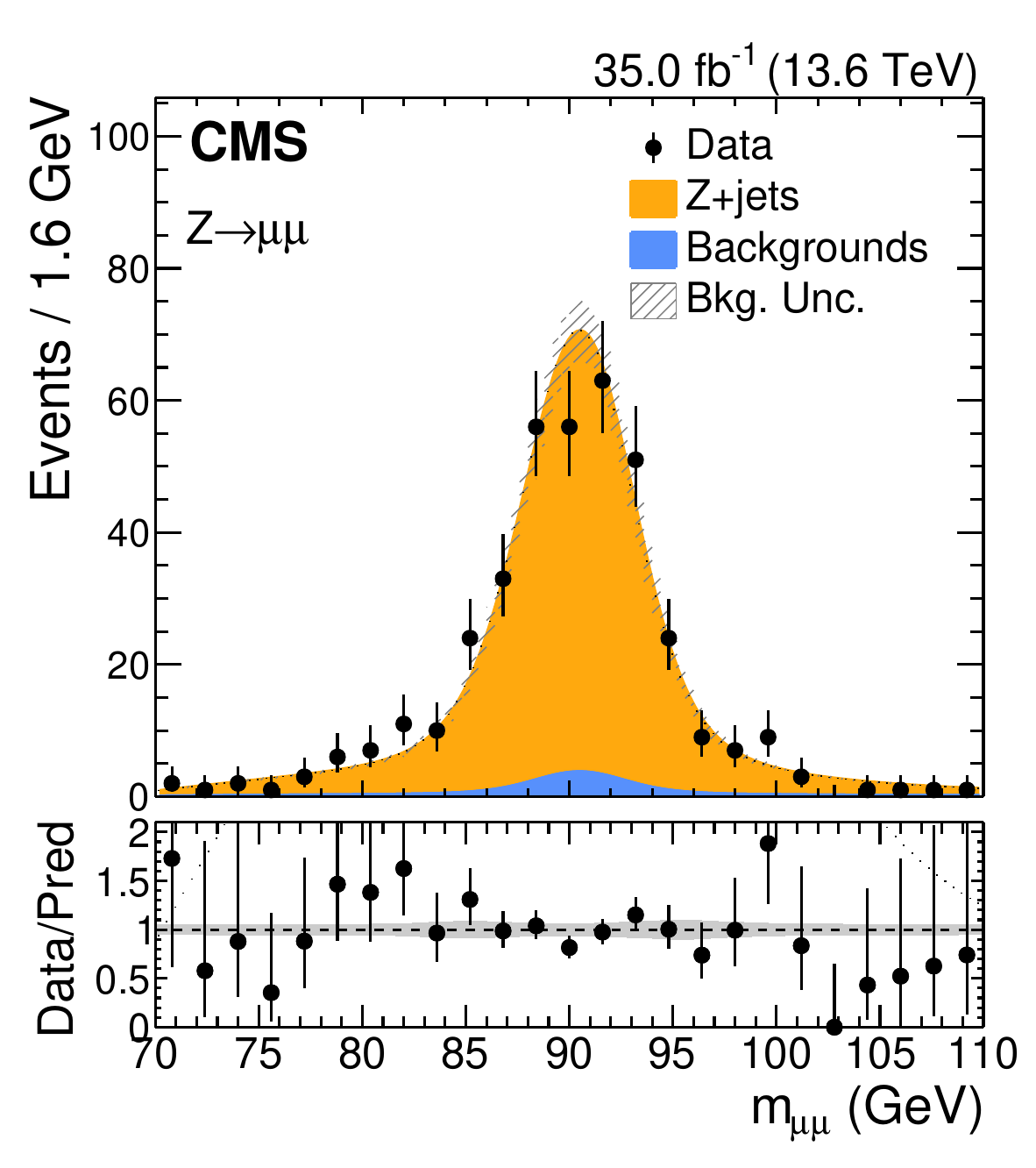}
    \caption{Comparison between data and fit prediction from a simultaneous signal-plus-background fit to the \GloParT regressed jet mass distributions in \zbb VHP (left) and HP (middle) categories. Data corresponds to the full integrated luminosity collected with the CMS detector in 2022. The same comparison is performed for the $m_{\PGm\PGm}$ distributions in the \zmm region (right). Each process considered in the fit is modeled via a parametric function, as described in the text.}
    \label{fig:sf_glopart_zbb_zmm_distributions}

\end{figure*}

Figure~\ref{fig:sf_glopart_zbb_zmm_distributions} shows data and post-fit distributions obtained with 2022 data for events in the \zbb VHP (left), HP (middle), and \zmm (right) categories. The \GloParT \xbb efficiency SFs for both 2022 and 2023 data-taking periods are found to be compatible with unity, within an uncertainty ranging from 10\% to 25\%. The \GloParT regressed mass scale (JMS) and resolution (JMR) corrections are found to be compatible with unity, with an uncertainty of 0.5--2\% for JMS and 7--10\% for JMR depending on the purity category. An alternative measurement of JMS and JMR corrections was also performed using high-\pt ${\PW\to\mathrm{\PQq\PQq'}}$ samples selected from \ttbar events including one semileptonic top quark decay. Results are consistent with the \zbb measurement within the uncertainties. 

The \HHbbbb merged analysis explores final states in which the AK8 jets from \hbb decays have ${\pt>300\GeV}$. Unfortunately, measuring corrections for jets with ${\pt<450\GeV}$ via this method is not possible because of trigger limitations. Therefore, corrections for lower-\pt jets in each \xbb purity category are fixed to unity with a relative uncertainty equal to that measured with ${\pt>450\GeV}$ jets. The uncertainties are uncorrelated between the high- and low-\pt regions and among the purity based WPs. For jets with \TXbb score outside the range explicitly calibrated, the correction is fixed to unity and the relative uncertainty is set to twice that measured for the closest calibrated \TXbb interval. 

\section{The resolved topology}\label{sec:resolved_analysis}

\subsection{Overview} \label{sec:resolved_common}

Despite the major Run 3 advancements in trigger selection, \PQb jet identification, and jet \pt reconstruction, it remains very difficult to disentangle the small number of expected \HHbbbb events from the background. A powerful tool in separating signal from background is ML-based event classification, which aims to order the selected events by the expected signal-to-background (S/B) ratio. Each event is assigned a score, \psig, ranging from zero to one, with the most signal-like events concentrated near scores of one. The presence of the \HHbbbb signal then manifests as an excess of events in data over the background-only expectation at high values of \psig. Because of the enormous background, measuring the \HHbbbb process requires a highly precise and accurate background prediction, particularly for the high-score region of the \psig distribution. The aim of the resolved analysis is to control the background prediction to within about 5\%, approximately the size of the expected signal contribution, in order to ensure sensitivity to the \HHbbbb signal. It is very difficult to achieve the accuracy and precision required for this analysis using simulated samples for the background processes, particularly for the dominant QCD multijet background. It is thus necessary to employ a background estimation method fully based on CRs in data. This technique is described in detail in this section, after an overview of the analysis strategy and a summary of the event selections applied. 

The resolved analysis targets both the dominant \gghh and subdominant \qqhh production modes. Thresholds on the reconstructed AK4 jet \pt are applied similar to the trigger-level selections, ensuring that the selected events lie near the trigger efficiency plateau. The average of the two highest \PQb tag scores among the selected jets is required to be at least that of the equivalent trigger-level selection. Events with a muon or electron candidate with ${\pt>10\GeV}$ passing loose identification and isolation criteria are rejected~\cite{CMS:2018rym, CMS:2020uim}. After applying these criteria, the selected events are dominated by the QCD multijet processes, with a subdominant \ttbar contribution of about 10\%. Other backgrounds, such as single \PH production, {\PV}+jets production, and \VV processes contribute less than 1\% to the total expected background.

A key feature in discriminating the \HHbbbb signal from the QCD and \ttbar backgrounds is the invariant mass of \hbb candidates in signal events, which peaks near 125\GeV, compared to the smoothly falling distribution from the backgrounds. Discerning this information depends on accurately reconstructing the two \hbb candidates from the jets present in the selected events.
In the resolved topology, at lower \pt the angular separation between the \PQb jets from \hbb decay increases and correctly pairing the \PQb jets is more difficult.
It is important to design a pairing algorithm that allows for a correct reconstruction of the \hbb candidates in signal events and does not artificially shape the mass distribution in background events.

The \hbb candidates are constructed from the four jets in the event with leading \PQb tagging score. Of the three possible pairing options, the pairing that minimizes the absolute difference in reconstructed mass, $\Delta\PH$, between the two \hbb candidates is chosen~\cite{CMS:2022cpr}. When $\Delta\PH$ is smaller than 30\GeV, which is roughly twice the experimental mass resolution, from the best two pairing options the one with the highest \PH candidate \pt (\pTH) is chosen. This pairing algorithm has an accuracy of about 94\% for the SM \HHbbbb signal. The average pairing accuracy for anomalous \kappal signals is typically lower than for the SM signal because the average \mhh is lower, with a pairing accuracy of 74 (92)\% for the $\kappal=5$ ($\kappal=0$) \gghh signals. 

A radial distance in the $m_{\Hone}$-- $m_{\Htwo}$ mass plane is defined as 
\begin{equation*}
  R_{\PH\PH} = \sqrt{(m_{\Hone}-125\GeV)^2 + (m_{\Htwo}-120\GeV)^2}.
\end{equation*}

\begin{figure*}[!htb]
        \centering
        \includegraphics[width=0.425\textwidth]{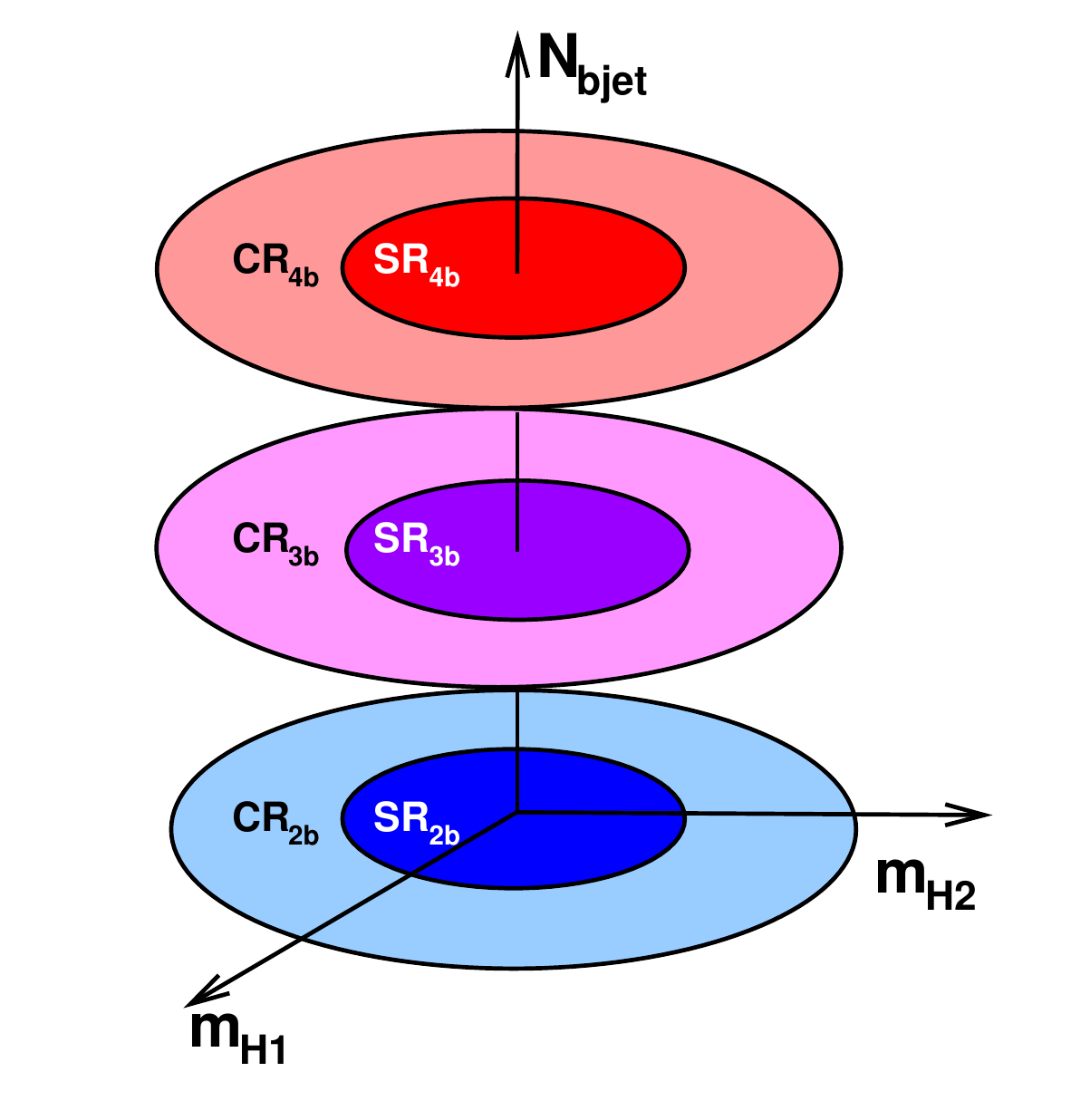}
        \includegraphics[width=0.425\textwidth]{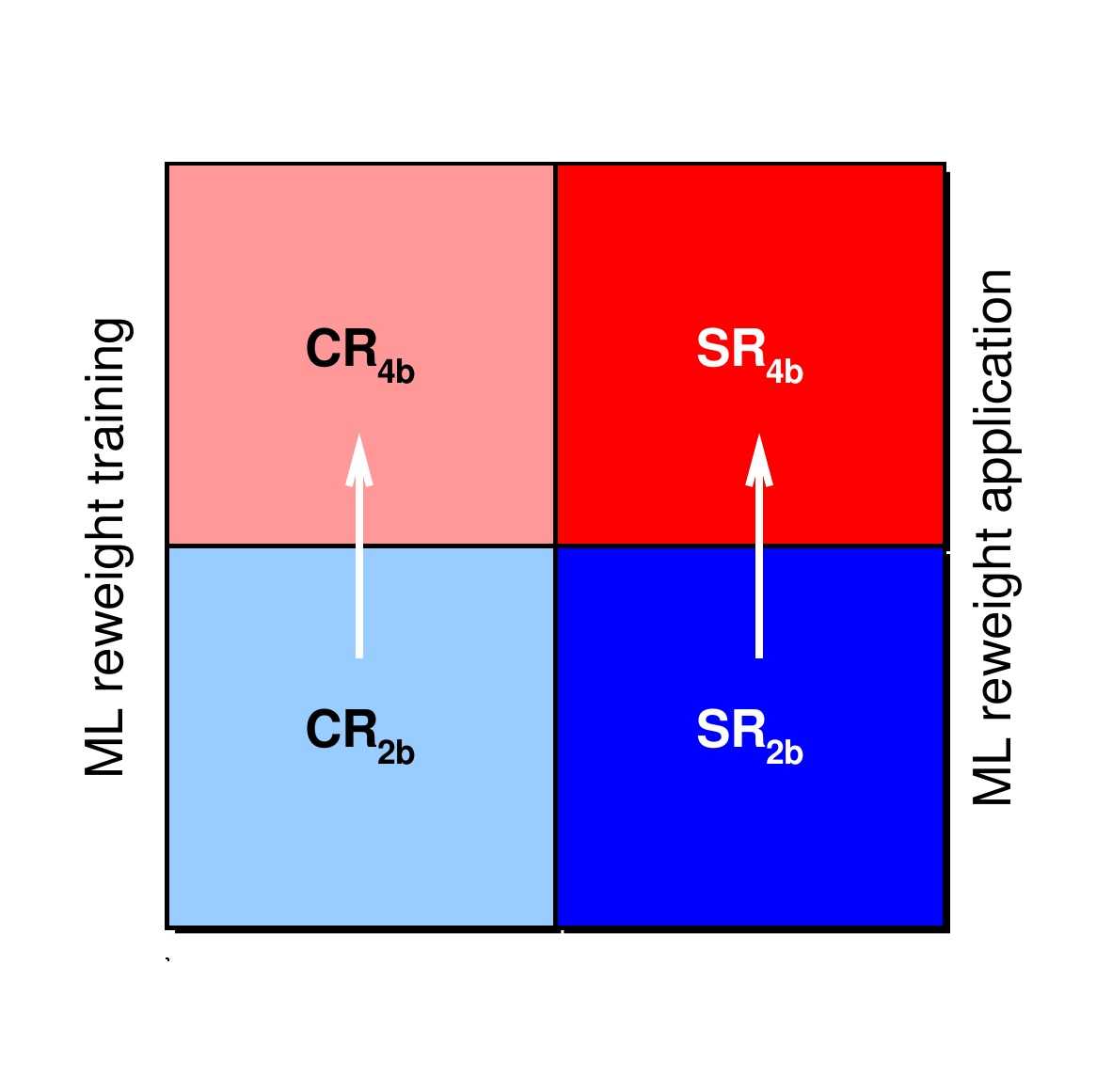}
        \caption{Left: schematic diagram of the signal regions (dark shaded circles) and control regions (annular regions) in the $m_{\Hone}$--~$m_{\Htwo}$ mass plane as a function of \Nbjet. Right: schematic diagram showing the background estimation strategy, which applies a multidimensional reweighting of events from \regSRTwob to \regSRFourb.}
        \label{fig:schematic_resolved_cartoon_SR_CR}
\end{figure*}

The centroid in $m_{\Htwo}$ of 120\GeV is chosen to account for flavor-specific scale differences in the reconstructed jet \pt not accounted for by the flavor-inclusive jet energy scale corrections described in Section~\ref{sec:ak4_btag_pt_regression}. These jet \pt scale differences mostly affect the $m_{\Htwo}$ reconstruction because they are primarily present for low jet \pt. Events are partitioned according to the number of jets tagged passing the medium \PQb tagging WP (\Nbjet)--2\PQb, 3\PQb, or 4\PQb--and the $R_{\PH\PH}$ value, as illustrated in Fig.~\ref{fig:schematic_resolved_cartoon_SR_CR} (left). Events with ${R_{\PH\PH}<30\GeV}$ are assigned to the SRs, while events with $30 < {R_{\PH\PH} < 55\GeV}$ are assigned to the CRs. Events with $R_{\PH\PH}>55\GeV$ are rejected. These regions, some of which (\regSRTwob, \regCRTwob) are newly accessible for the analysis because of the Run 3 triggers, serve as the basis for the background prediction (\regCRTwob, \regSRTwob, and \regCRFourb) and the assessment of the uncertainties in the background model (\regCRThreeb, \regSRThreeb).

The core task of the background model is to learn how to map the distributions of physical observables, particularly those related to separating signal from background, from lower to higher \Nbjet based on data events in the mass sideband CRs. This mapping, from \regCRTwob to \regCRFourb data events, is then applied to lower \Nbjet data events in the signal region (\regSRTwob), providing a background prediction in the \regSRFourb region fully based on CRs in data, from which the \HHbbbb signal is extracted. Figure~\ref{fig:schematic_resolved_cartoon_SR_CR} (right) shows a schematic of this method, which is a multidimensional application of an ABCD technique~\cite{D0:1994wmk}.

In this paper, we present two approaches to the background model, event categorization, and signal-vs-background classifiers. Each approach uses different ML algorithms to map the differences in physical observables between lower and higher \Nbjet regions in data. The \HHbbbb signal is then extracted via a binned maximum likelihood fit to the output of an ML classifier trained to differentiate the \HHbbbb signal from the background prediction in exclusive event categories enriched in either \gghh or \qqhh signal production. The two approaches also use different methods to assess the systematic uncertainties affecting the background prediction and perform different, complementary validation tests of the background model, thereby building confidence in its accuracy and in the substantial sensitivity gain achieved in Run~3 with respect to previous searches. The following sections will describe the details of the two approaches. Finally, we present in Section~\ref{sec:resolved_run2} an updated analysis of Run~2 data in the resolved topology that incorporates several key improvements described in Ref.~\cite{CMS:2024tdk}. Sections~\ref{sec:resolved_systematics} and~\ref{sec:resolved_results} summarize the systematic uncertainties and results, respectively.

\subsection{The first approach: transformer-based classification and validation in \texorpdfstring{$\PH\PH$}{HH} mass sidebands} \label{sec:resolved_transformer}

\subsubsection{Background model overview}\label{sec:resolved_bkgmodel}

The first approach derives the background prediction from a DNN with four hidden layers, trained to differentiate \regCRTwob from \regCRFourb data events based on a variety of input features: a) kinematic properties and the angular correlations among the five AK4 jets in the event with the highest \PQb tag score, b) the \hbb candidate properties and estimated mass resolution, and c) properties of the $\PH\PH$ system. The DNN explicitly takes into account changes in reconstruction conditions during the various CMS data-taking periods via a categorical input feature. The output of the DNN is used to define a ratio of class probabilities as:
\begin{equation*}
\mathrm{weight} = \frac{\text{score}}{1-\text{score}} = \frac{\mathcal{P}_{\regCRFourb}}{1-\mathcal{P}_{\regCRFourb}} = \frac{\mathcal{P}_{\regCRFourb}}{\mathcal{P}_{\regCRTwob}},
\end{equation*}
which estimates the likelihood ratio that a data event derives from the \regCRTwob vs. \regCRFourb region. Applying this weight to \regCRTwob data events is equivalent to a multidimensional reweighting to match the event features in \regCRFourb data. The background prediction is then obtained by inferring the DNN on \regSRTwob data events and performing a per-event reweighting following the sketch shown in Fig.~\ref{fig:schematic_resolved_cartoon_SR_CR} (right). The baseline assumption of this method is that the masses of the \hbb candidates and \Nbjet are largely uncorrelated. This implies that the reweighting function learned in the CRs (${30 < R_{\PH\PH} < 55\GeV}$) can be applied to events in the SR (${R_{\PH\PH} < 30\GeV}$). The validity of this assumption is tested later in this section. The normalization of the background prediction in the \regSRFourb is fit to data without any prior.

It is critical in this analysis that the background prediction achieves a precision of a few percent, considering both statistical and systematic uncertainties, in modeling the distribution of many different physical observables in the \regSRFourb in order to be sensitive to the small expected signal contribution relative to the total background. With this goal in mind, the reweighting is derived following a $k$-fold ensembling method~\cite{10.5555/1643031.1643047} in which the reweighting factor is obtained from the mean of $k=20$ individual DNN reweighting estimates. In each of the $k$ trainings, \regCRTwob and \regCRFourb data events are randomly split into training and validation sets and the initial values of the DNN hyper-parameters are resampled randomly from a uniform distribution in the range $[-1,1]$. Such an ensembling technique reduces the variance in the background prediction that arises from the convergence of the DNN to different minima of the loss function, and the spread among the individual $k$-fold background predictions provides a quantitative measure of the residual variance in the background prediction. Background predictions from ensembles of ${k>20}$ trainings were also attempted, leading to similar uncertainties in the prediction but with a larger computational cost. It is worth noting that the training sets for each of the ${k=20}$ individual DNN trainings significantly overlap and this should be considered when assessing the variance. These observations will be described in more detail in Section~\ref{sec:resolved_bkgpred}.

\subsubsection{Event selections}\label{sec:resolved_selections}

Data events are separated into pre- and post-ParkingHH samples, corresponding to the different trigger strategies described in Section~\ref{sec:trigger_resolved}. In pre-ParkingHH data, the four jets with highest \pt and ${\abs{\eta}<2.5}$ are required to satisfy ${\pt >80}$, 60, 45, and 35\GeV, while in post-ParkingHH data the jets are required to have ${\pt >35}$, 35, 35, and 30\GeV. The regions with different \Nbjet are partitioned according to the number of \PQb jets passing the \PNET medium WP, with a light-flavor quark or gluon jet misidentification rate of 1\%. Events are further split into exclusive categories targeting either the \gghh or \qqhh signal.

Events in the \qqhh category are required to contain at least two additional AK4 jets with ${\pt>25\GeV}$ and ${\abs{\eta}<4.7}$. The maximum dijet invariant mass (\mjj) among pairs of those additional jets is required to be greater than 400\GeV and the absolute value of the $\eta$ separation of that jet pair ($\abs{\detajj}$) must exceed 2.5. Finally, events in the \qqhh category must also pass a selection on the output of a DNN trained to separate the \qqhh and \gghh signals, as will be described in Section~\ref{sec:vbf_hh4b_resolved}. Remaining events not satisfying all the \qqhh category criteria above are assigned to the ggHH category.  

\subsubsection{Signal separation with a transformer architecture}
\label{sec:resolved_signal_vs_bkg}

In the \regSRFourb of the \gghh category, the inclusive expected S/B is about 0.1\%. In order to improve the signal purity, an ``SvsB'' multivariate classifier is trained to separate selected \HHbbbb simulated events from the total background prediction, estimated via the reweighting technique described in Section~\ref{sec:resolved_bkgmodel}. The network is composed of two feature encoders, followed by a common decoder that solves the classification problem. This network design provides an efficient way to include not only event-level features, but also input features related to the individual jets and their pairwise properties, improving the signal separation power. The first encoder is a feed-forward DNN with three consecutive hidden layers that takes as input a variety of event-level features similar to those used in the background reweighting DNN. The second encoder is a series of eight transformer blocks that take as input the four-vectors of the five leading \PQb-tagged jets in the event as well as various properties of every possible two-jet pairing among these jets such as \mjj, \detajj, etc. The self-attention mechanism~\cite{vaswani2023attentionneed} of the transformer is modified to account for these pairwise features, following an approach used in \textsc{ParT}~\cite{Qu:2022mxj}. The linear embeddings produced by each encoder are then passed to a DNN with three hidden layers that aims to separate signal from background. The resulting ``SvsB classifier'' is able to reject the background in the \regSRFourb of the \gghh category by an additional two (three) orders of magnitude while retaining a SM \HHbbbb signal efficiency of about 40 (10)\%. As for the DNN used in the background estimation, the classifier accounts for differences in data-taking conditions. 

Although the \hbb candidate properties are based on a single \PQb jet pairing choice, the properties of \PQb jet pairs under alternative pairing choices are very useful in discriminating the signal from the background. Because the jet pairing in background events is a random combination of two jets without a common underlying origin, the properties of these alternative pairs tend to be quite similar. The correct pairing choice in signal, in contrast, tends to have very distinctive features compared to other pairing options. The transformer blocks in the SvsB classifier were engineered to leverage this information directly from the pairwise jet properties. The inclusion of the individual jet four-vectors and the pairwise jet information, as opposed to the \hbb candidate properties alone, significantly improves the separation performance of the SvsB classifier, enhancing the sensitivity of the analysis by about 25\%.

The kinematic event features in \HHbbbb events strongly depend on the value of \kappal, which determines the relative importance of the trilinear coupling and the box diagram in the \gghh production~\cite{DiMicco:2019ngk}. As discussed in Section~\ref{sec:intro}, current measurements from the LHC provide limited constraints on \kappal. In order to optimize the analysis performance throughout the experimentally allowed range of \kappal hypotheses, the \HHbbbb signal events used in the training are obtained from an equal mixture of four simulated samples with $\kappal=(0,1,2.45,5)$. This configuration retains equivalent performance for the SM \HHbbbb compared to a training performed using only $\kappal=1$ signal events. At the same time, it retains good sensitivity to non-SM \kappal values. The background events considered in the training are \regSRTwob data events, with weights assigned following the method described in Section~\ref{sec:resolved_bkgmodel}.

\subsubsection{Variance uncertainties in the background prediction}\label{sec:resolved_bkgpred}

Variance in the background model prediction arises because of the finite number of \regCRFourb and \regCRTwob data events in the DNN training and the convergence of the DNN to different loss function minima. In differentiating between \regCRTwob and \regCRFourb data events, the DNN is limited by the number of data events in training, particularly when focused on the tails of distributions where the amount of training data is most limited. This variance is particularly relevant for the background prediction in the region where the SvsB classifier predicts the highest signal purities, since this is a region where the reweighting DNN training data is particularly scarce. It is relevant to note that the variance is statistical in nature--with a larger \regCRFourb data sample, the variance in the background prediction decreases. The techniques described in this section can therefore be used to control the background to an even better precision in the future with larger data samples.

The variance is quantified from the spread, defined as the half-difference between 16\% and 84\% quantiles, among the ${k=20}$ individual background predictions in the $k$-fold ensemble and assessed independently per bin for each observable. For the distribution of the SvsB classifier score, \psig, used to extract the \HHbbbb signal, this variance uncertainty ranges from 0.1\% to about 5\%, increasing with increasing values of \psig. It is also important to account for variance effects that give rise to differences in the background prediction that are correlated across bins of the distribution--for example, one individual background model among the $k$-fold ensemble may systematically predict more background at high \psig than another. In some cases, such differences in the background prediction can look similar to the expected profile of the \HHbbbb signal in data. This effect must therefore be precisely understood and quantified in order to ensure an unbiased measurement of the \HHbbbb signal. 

Because the training data sets used among the $k$ reweighting DNNs significantly overlap, shape differences among the $k$-fold background predictions in the \regSRFourb may not necessarily cover the full statistical variance of the background prediction. It is therefore necessary to assess the scale of the correlated variance effects by either partitioning the \regCRFourb data, which further increases the variance, or by employing another data sample. We use \regCRThreeb data events, as a proxy of the \regCRFourb, to derive shape uncertainties in the background prediction for the \psig distribution. This method is possible because of the expanded acceptance of the Run 3 triggers, which allows for the 3\PQb data events to be used purely for validation and uncertainty assessment. The Run~2 analysis instead relied on 3\PQb data events as the basis of the background model, which was necessary because of the Run~2 trigger-level requirement of at least three \PQb-tagged jets. The \regCRThreeb data events are split into ten statistically independent sub-samples (split data sets), each with a statistical precision comparable to the \regCRFourb. For each of the ten 3\PQb split data sets, an ensemble of $k$-fold reweighting DNNs are trained following exactly the same methods as described in Section~\ref{sec:resolved_bkgmodel}. The ten independent $k$-fold ensemble background model predictions differ at high \psig score by as much as 5--6\%, and these shape differences can sometimes manifest themselves with a very similar signature to that of the \HHbbbb signal. It is therefore crucial to accommodate such variance effects in the prediction of the background \psig distribution.

The shape differences in the background \psig prediction are parameterized through a Fourier series, following the procedure detailed in Ref.~\cite{CMS:2024tdk}. The Fourier basis is defined as:
\begin{equation}\label{eq:fourier_basis_original}
b(i) = \begin{cases}
a_i \sin{\frac{i+1}{2} \pi x},& i=1, 3, 5,\dotsc \\
a_i \cos{\frac{i}{2} \pi x},& i=0, 2, 4,\dotsc,
\end{cases}
\end{equation} 
where $b(i)$ is the individual basis function of order $i$, as a function the SvsB DNN score $x$, and $a_i$ is the amplitude. A modified Gram--Schmidt decomposition is performed such that each basis function defines an independent shape variation in the \psig background prediction. A fit is then performed to the difference between each of the ten 3\PQb split data set independent background predictions and the average of all background predictions, allowing the basis functions to adjust for the shape differences. The number of basis functions necessary to model, with sufficient precision, the observed variations is determined via a Fisher F-test, and the spread of the fit adjustments made by the basis functions across the fits to the ten 3\PQb split data sets is used to determine the amplitude, $a_i$, of the basis functions, corresponding to the size of the shape uncertainties in the background prediction. 

\subsubsection{Background model validation}

\begin{figure*}[!htb]
  \centering
    \includegraphics[width=0.425\textwidth]{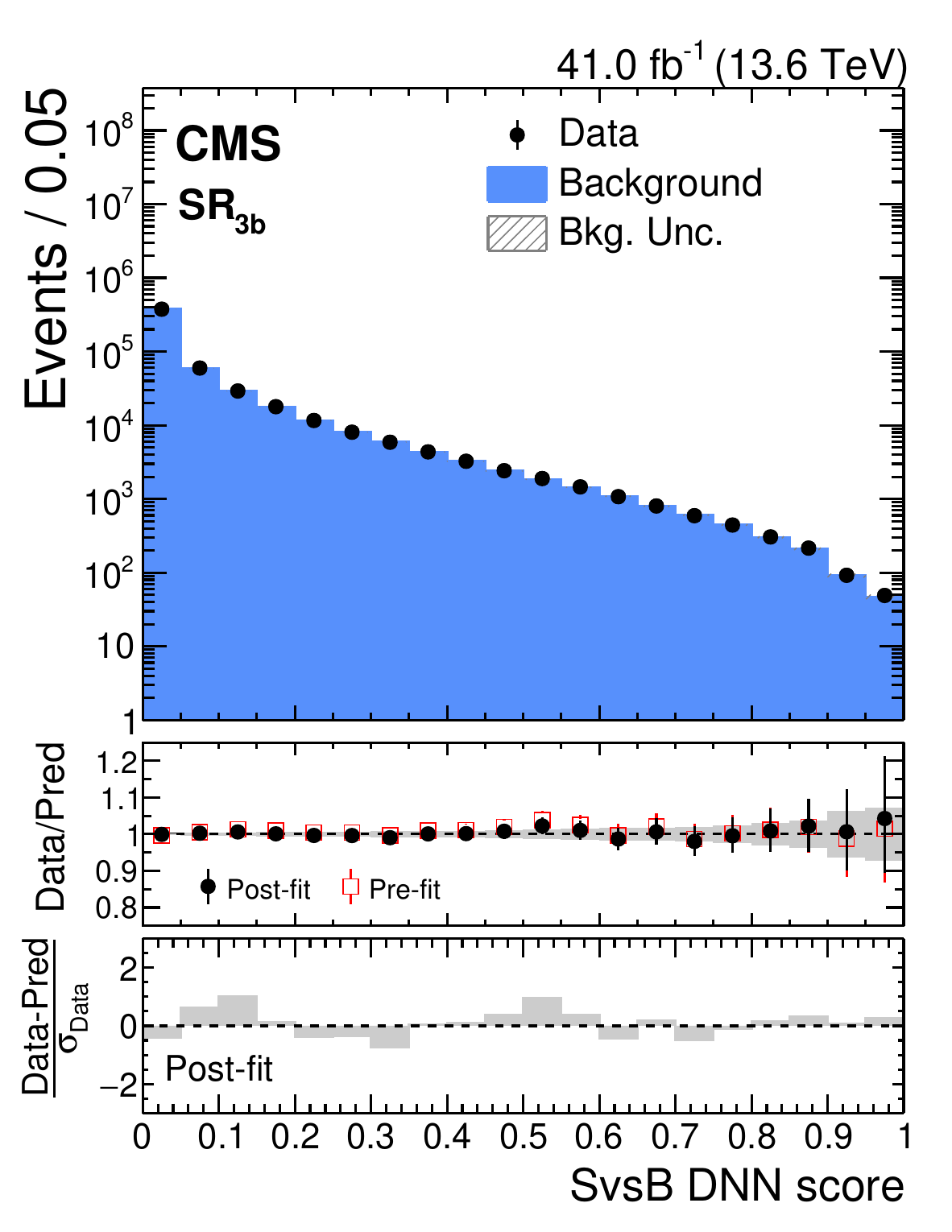}
    \includegraphics[width=0.425\textwidth]{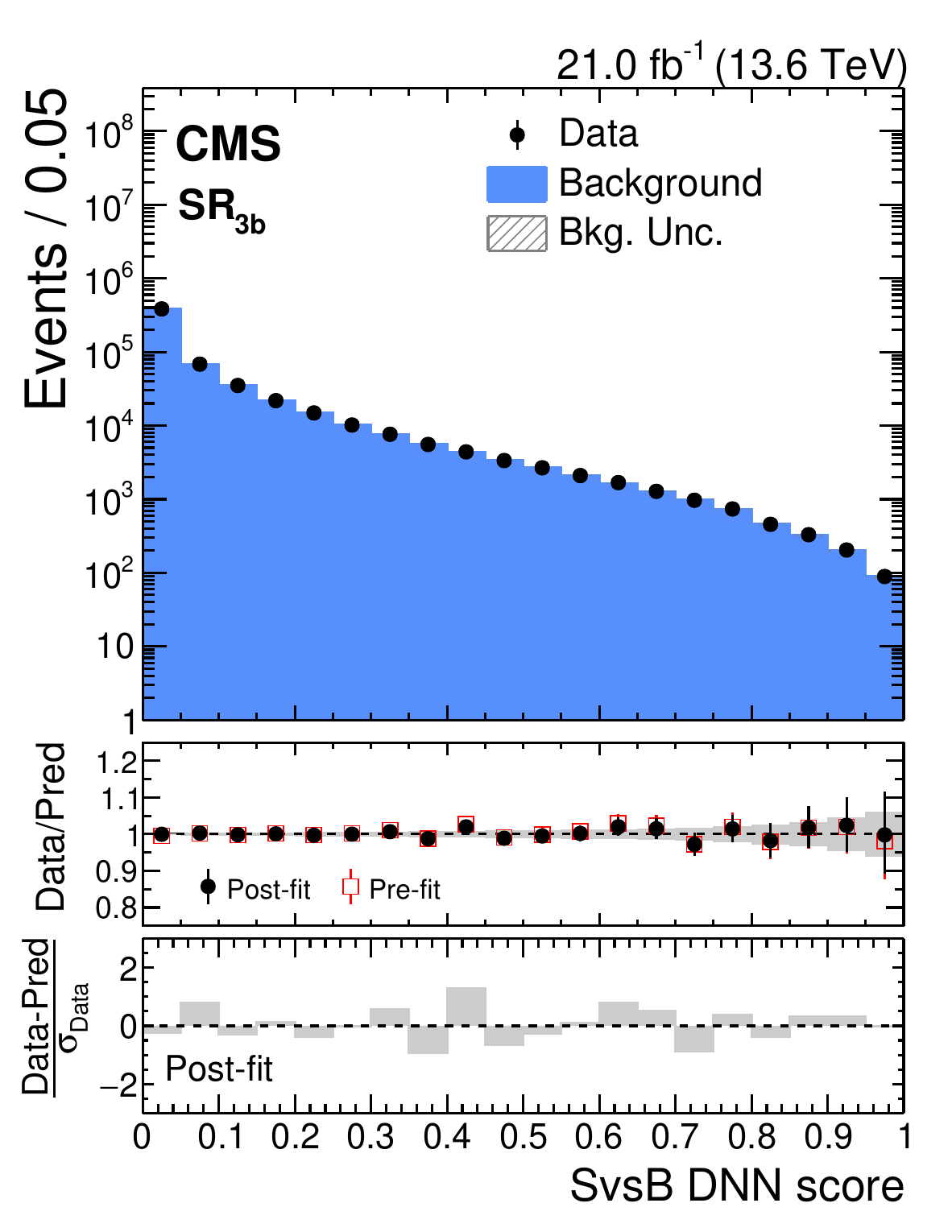}
    \caption{Distribution of the SvsB classifier score in the \gghh category for \regSRThreeb data (black points) compared to the background prediction (blue histogram), for the pre-ParkingHH (left) and post-ParkingHH (right) data sets. The middle panel shows the ratio of data to the pre-fit (red open markers) and post background-only fit (black solid markers) background prediction, with the gray band indicating the background post-fit uncertainty. The lower panel shows the distribution of the pulls, defined as the difference between the data and the post-fit background prediction, divided by the statistical uncertainty in the data.}
    \label{fig:dnn_3b_control_region_fitted}

\end{figure*}

The background model is validated via a variety of closure tests in data regions similar to the \regSRFourb. Since the signal purity in the \regSRThreeb is an order of magnitude smaller than in the \regSRFourb, the background model can be tested over the full range of the SvsB score distribution in \regSRThreeb data. Moreover, the large number of expected background events in the \regSRThreeb compared to the \regSRFourb enables a closure test of the background model with much better statistical precision than is possible with \regSRFourb data. Figure~\ref{fig:dnn_3b_control_region_fitted} shows the SvsB score distribution for \regSRThreeb data in the \gghh category, compared to the background prediction, in the pre-ParkingHH (left) and post-parkingHH (right) data sets. In order to test for any bias in the background model due to the extrapolation from mass sideband (${30 < R_{\PH\PH} < 55\GeV}$) CR events to the SR (${R_{\PH\PH} < 30\GeV}$), we fit this distribution including the shape uncertainties described in Section~\ref{sec:resolved_bkgpred}. The resulting goodness-of-fit test, using the saturated $\chi^2$ method~\cite{BAKER1984437}, yields a $p$-value greater than 80\% for all data-taking periods. The fit makes a small adjustment to the predicted background shape, which is considered in the later \HHbbbb signal extraction as an additional shape uncertainty in the background prediction and measured separately for each data-taking period. This accounts for any potential bias in the extrapolation of the background prediction from CR to SR, and its impact on the result is limited compared to the variance uncertainties measured from the 3\PQb split data set tests, as described in Section~\ref{sec:resolved_bkgpred}. 

\begin{figure}[!htb]
  \centering
    \includegraphics[width=0.5\textwidth]{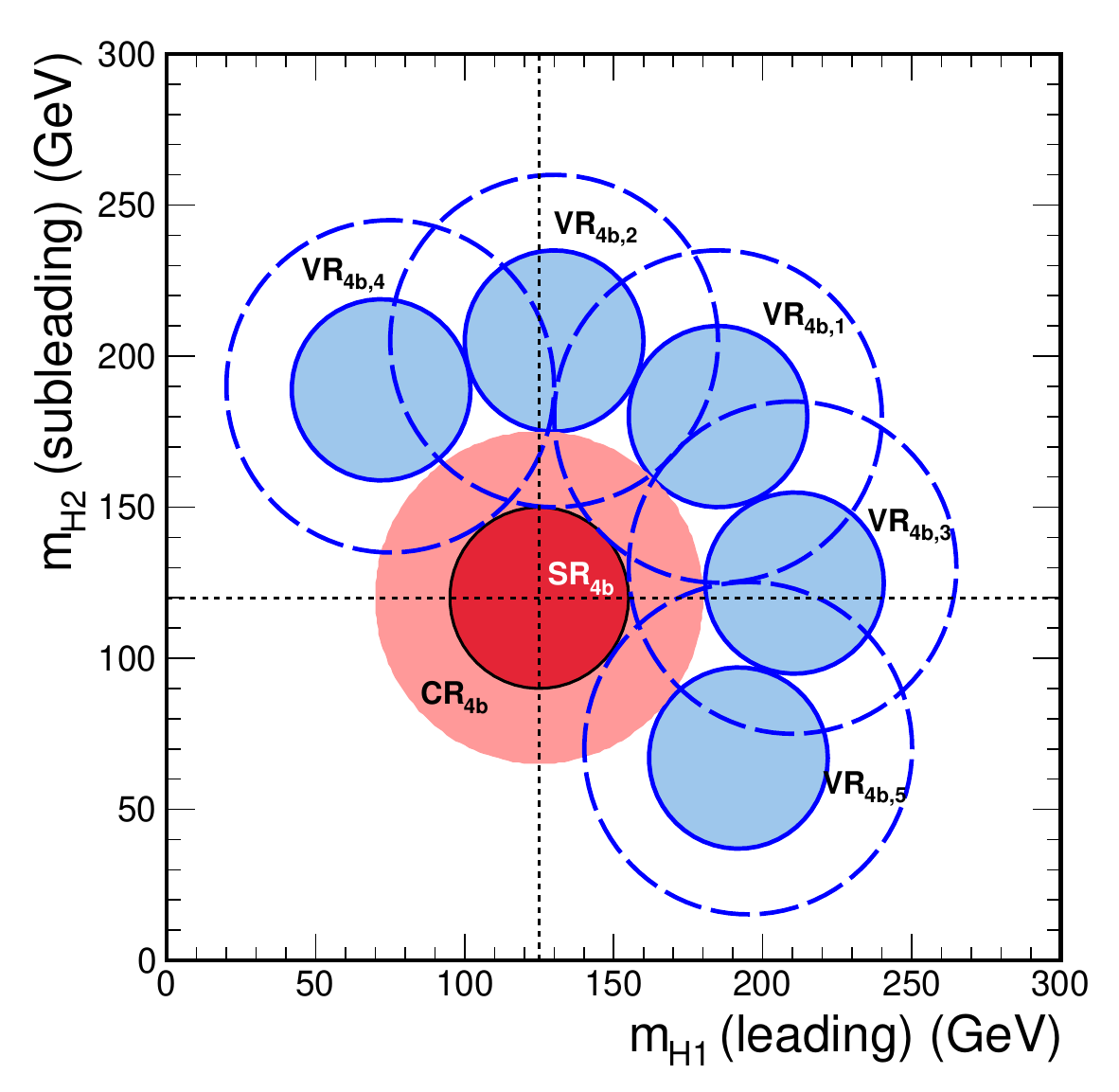}
    \caption{A schematic diagram of the 4\PQb validation regions defined in the {$m_{\Hone}$--~$m_{\Htwo}$} mass plane and orthogonal to the \regSRFourb. In each validation region, the solid blue area identifies the signal region, while the dashed blue lines indicate the corresponding control region. The ``leading'' \PH candidate is the one with largest $\pt(\PH)$.}
    \label{fig:vr_cartoon}

\end{figure}

\begin{figure*}[!htb]
\centering
\includegraphics[width=0.425\textwidth]{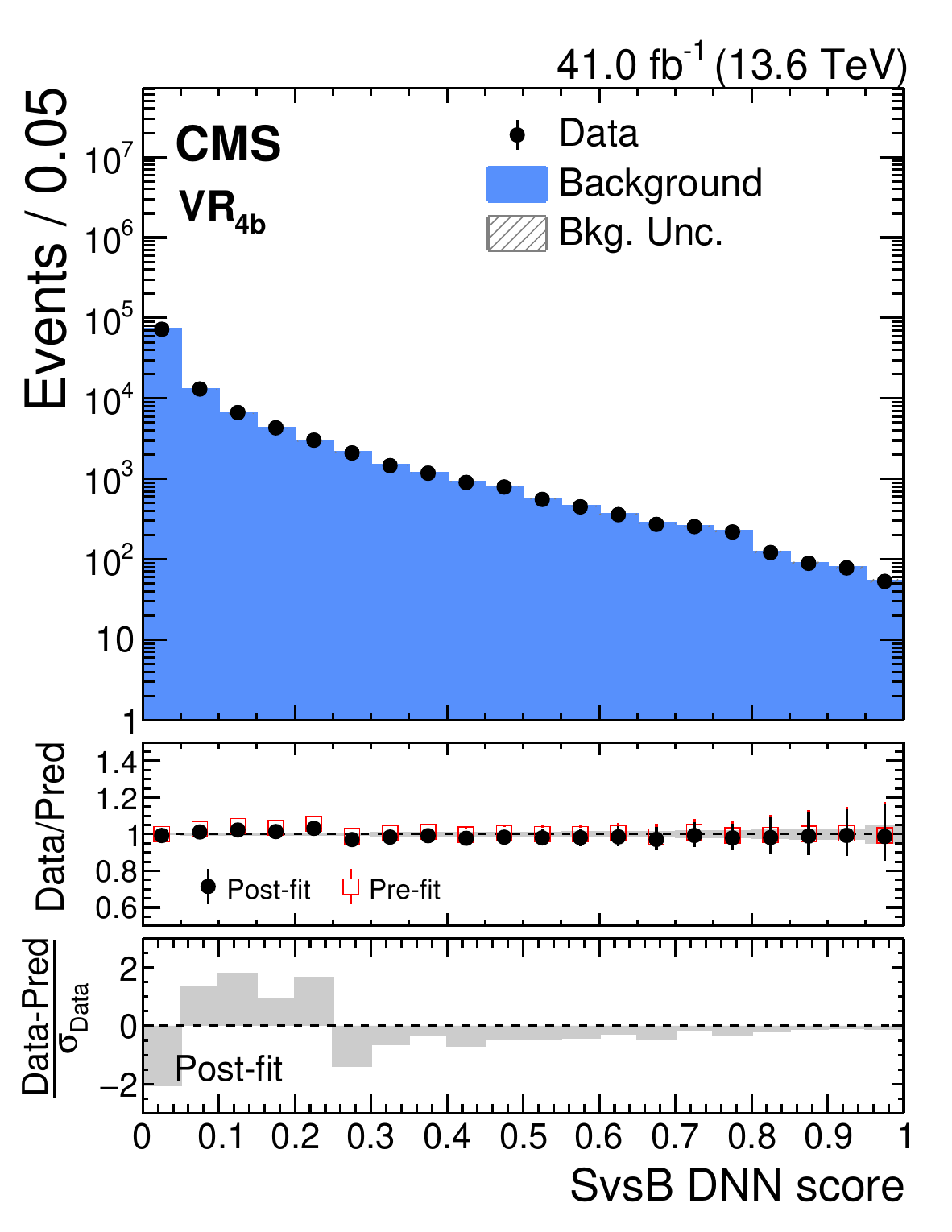}
\includegraphics[width=0.425\textwidth]{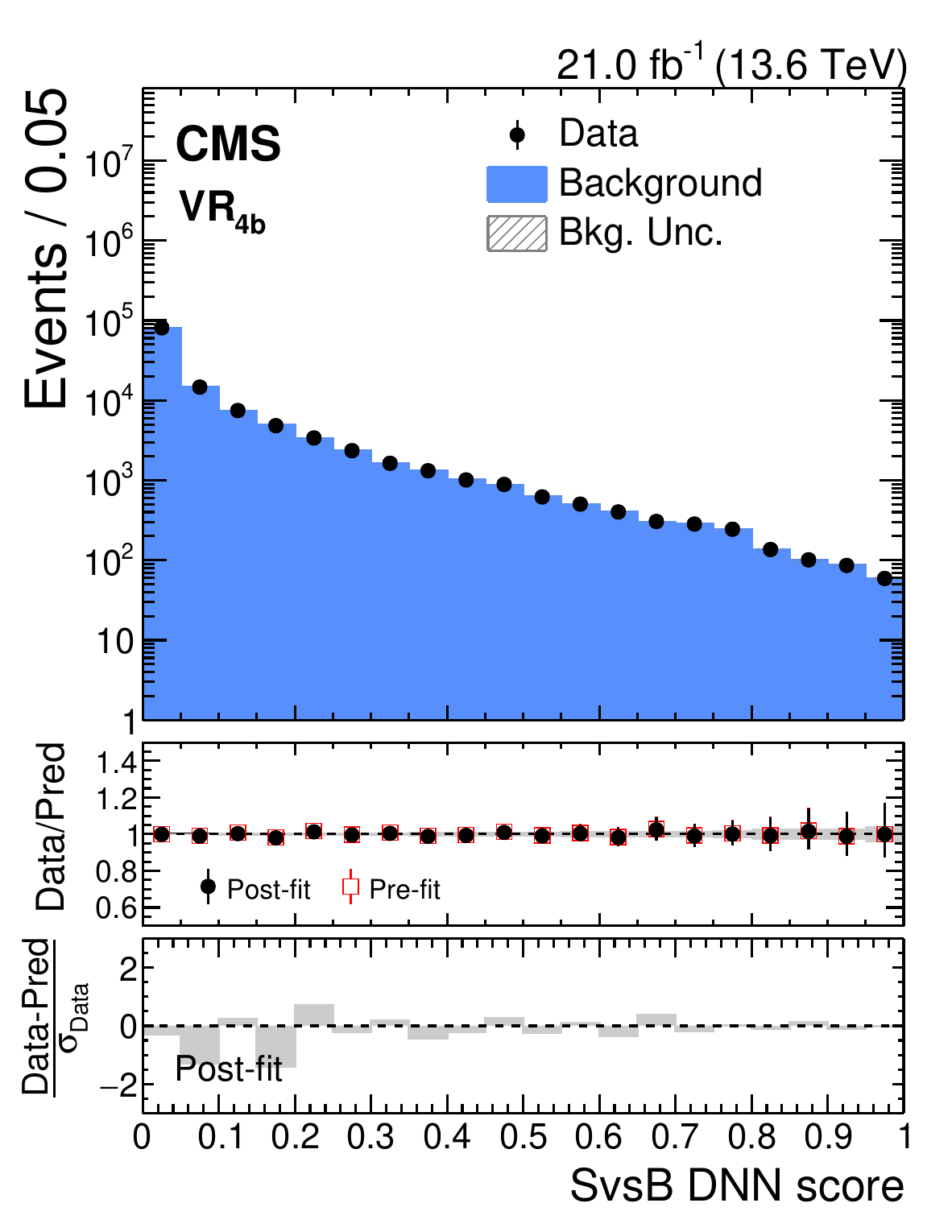}
\caption{Pre-fit and post background-only fit distributions of the SvsB classifier output for the sum of all 4\PQb validation regions in pre-ParkingHH (left) and post-ParkingHH (right) data. Notations are as in Fig.~\ref{fig:dnn_3b_control_region_fitted}.} 
\label{fig:resolved_4bvr_closure}
\end{figure*}

The \regSRThreeb closure test is an important validation of the background model, but it is also necessary to further validate the background model on 4\PQb data events. With this goal, we define five independent validation regions in the {$m_{\Hone}$--~$m_{\Htwo}$} plane, ${\mathrm{VR_{4\PQb,[1-5]}}}$, as depicted in Fig.~\ref{fig:vr_cartoon}. Events in the lower left quadrant of the {$m_{\Hone}$--~$m_{\Htwo}$} plane were not considered for the validation regions because of significant variations in the trigger efficiency and the presence of the signal-like \ZZ and \ZH processes. Each validation region is composed of an annular CR and a mass circle, mimicking the definitions of the \regCRFourb and \regSRFourb but with shifted centers in the {$m_{\Hone}$--~$m_{\Htwo}$} mass plane. 
In each validation region, we train an ensemble of $k=20$ background models using the same methods as described in Section~\ref{sec:resolved_bkgmodel} and assess the closure of the background model prediction with the observed 4\PQb data. Due to differences in the SvsB classifier input feature distributions compared to those of the SR induced by the different {$m_{\Hone}$--~$m_{\Htwo}$} selections, the SvsB classifier score, as evaluated on the raw input features in the validation regions, is skewed towards low \psig values, limiting the utility of the closure test. With the aim of making each validation region as close a proxy as possible to the SR, we perform a quantile matching procedure~\cite{quantileMatchingBook} to the input features, based on the cumulative distribution of each feature in each region. The quantile matching technique ensures that the \psig distributions in the \regVRFourb and \regSRFourb regions are as similar as possible. Figure~\ref{fig:resolved_4bvr_closure} shows both the pre-fit and post-fit closure of the background prediction with the observed data for the sum of the five validation regions in pre-ParkingHH (left) and post-ParkingHH (right) data. The background model fits the 4\PQb data well in all five validation regions, with goodness-of-fit $p$-values larger than 5\% for all regions and data sets, and with no significant pulls in the background shape uncertainties. These tests, along with the \regSRThreeb fit mentioned above, demonstrate the ability of the background model to describe the \psig distribution for background events in data in all analysis regions. 

\subsubsection{Exclusive VBF \texorpdfstring{\HHbbbb}{HHbbbb} category}\label{sec:vbf_hh4b_resolved}

The selection criteria specific to the \qqhh category, described in Section~\ref{sec:resolved_selections}, have an efficiency of about 10--15\% for the \gghh signal. Because the \qqhh cross section is about fifteen times smaller than that of \gghh, \gghh events constitute the majority of the expected signal yield after VBF selections. A DNN is trained to differentiate among the SM \qqhh and \gghh signals, as well as from the background, using the input features described in Section~\ref{sec:resolved_signal_vs_bkg} and additional features related to the ``VBF jets'', corresponding to the jet pair with largest \mjj (\detajj, \mjj, and centrality of \PH candidates relative to the VBF jets) and the four-vectors of the two VBF jets. The DNN has three output nodes ($\mathcal{P}_{\gghh}$, $\mathcal{P}_{\mathrm{VBF}}$, and $\mathcal{P}_{\mathrm{bkg}}$), proportional to the likelihood of that event corresponding to the \gghh, \qqhh, or background classes, respectively. A selection is then applied in order to significantly reduce the \gghh contamination using a \qqhh vs. \gghh discriminant, defined as $\mathcal{D}_{\gghh}=\mathcal{P}_{\mathrm{VBF}}/(\mathcal{P}_{\mathrm{VBF}}+\mathcal{P}_{\gghh})$. The requirement on $\mathcal{D}_{\gghh}$ retains 80\% of the preselected SM \qqhh signal while rejecting about 90\% of the expected SM \gghh contribution. After these selections, \qqhh constitutes about 45\% of the total expected signal. Any events not satisfying the $\mathcal{D}_{\gghh}$ requirement are assigned to the inclusive, \gghh-dominated, category.

The \qqhh signal is separated from the remaining background via a ``VBF-vs-background'' score, defined as $\mathcal{D}_{\mathrm{VBF}}=\mathcal{P}_{\mathrm{VBF}}/(\mathcal{P}_{\mathrm{VBF}}+\mathcal{P}_{\mathrm{bkg}})$. The $\mathcal{D}_{\mathrm{VBF}}$ discriminant is able to reject the background in the VBF \regSRFourb by two orders of magnitude, while retaining a SM \qqhh signal efficiency of about 55\%. The background prediction is obtained via the same methodology of reweighting of \regSRTwob events, including the VBF category selections described in Section~\ref{sec:resolved_bkgmodel}. In the VBF category, the statistical precision of the reweighted \regSRTwob template is much less than that of the inclusive \gghh category and this dominates the total uncertainty in the background prediction. It is, therefore, sufficient to assess the background variance with bin-by-bin uncertainties derived from the spread among $k$-fold background predictions alone, without introducing shape uncertainties as defined in Eq.~(\ref{eq:fourier_basis_original}), which would have a negligible impact. The background model is validated on \regSRThreeb data events including all VBF category selections and is found to model the observed data adequately.

\subsubsection{Signal extraction}\label{sec:resolved_transformer_sigex}

\begin{figure*}[!htb]
    \centering
    \includegraphics[width=0.425\textwidth]{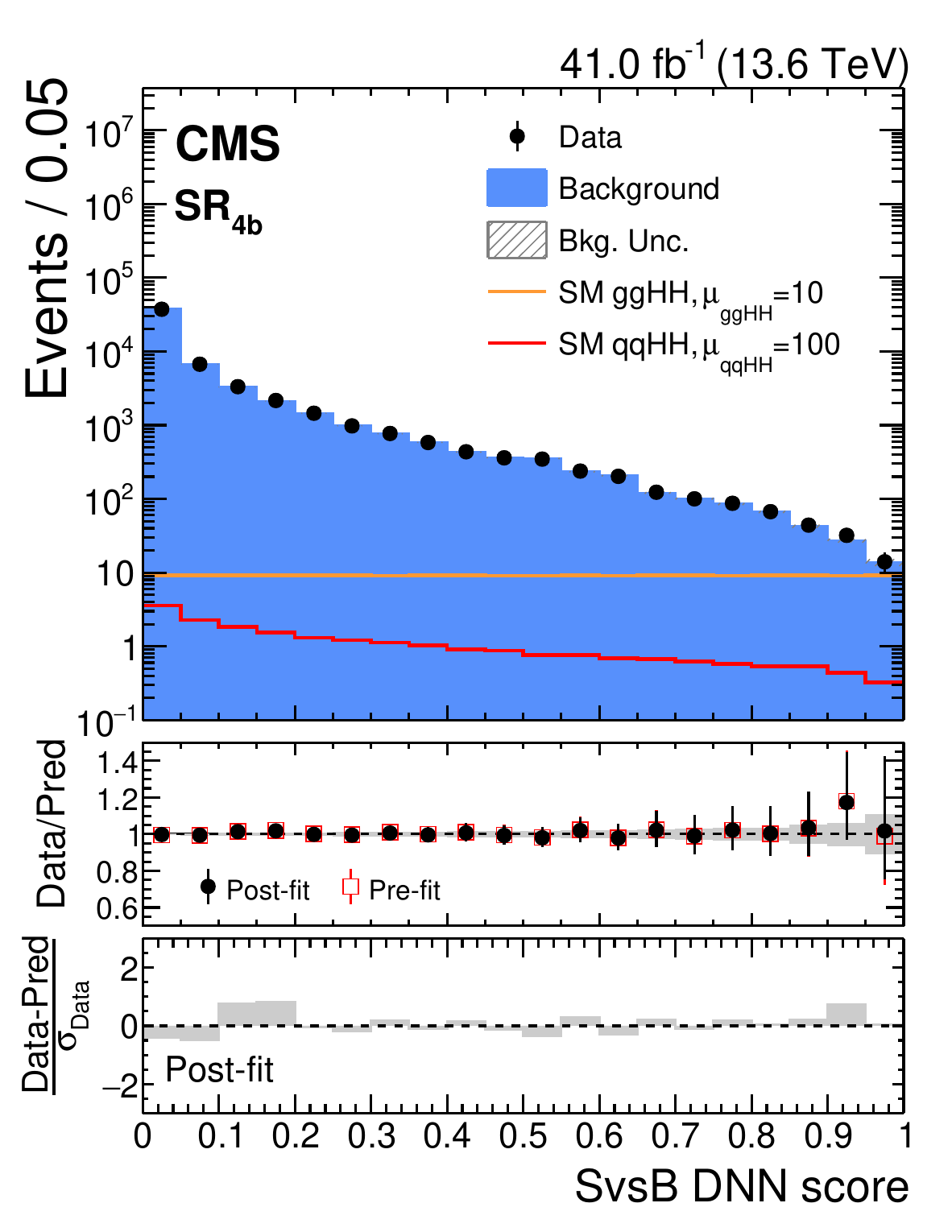}
    \includegraphics[width=0.425\textwidth]{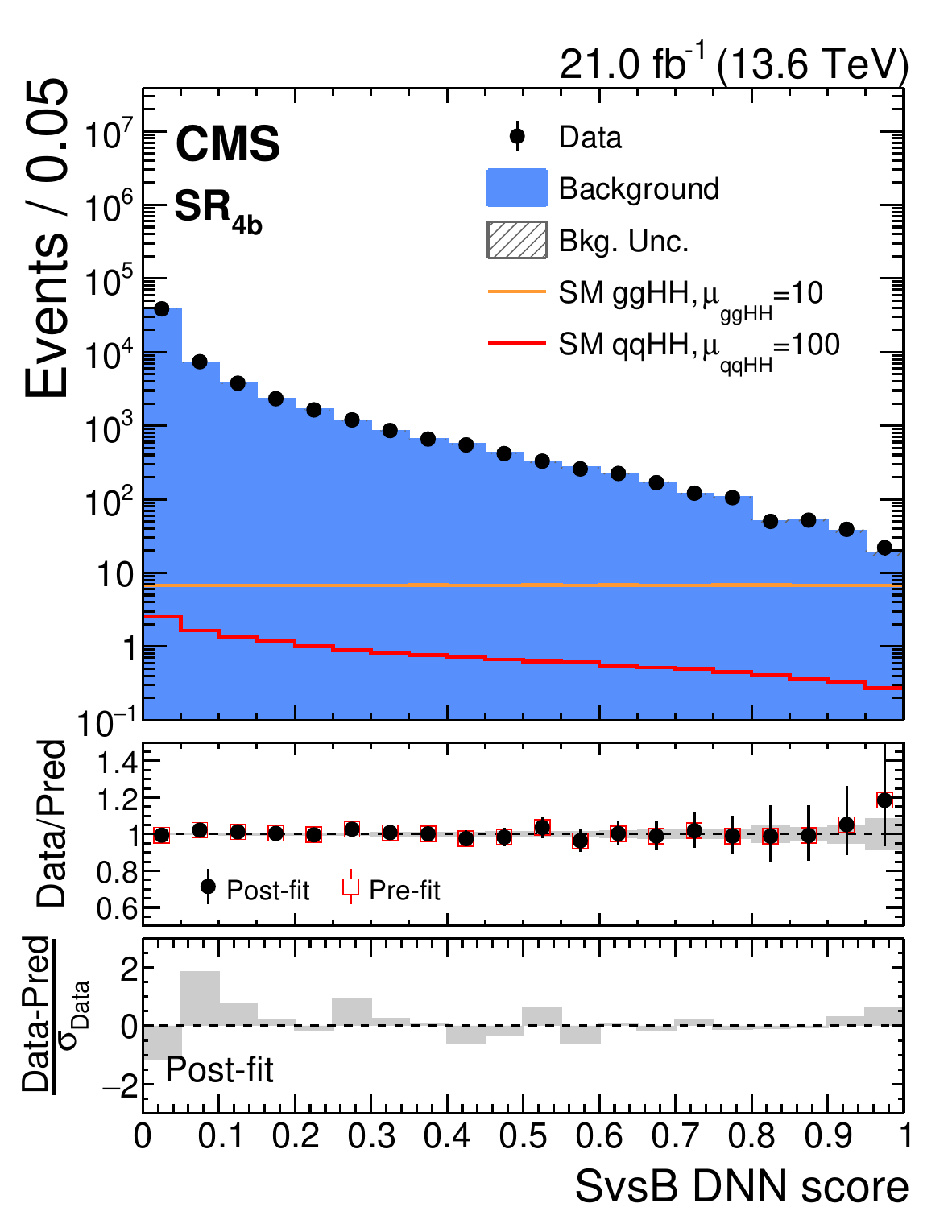}
    \caption{Post-fit distributions of the SvsB classifier score in the \regSRFourb of the \gghh resolved analysis for data (black points) and the predicted background (blue filled histograms), for pre-ParkingHH (left) and post-ParkingHH (right) data. The distributions of the SM \gghh (orange line) and \qqhh (red line) signals, scaled to improve their visibility, are overlaid. Notations are as in Fig.~\ref{fig:dnn_3b_control_region_fitted}.}
    \label{fig:dnn_4bSR_postfit_ggHH}
  
\end{figure*}

In the \gghh category, a monotonic transformation of the SvsB classifier score, \psig, distribution is performed, such that the expected distribution for the \gghh signal in the \regSRFourb is uniform. The \HHbbbb signal is then measured via a binned maximum likelihood fit to the transformed SvsB classifier score distribution in the \regSRFourb, as shown in Fig.~\ref{fig:dnn_4bSR_postfit_ggHH} for the pre-ParkingHH (left) and post-ParkingHH (right) data. The simple equidistant binning in the transformed SvsB classifier score used in the fit provides a measurement sensitivity compatible with results obtained from more complex binning optimization procedures. 

\begin{figure*}[!htb]
  \centering
    \includegraphics[width=0.425\textwidth]{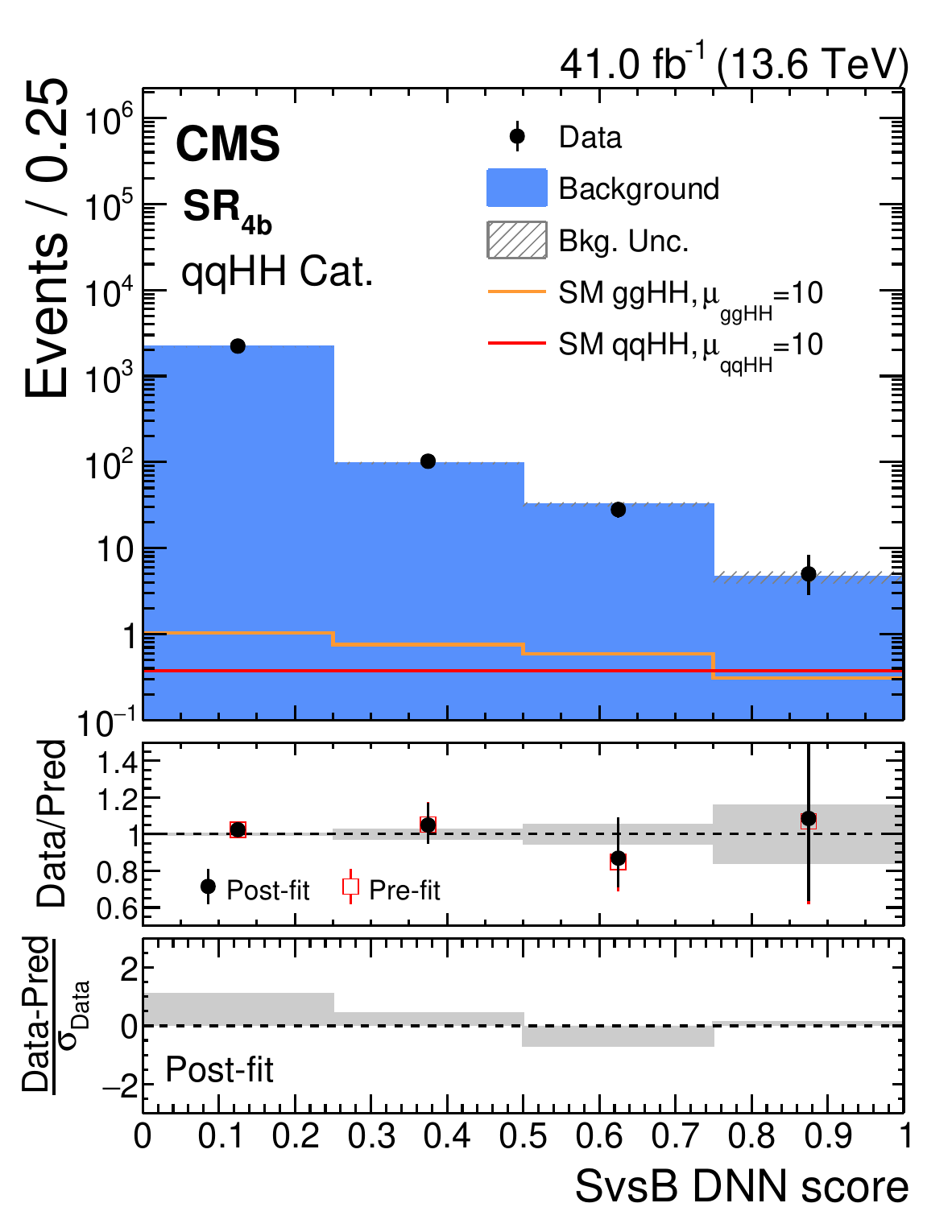}
    \includegraphics[width=0.425\textwidth]{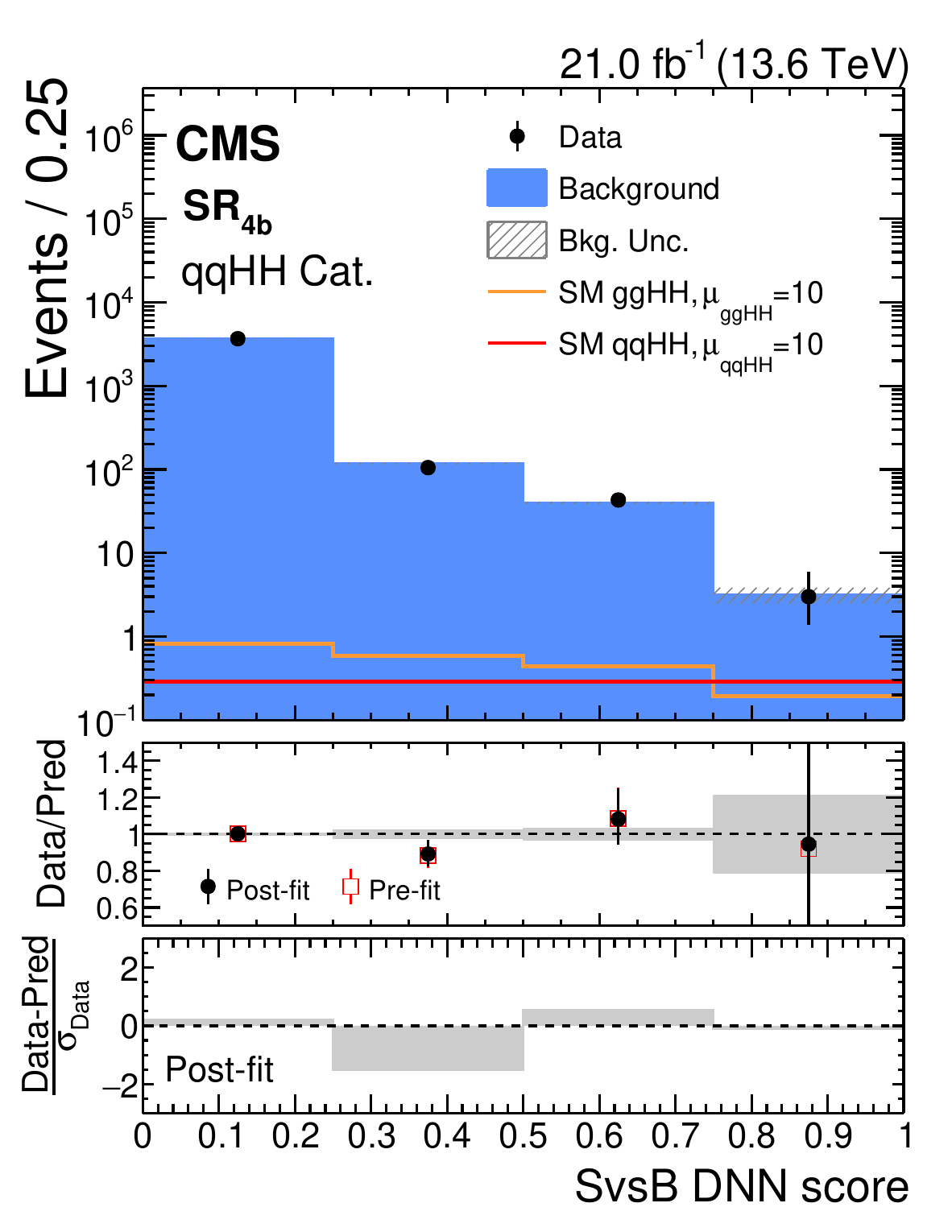}
    \caption{Post-fit distributions of the SvsB output score in the \regSRFourb of the \qqhh resolved analysis reported for data (black points) and the predicted background (blue filled histograms) for pre-ParkingHH (left) and post-ParkingHH (right) data. The distributions of the SM \gghh (orange line) and \qqhh(red line) signals, scaled to improve their visibility, are overlaid. Notations are as in Fig.~\ref{fig:dnn_3b_control_region_fitted}.}
    \label{fig:dnn_4bSR_postfit_qqhh}

\end{figure*}

In the \qqhh exclusive category, Fig.~\ref{fig:dnn_4bSR_postfit_qqhh} shows the distribution of the $\mathcal{D}_{\mathrm{VBF}}$ discriminant in pre-ParkingHH (left) and post-ParkingHH (right) data. Similarly to the \gghh analysis, a transformation is applied to the $\mathcal{D}_{\mathrm{VBF}}$ distribution such that the predicted \qqhh distribution is uniform. The number of bins was chosen to maintain a sensitivity within 5\% of more aggressive binning options while keeping the total number of bins in the fit as minimal as possible, ensuring that shape uncertainties in the background prediction as defined in Eq.~(\ref{eq:fourier_basis_original}) would be negligible.

\subsection{The second approach: GNN-based classification including \texorpdfstring{$\PZ\PZ/\PZ\PH$}{ZZ/ZH} regions} \label{sec:resolved_graph}

Another approach to the search for \HHbbbb in the resolved topology is presented below. An important difference with the approach described in Section~\ref{sec:resolved_transformer} is in the treatment of the SRs and CRs, which are modified to target the \HH, \ZZ, and \ZH processes, following the strategy in Ref.~\cite{CMS:2024tdk}. Due to their larger cross sections, the SM \ZZ and \ZH processes are expected to be observed in the 4\PQb final state before the \HH process. Since they share the same experimental challenges as \HHbbbb, it is important to pursue them in tandem as they serve as valuable validation tools for all analysis methodologies, and in particular for the background estimation procedure. Another key difference is that at the core of this approach is a custom graph-based neural network architecture, referred to as \FeynNet, used for the derivation and validation of the background model, for process classification, and for signal-vs-background discrimination.

Events with at least four AK4 jets with ${\pt > 70}$, 50, 40, and 35\GeV in pre-ParkingHH and 30\GeV in post-ParkingHH data, ${\abs{\eta}<2.5}$, and satisfying tight identification criteria are selected. Events are further required to have at least two \PQb-tagged jets using the \PNET medium WP, and the average \PQb tagging score of the leading two is required to satisfy the online selection. Additional jets with ${\pt > 30\GeV}$ and ${\abs{\eta}<4.7}$, not \PQb tagged, are considered as VBF candidate jets.

\begin{figure}[!htb]
    \centering
    \includegraphics[width=0.45\textwidth]{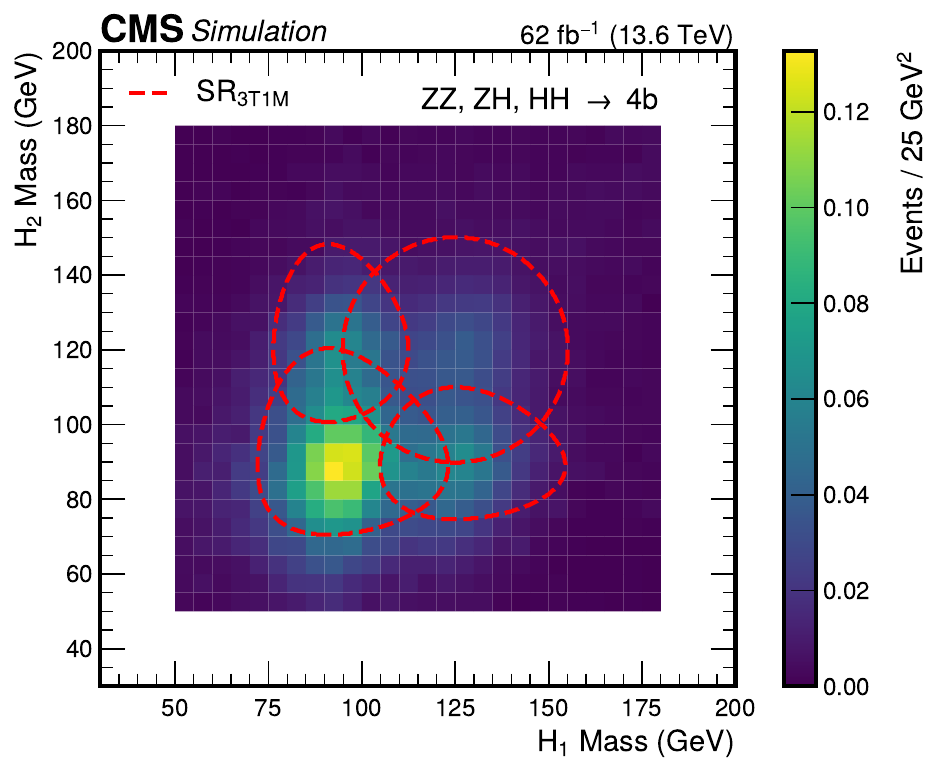}
    \includegraphics[width=0.45\textwidth]{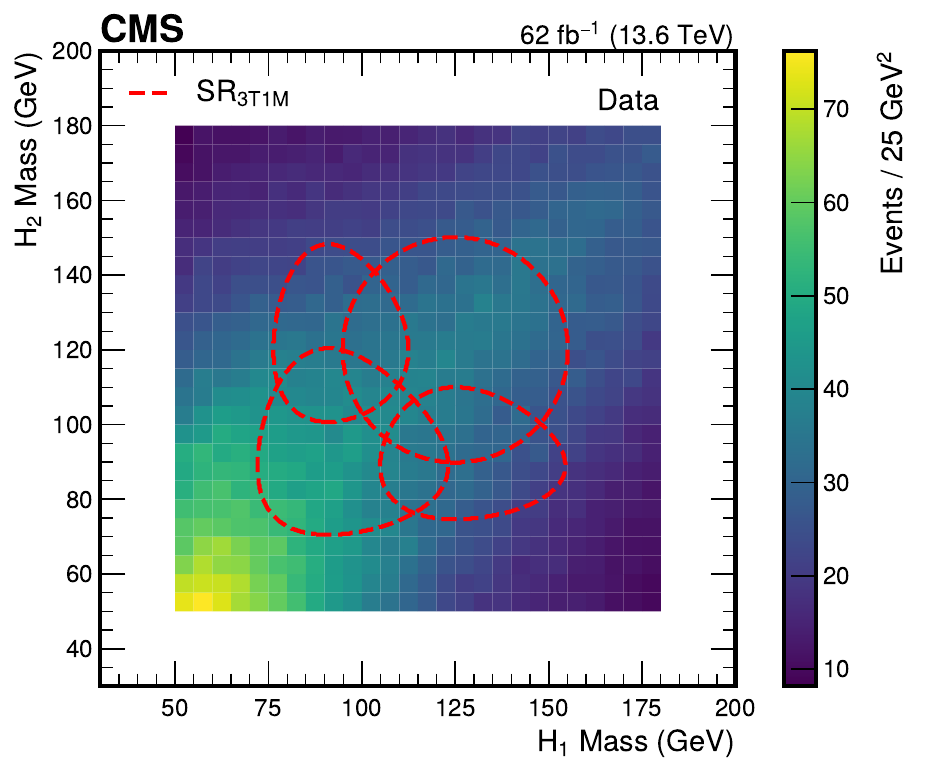}
    \caption{The expected \HH, \ZH, and \ZZ signal yields as estimated from simulation (\cmsLeft) and the observed data (\cmsRight) for the Run~3 data set, in the 3T1M region, as a function of the reconstructed masses of the leading and subleading in \pt \PH candidates. The signal region is defined by the union of the regions enclosed by the dashed red lines.}
    \label{fig:run3_feynnet_SRCR}
\end{figure}

The analysis uses four mutually exclusive regions, defined by the number of jets satisfying the loose, medium, and tight \PNET WPs, and optimized for this approach. These regions include 2\PQb events; events in which the three leading in \PQb tagging score jets satisfy the tight WP while the fourth one can be either medium (3T1M) or loosely (3T1L) \PQb tagged; and events in which the three leading in \PQb tagging score jets satisfy the medium but failing the tight WP (3MnT). For all of  these regions, the four jets leading in \PQb tagging score are considered as the boson candidate jets, and the pairing is performed with the method described in Section~\ref{sec:resolved_common}. Once the two boson candidates are paired, four overlapping regions in the plane defined by the reconstructed dijet masses of the leading ($m_{\Hone}$) and subleading ($m_{\Htwo}$) candidates in boson \pt, are used to target the \HH, \ZZ, and \ZH signals in the 4\PQb final state. Defining the variables $X_{\HZ}$, $X_{\ZH}$, and $X_{\ZZ}$ as:
\begin{equation}
    X_{\HZ} = \sqrt{\left[\frac{m_{\Hone}-125\GeV}{\sigma_{m_{\Hone}}}\right]^2 + \left[\frac{m_{\Htwo}-89\GeV}{\sigma_{m_{\Htwo}}}\right]^2}
    \label{eq:extended_hz}
\end{equation}
\begin{equation}
    X_{\ZH} = \sqrt{\left[\frac{m_{\Hone}-91\GeV}{\sigma_{m_{\Hone}}}\right]^2 + \left[\frac{m_{\Htwo}-120\GeV}{\sigma_{m_{\Htwo}}}\right]^2}
    \label{eq:extended_zh}
\end{equation}
\begin{equation}
    X_{\ZZ} = \sqrt{\left[\frac{m_{\Hone}-91\GeV}{\sigma_{m_{\Hone}}}\right]^2 + \left[\frac{m_{\Htwo}-89\GeV}{\sigma_{m_{\Htwo}}}\right]^2}, 
    \label{eq:extended_zz}
\end{equation}
the SRs take the form:
\ifthenelse{\boolean{cms@external}}{
\begin{multline}
    \text{SR:} \ (R_{\HH} < 30) \ \cup \ (X_{\ZZ} < 2.6) \ \cup \\
    \ (X_{\ZH} < 1.9) \ \cup \ (X_{\HZ} < 1.9).
    \label{eq:extended}
\end{multline}
}
{
\begin{equation}
    \text{SR:} \ (R_{\HH} < 30) \ \cup \ (X_{\ZZ} < 2.6) \ \cup \ (X_{\ZH} < 1.9) \ \cup \ (X_{\HZ} < 1.9).
    \label{eq:extended}
\end{equation}
}
The centroids in $m_{\Htwo}$ in Eqs.~(\ref{eq:extended_hz}) and (\ref{eq:extended_zz}) are chosen to be 89\GeV to reflect a residual \pt dependence in the reconstructed mass after the application of the flavor-inclusive JES corrections described in Section~\ref{sec:ak4_btag_pt_regression}. The denominator in the \ZZ, \ZH, and \HZ terms corresponds to the approximate mass resolution of the reconstructed boson candidates, as estimated from simulation to be about 10\% of the reconstructed dijet mass. The CRs surround the SRs, with their outer bounds defined as:
\begin{equation*}
    \text{CR:} \ (52 < m_{\Hone} < 180\GeV) \ \cup \ (50 < m_{\Htwo} < 173\GeV).
\end{equation*}
The signals are extracted in two regions, \SRone and \SRtwo, where the inclusion of \SRtwo improves the results by approximately 6\%. The boundaries of the SR and CR in the 3T1M region are shown in Fig.~\ref{fig:run3_feynnet_SRCR} for the total \HH, \ZZ, and \ZH yield as taken from simulation and normalized to the expected yield (\cmsLeft) and the observed data (\cmsRight) as a function of $m_{\Hone}$ and $m_{\Htwo}$.

The \FeynNet algorithm structures the four-vectors of the four leading in \PQb tagging score jets ($j$) as well as the two leading in \pt VBF candidates ($f$) with opposite-$\eta$, if present, as a collection of nodes in a graph. Subsequently, it constructs all unique combinations of dijets and builds composite nodes for higher-level structures. These nodes represent the possible decay topologies of the \HH-quadjet and VBF-dijet ($ff$) systems. The input features used include: the four-vectors of all six jets considered, the minimum distance between an \PH and a VBF candidate jet ($\min\Delta R_{jf}$), the distances between each \PH candidate jet ($\Delta R_{jj}$) and VBF candidate jets ($\Delta R_{ff}$), the \PH-dijet candidates ($\Delta R_{\HH}$), and the \HH-quadjet and VBF-dijet system ($\Delta R_{\HH,ff}$). In the case of the boson candidate jets, a flag is provided as an additional input feature to denote the pairing assignment. The \FeynNet algorithm returns a set of event classification probabilities derived from the linear combination of the different objects.

Three sets of trainings of the \FeynNet algorithm are performed. The first one is used to estimate the total background in the SRs using CRs in data. The second training is used to discriminate the \gghh and \qqhh, \ZH, and \ZZ processes from the derived background model and its output scores construct the variables used in the final signal extraction. The third training is used to validate the background estimation procedure in a signal-depleted region. 

To estimate the background in the \SRone (\SRtwo), a \FeynNet classifier is trained using data in CRs to model the kinematic differences between them. For the \SRone, the classifier is trained with data from the most populated CR, \CRtwob, as well as from \CRone. For the \SRtwo, the training uses data from the \CRthree and \CRtwo. Based on the classifier's output probability scores, weights, defined as $w_{\text{3T1M}}=\Prob(\CRone)/\Prob(\CRtwob)$ and $w_{\text{3T1L}}=\Prob(\CRtwo)/\Prob(\CRthree)$, are applied to the data in the signal-depleted \SRtwob and \SRthree, respectively. The reweighted \SRtwob and \SRthree data provide the background estimate for the \SRone and \SRtwo, respectively. For both cases, a $k$-fold ensembling method is applied with $k=30$, where 75 (25)\% of the data events in the CRs are used for the training (evaluation). The choice of $k$ ensures that the estimated variance has converged and remains stable as the number of trained models increases.

\begin{figure}[!htb]
    \centering
    \includegraphics[width=0.40\textwidth]{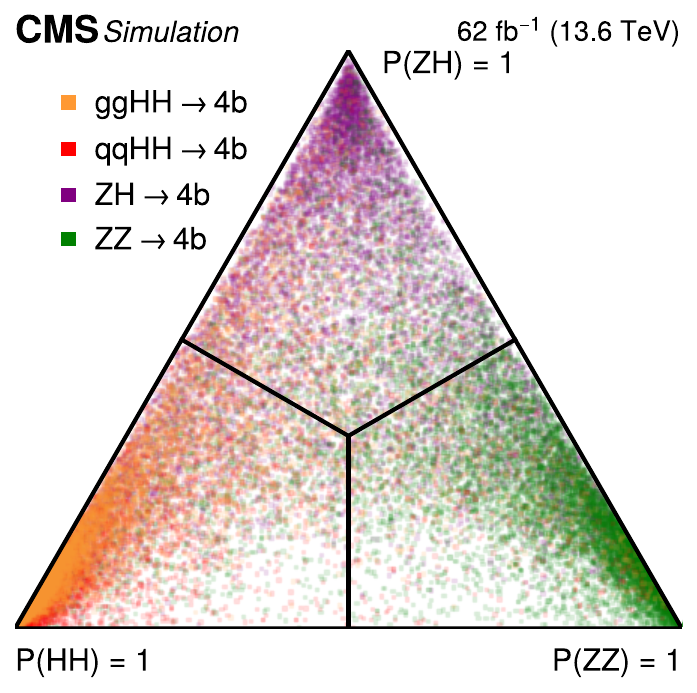}
    \caption{Distribution of the \gghh, \qqhh, \ZZ, and \ZH signal processes, normalized to unity, as a function of the three \FeynNet probability scores.}
    \label{fig:run3_feynnet_probs}
\end{figure}

A second \FeynNet model is trained as a multiclassifier in the \SRone to discriminate the \gghh and \qqhh, \ZH, and \ZZ processes from the background model derived in the previous step. Three nonoverlapping regions targeting \HH, \ZZ, and \ZH production, respectively, are defined according to which of the output signal probabilities is the largest. Figure~\ref{fig:run3_feynnet_probs} shows the distribution of the \gghh, \qqhh, \ZZ, and \ZH signal processes in the plane defined by the three probability scores. About 85\% and 93\% of the \gghh and \qqhh signal events, respectively, are correctly assigned to the \HH region. Small contaminations of the \gghh (\qqhh) events exist in the \ZZ and \ZH regions, with 5\% and 10\% (3\% and 4\%), respectively.

The \HH region is further divided into \gghh and \qqhh regions based on the presence of two VBF candidate jets in the event. To improve the separation of \gghh events in the \qqhh region, a selection is applied on the \gghh-vs-\qqhh score, built from the probabilities:
\begin{equation*}
  D(\text{\gghh-vs-\qqhh}) = \frac{\Prob(\gghh)}{\Prob(\gghh) + \Prob(\qqhh)}, 
\end{equation*}
which returns about 92 (27)\% of the \gghh (\qqhh) events in the \gghh region and keeps about 8 (73)\% of the \gghh (\qqhh) events in the \qqhh region. For each category, an SvsB score is defined as:
\begin{equation}\label{eq:svb-feynnet}
  D(\text{signal-vs-bkg}) = \frac{\Prob(\text{signal})}{\Prob(\text{signal}) + \Prob(\text{bkg})}, 
\end{equation}
and used in the final signal extraction. 

\begin{figure*}[!htb]
\centering
\includegraphics[width=0.35\textwidth]{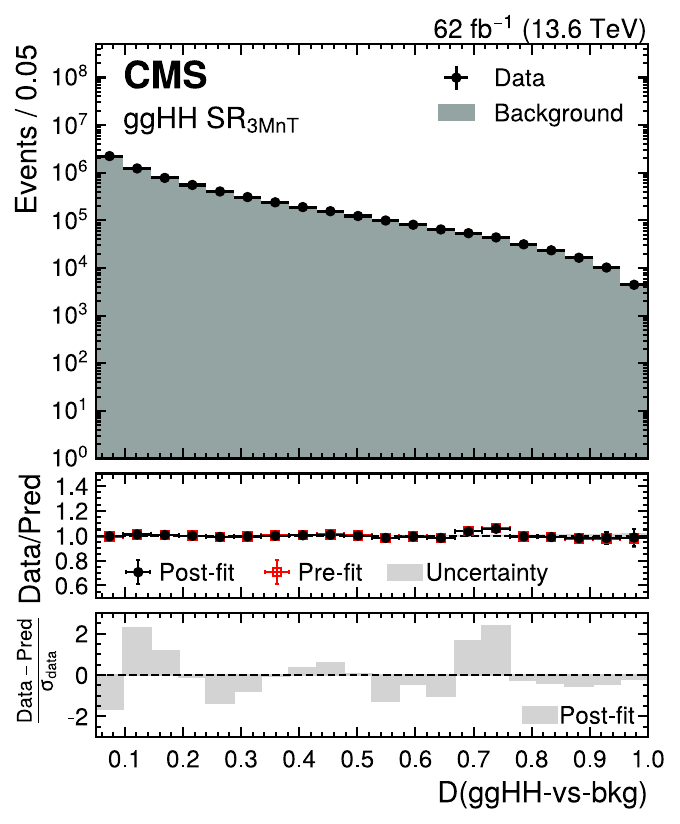}
\includegraphics[width=0.35\textwidth]{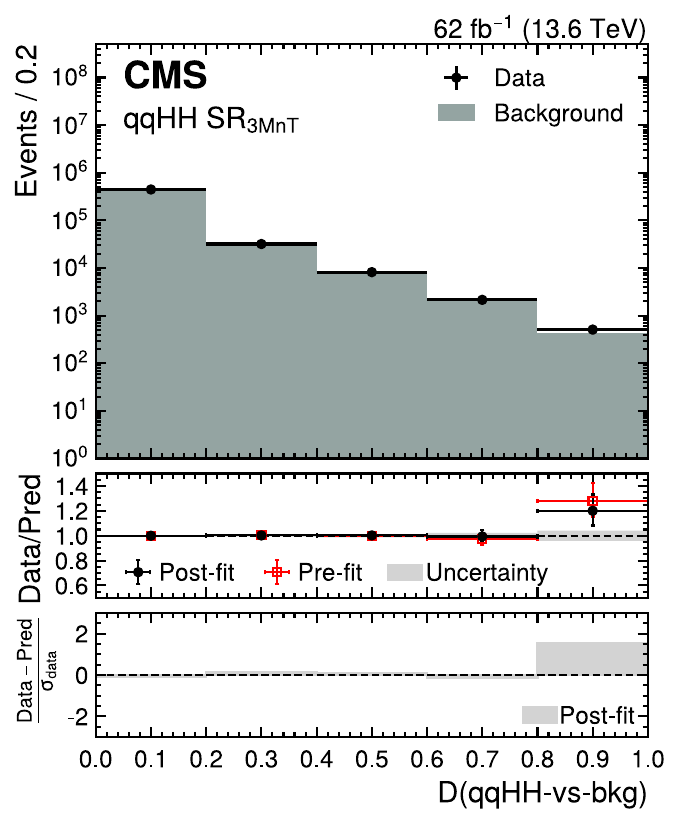}\\
\includegraphics[width=0.35\textwidth]{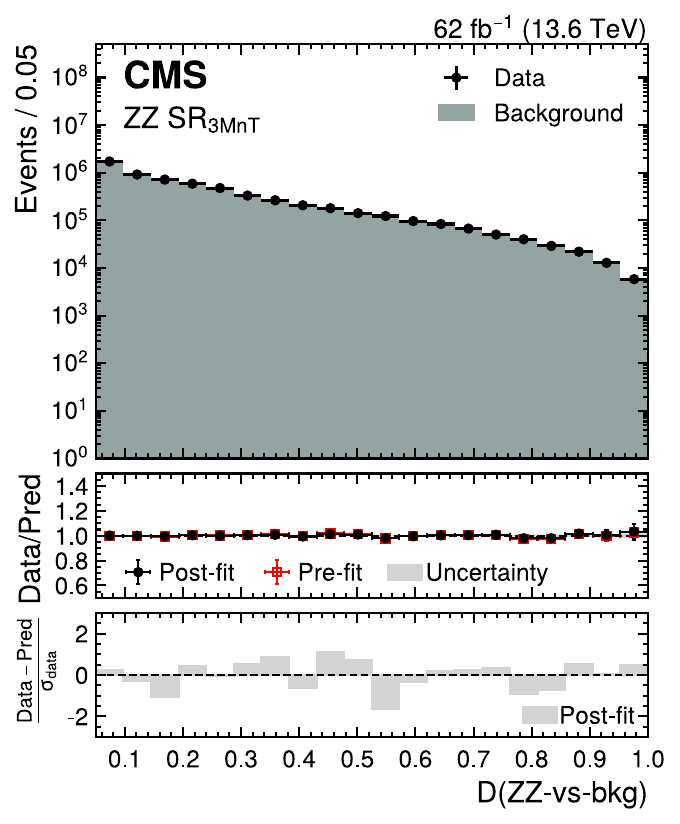}
\includegraphics[width=0.35\textwidth]{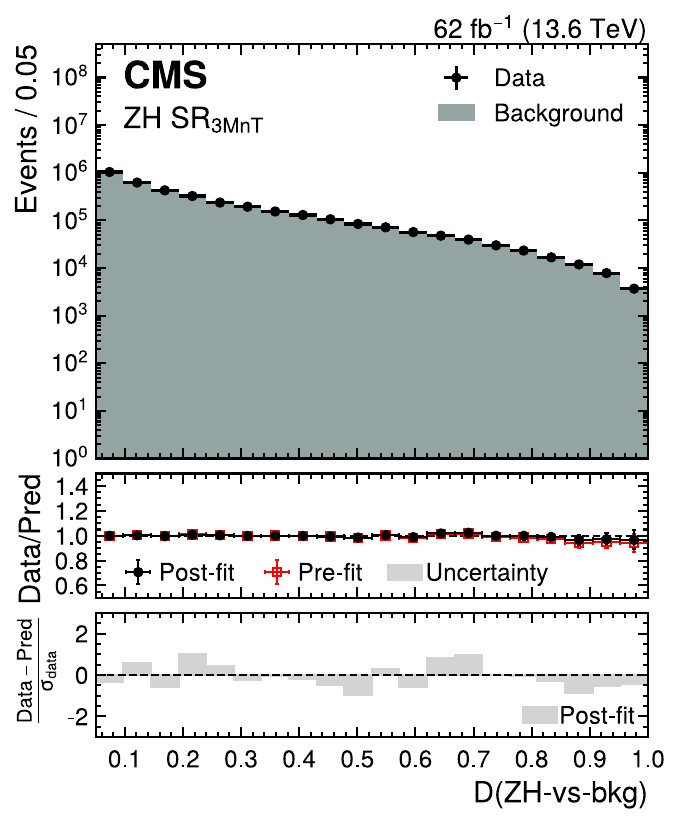}
\caption{Post-fit distribution of the \Dgghh (upper left), \Dqqhh (upper right), \DZZ (lower left), and \DZH (lower right) scores in the validation region \SRthree for data (black points) and the predicted background (gray filled histograms) with the Run~3 data set. Notations are as in Fig.~\ref{fig:dnn_3b_control_region_fitted}.}
\label{fig:feynnet-scores-validation}
\end{figure*}

The background estimation procedure is validated with a third \FeynNet training  in \SRthree. The 3MnT data set is randomly split into 10 subsets and the background estimation procedure is repeated in each of them. The kinematic weights, derived from a training in \CRtwob and \CRthree and  defined as $w_{\text{3MnT}}=\Prob(\CRthree)/\Prob(\CRtwob)$, are applied on the observed data in \SRtwob to estimate the background in the \SRthree. The distributions of the \Dgghh, \Dqqhh, \DZZ, and \DZH scores are shown in Fig.~\ref{fig:feynnet-scores-validation} for the average of the 10 background models and the observed data in the \gghh, \qqhh, \ZZ, and \ZH \SRthree, respectively. A selection on the SvsB scores above 0.05 is applied in the \gghh, \ZZ, and \ZH SRs. This selection is found to improve the agreement between the background model and data in the \SRthree and reduce the systematic uncertainties related to the background estimation without affecting the \HH sensitivity in the \gghh \SRone and \SRtwo. Due to the limited number of events in the \qqhh \SRone, no such selection is applied.

Systematic uncertainties in the background model arise, first, from the finite size of data sample in the CRs used to train the \FeynNet classifier, and second, from the extrapolation to the SR from a different region of phase space. The variance of the background models in \SRone and \SRtwo is estimated by comparing the average of the background models to each of the thirty models trained, as a function of the discriminant scores, defined in Eq.~(\ref{eq:svb-feynnet}). The differences among the models are parameterized with a set of three orthogonal Fourier basis functions, following the procedure used in Ref.~\cite{CMS:2024tdk}, resulting in three uncorrelated shape uncertainties. The same method is repeated separately for the \Dgghh, \DZZ, and \DZH scores, and for the \SRone and \SRtwo. Due to the limited number of events in the \qqhh \SRone, an uncorrelated uncertainty is derived from the root-mean-square of all background model variations. Any residual difference between the average model and the variations is included as an additional uncorrelated uncertainty (up to 2\%).  The extrapolation uncertainty is assessed in the \SRthree, by fitting the observed data in the \SRthree  with the average of the 10 background models. The post-fit pulls of each of the three basis-function coefficients give an estimate of the bias of each component. The magnitude of the pulls is used to assign an additional systematic uncertainty for each of the basis-function coefficients.

\begin{figure*}[!htb]
\centering
\includegraphics[width=0.32\textwidth]{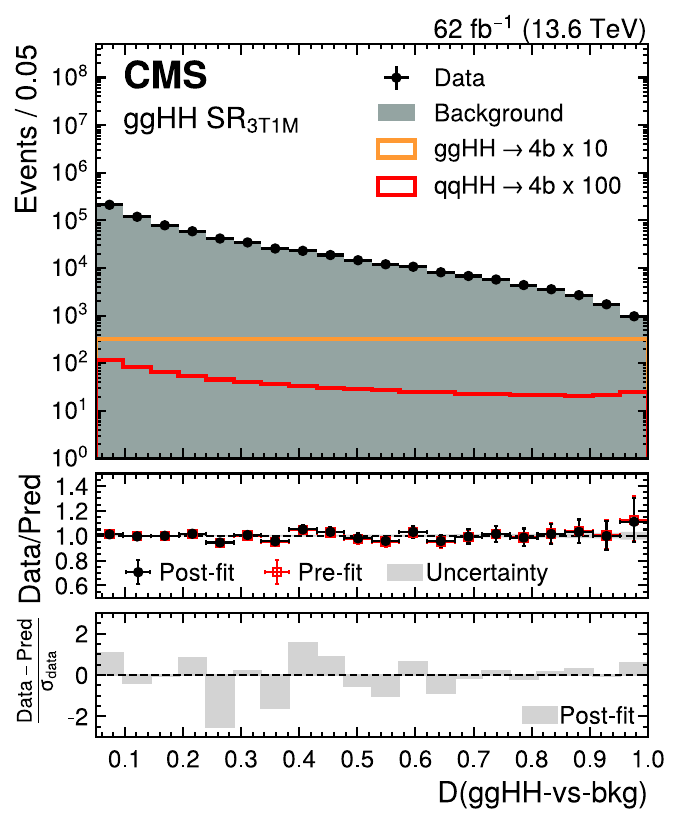}
\includegraphics[width=0.32\textwidth]{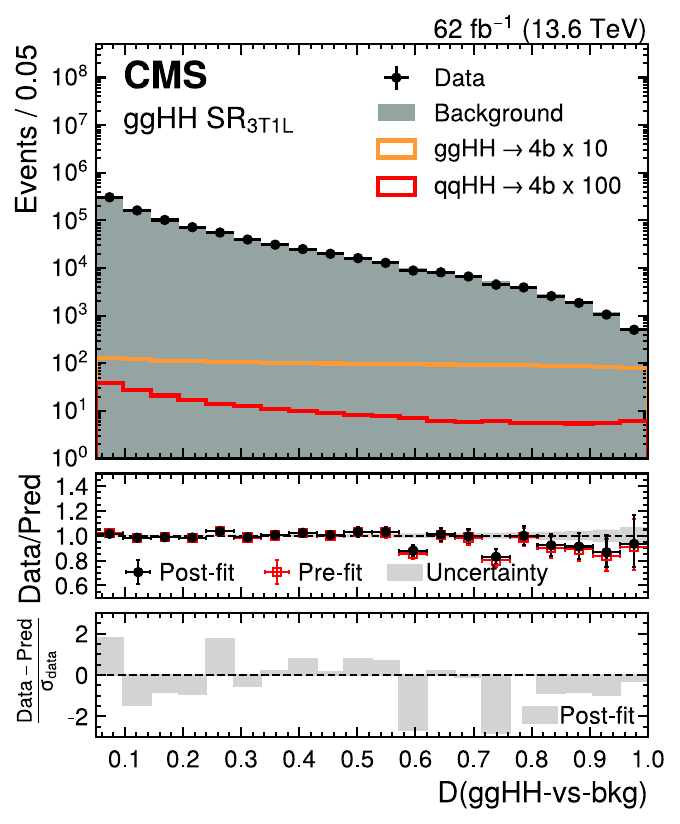}
\includegraphics[width=0.32\textwidth]{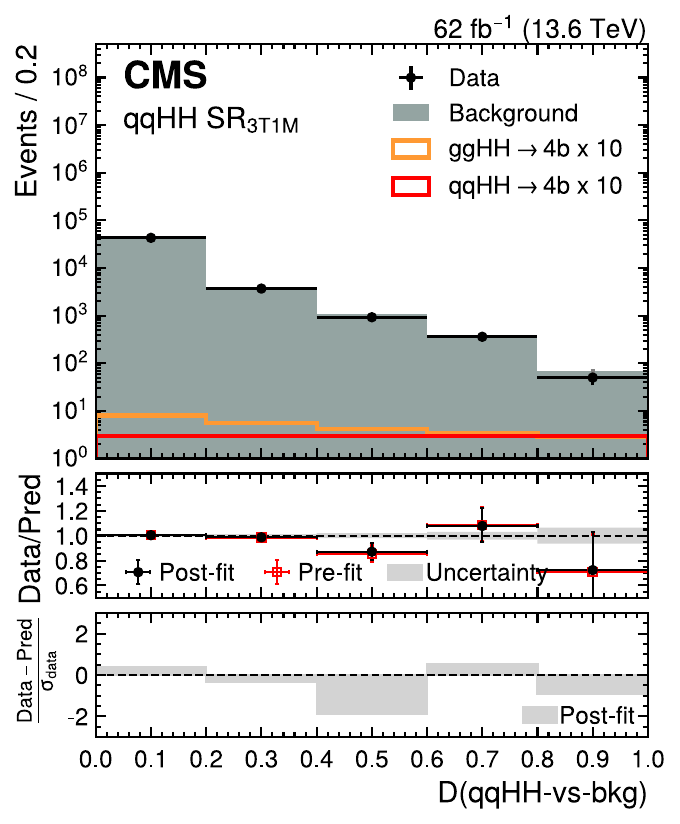}
\caption{Post-fit distributions of the transformed \Dgghh score in the \gghh \SRone (left) and \gghh \SRtwo (middle) categories, and \Dqqhh score in the \qqhh \SRone (right) category for data (black points) and the predicted background (gray filled histograms) for the Run~3 data set. The distributions of the SM \gghh (orange line) and \qqhh (red line) signal processes, scaled to improve their visibility, are also overlaid. Notations are as in Fig.~\ref{fig:dnn_3b_control_region_fitted}.}
\label{fig:feynnet-hh-results}
\end{figure*}

The transformed distributions of the \Dgghh and \Dqqhh scores used in the maximum likelihood fit to extract the \HH signal are shown in Fig.~\ref{fig:feynnet-hh-results} for the \gghh \SRone, \gghh \SRtwo, and \qqhh \SRone categories. 

\subsubsection{Validation with \texorpdfstring{\ZZ}{ZZ} and \texorpdfstring{\ZH}{ZH} processes}

\begin{figure*}[!htb]
\centering
\includegraphics[width=0.32\textwidth]{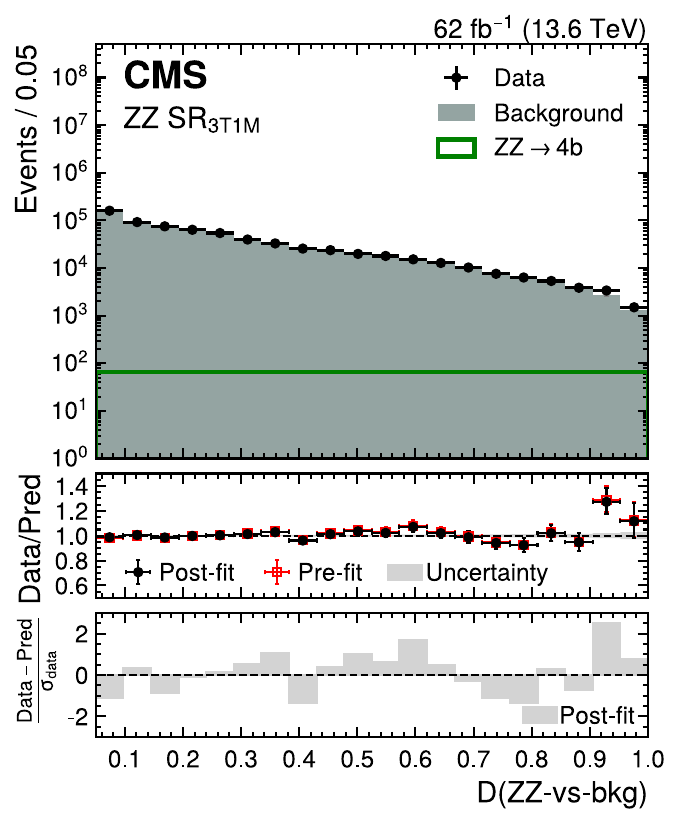}
\includegraphics[width=0.32\textwidth]{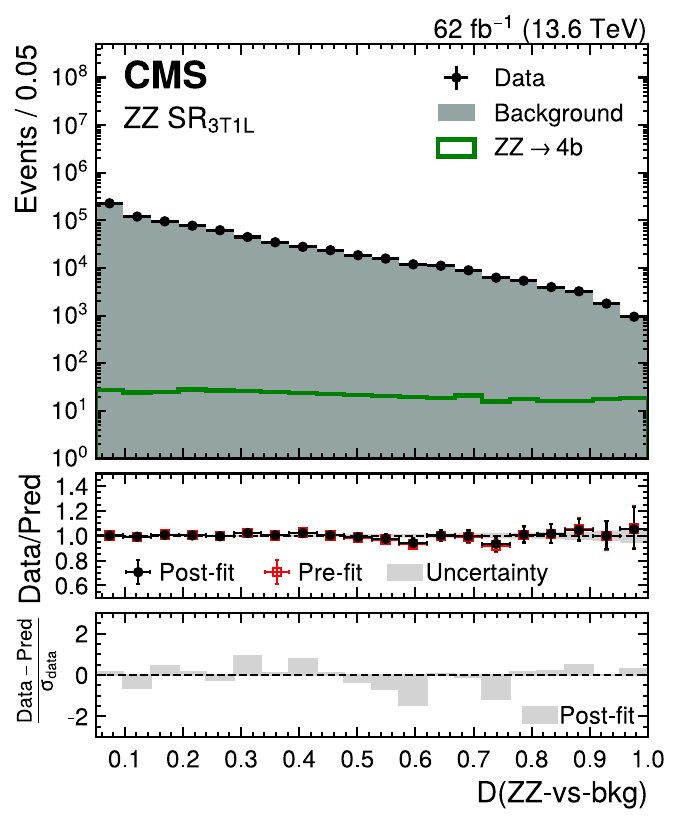}\\
\includegraphics[width=0.32\textwidth]{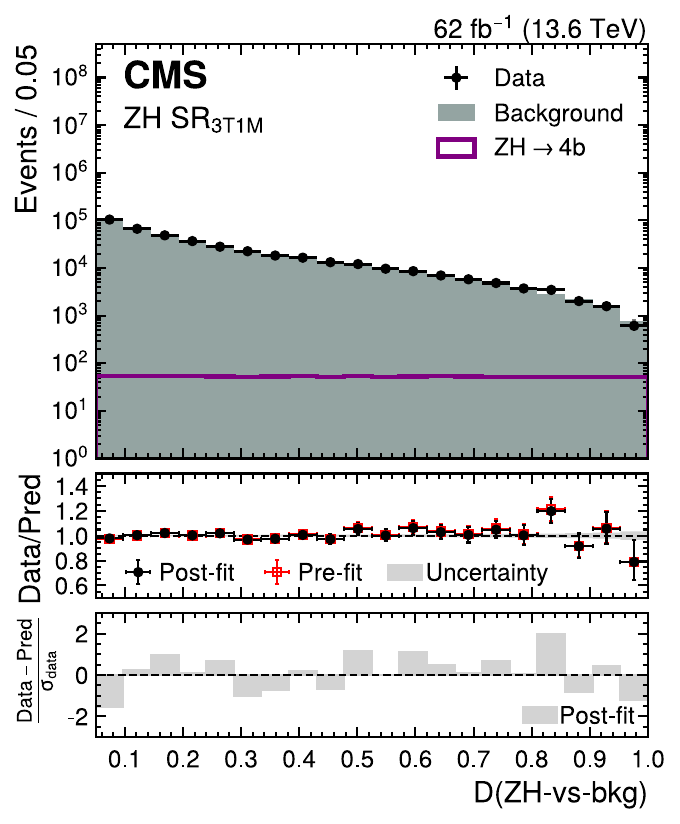}
\includegraphics[width=0.32\textwidth]{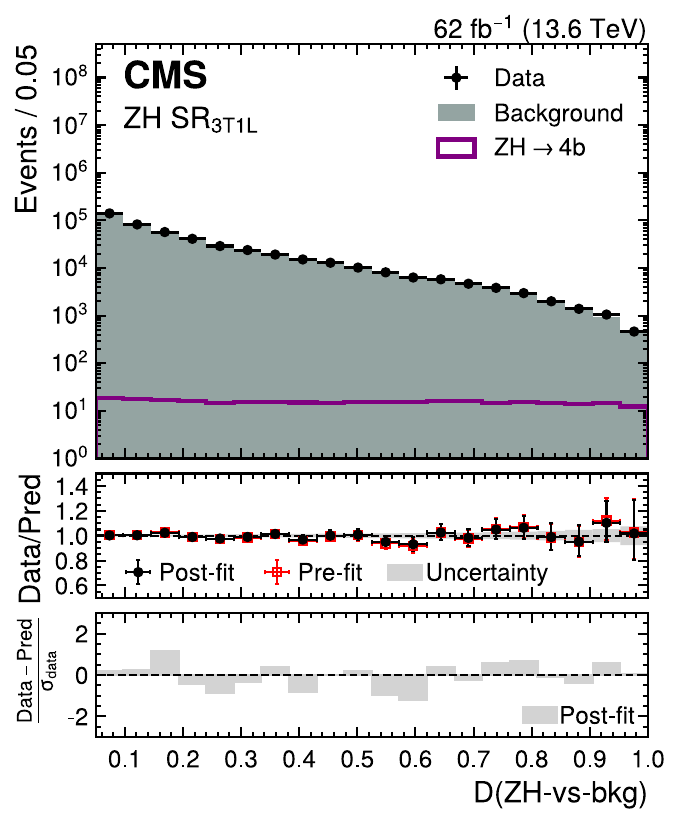}
\caption{Post-fit distribution of the transformed \DZZ score in the \ZZ \SRone (upper left) and \ZZ \SRtwo (upper right) categories, and \DZH score in the \ZH \SRone (lower left) and \ZH \SRtwo (lower right) SR for data (black points) and the predicted background (gray filled histograms) for the Run~3 data set. The distributions of the SM \ZZ (green line) and \ZH (purple line) processes are also overlaid. Notations are as in Fig.~\ref{fig:dnn_3b_control_region_fitted}.}
\label{fig:feynnet-zz-zh-results}
\end{figure*}

A crucial test before extracting the \HH signal is the validation of the analysis strategy using the SM \ZZ and \ZH processes. A search for the SM \ZZ and \ZH processes is performed with a binned maximum likelihood fit on the \DZZ and \DZH  discriminant scores in the exclusive \ZZ and \ZH \SRone and \SRtwo regions, respectively. Figure~\ref{fig:feynnet-zz-zh-results} shows the \DZZ (left) and \DZH (right) discriminant scores, and compares the observed data to the predicted background, with the \ZZ and \ZH signals overlaid in each case, in the \SRone (upper row) and \SRtwo (lower row).

The observed \ZZ and \ZH signal strengths are reported in Table~\ref{tab:resolved-zz-zh-run3-run2}, in terms of the signal strength modifiers, \muzz and \muzh, defined as ${(\sigma\mathcal{B})_{\text{obs}}/(\sigma\mathcal{B})_{\text{SM}}}$, where $\sigma$ is the signal production cross section and $\mathcal{B}$ is the branching fraction. Upper limits at 95\% \CL on the signal strengths are also set, assuming no signal exists. Systematic uncertainties, described in Section~\ref{sec:resolved_systematics}, affecting the signal models and the background prediction are incorporated in the fit via nuisance parameters. Results are determined following the CMS statistical procedures, which are described in detail in Section~\ref{sec:resolved_results}.

The measured signal strength for the \ZZ process is ${\muzz=3.5^{+2.0}_{-1.8}}$, which is compatible with the SM expectation at the level of 2 standard deviations. For the \ZH process, the measured signal strength is  ${\muzh=0.2^{+1.6}_{-1.5}}$. Compared to previous results reported in Ref.~\cite{CMS:2024tdk} using the LHC Run 2 data, a similar sensitivity is achieved with almost half the integrated luminosity. This sensitivity gain can be attributed to a variety of improvements in the Run~3 methods, as described in Section~\ref{sec:objet}, and validates the relative improvements obtained for the \HHbbbb limits shown in Section~\ref{sec:resolved_results}. Expanding the analysis SR from the \HH-sensitive SR (${R_{\HH} < 30\GeV}$) to the SR defined in Eq.~(\ref{eq:extended}), allows for a simultaneous search for the \HH, \ZZ, and \ZH processes with a small effect on the \HH sensitivity (about 2\% loss). The combinations of the Run 2~\cite{CMS:2024tdk} and Run~3 data sets in the \ZZ and \ZH categories are summarized in Table~\ref{tab:resolved-zz-zh-run3-run2} and yield observed (expected) upper limits at 95\% \CL of 4.2 (2.6) for \muzz and 3.2 (2.1) for \muzh.

\begin{table*}[!htb]
  \centering
  \topcaption{The observed and expected upper limits at 95\% \CL on the \muzz and \muzh, observed signal strength and significance for the Run 2~\cite{CMS:2024tdk} and Run~3 data sets and their combination. The upper limits on \muzz (\muzh) are obtained from a fit on the \DZZ (\DZH) score under the hypothesis of no \ZZ (\ZH) signal, while the \gghh, \qqhh and \ZH (\ZZ) processes are fixed to their SM values. The uncertainties given correspond to the 68\% \CL intervals.}
  \renewcommand{\arraystretch}{1.2}
  \begin{scotch}{lcc{c}@{\hspace*{5pt}}cc{c}@{\hspace*{5pt}}cc}
  & \multicolumn{2}{c}{Run~2~\cite{CMS:2024tdk}} && \multicolumn{2}{c}{Run~3} && \multicolumn{2}{c}{Run~2 and Run~3} \\
	  \cline{2-3}\cline{5-6}\cline{8-9}
  & \muzz & \muzh && \muzz & \muzh && \muzz & \muzh \\ \hline
  Expected upper limit at 95\% \CL & 3.8 & 2.9 && 3.8 & 3.1 && 2.6 & 2.1 \\
  Observed upper limit at 95\% \CL & 3.9 & 4.8 && 7.1 & 3.5 && 4.2 & 3.2 \\
  Observed signal strength & $0.2^{+1.8}_{-1.7}$ & $2.1^{+1.6}_{-1.4}$ && $3.5^{+2.0}_{-1.8}$ & $0.2^{+1.6}_{-1.5}$ && $1.9^{+1.3}_{-1.3}$ & $1.2^{+1.1}_{-1.1}$ \\
  Observed significance & 0.1 & 1.5 && 2.0 & 0.2 && 1.5 & 1.2 \\
  \end{scotch}
  \label{tab:resolved-zz-zh-run3-run2}
\end{table*}

\subsection{Improved analysis with Run 2 data} \label{sec:resolved_run2}

This analysis targets the \gghh process in the resolved topology, and extends the search for \ZZbbbb and \ZHbbbb using Run~2 data, described in Ref.~\cite{CMS:2024tdk}.

Events are initially selected using the Run~2 triggers sensitive to three \PQb-tagged jets, identified with the \DeepCSV algorithm~\cite{CMS:2017wtu}. The offline selection further requires at least four AK4 jets with ${\pt >40\GeV}$ and ${\abs{\eta} < 2.4}$. The \DeepJet algorithm identifies the four jets with the highest \PQb tagging scores, designating them as the \HH candidates. A subsequent pairing scheme is optimized to reconstruct \ZZ, \ZH, and \HH events in the $m_{\Hone}$--~$m_{\Htwo}$ plane. To maximize the analysis sensitivity, the \PQb tagging WP used to define the 3\PQb and 4\PQb regions is optimized based on the expected limits on \HH production. 

A key difference from the approaches described in Sections~\ref{sec:resolved_transformer} and \ref{sec:resolved_graph}, as well as the previous result with Run~2 data~\cite{CMS:2022cpr}, lies in the background estimation strategy. In this approach, the \ttbar contribution is taken from simulation, while the dominant QCD multijet background is derived from data. The measurement of the QCD multijet background is performed using events in the 3\PQb region, defined to be exclusive of the events in the 4\PQb region. Due to the tighter \PQb tagging requirements of the Run~2 triggers, the 2\PQb region is inaccessible. To enhance the statistical power of the data in the 3\PQb region, the offline \DeepJet WP is relaxed to a WP corresponding to a 70--80\% efficiency, as measured in simulated events with \PQb jets. These 3\PQb events are then weighted using two factors derived from CRs in data to predict the QCD multijet background in the 4\PQb region. The first weight accounts for the differences in jet and \PQb jet multiplicities between the two samples. It is calculated by parameterizing a functional form based on the number of jets and \PQb jets in a signal-depleted region of the $m_{\Hone}$--~$m_{\Htwo}$ plane, as described in Ref.~\cite{CMS:2024tdk}. The second weight addresses residual kinematic differences during the extrapolation from the 3\PQb to the 4\PQb region.
Both weights are derived from a multivariate classifier based on the hierarchical combinatorial residual (HCR) architecture introduced in Ref.~\cite{CMS:2024tdk}, which is built on a series of convolutional neural networks. The HCR classifier is designed to separate the QCD multijet and \ttbar components in data events with three and four \PQb-tagged jets. The input features of the network are the four-momenta of the jets, subject to the aforementioned kinematic requirements.

\begin{figure}[!hbt]
\centering
\includegraphics[width=0.425\textwidth]{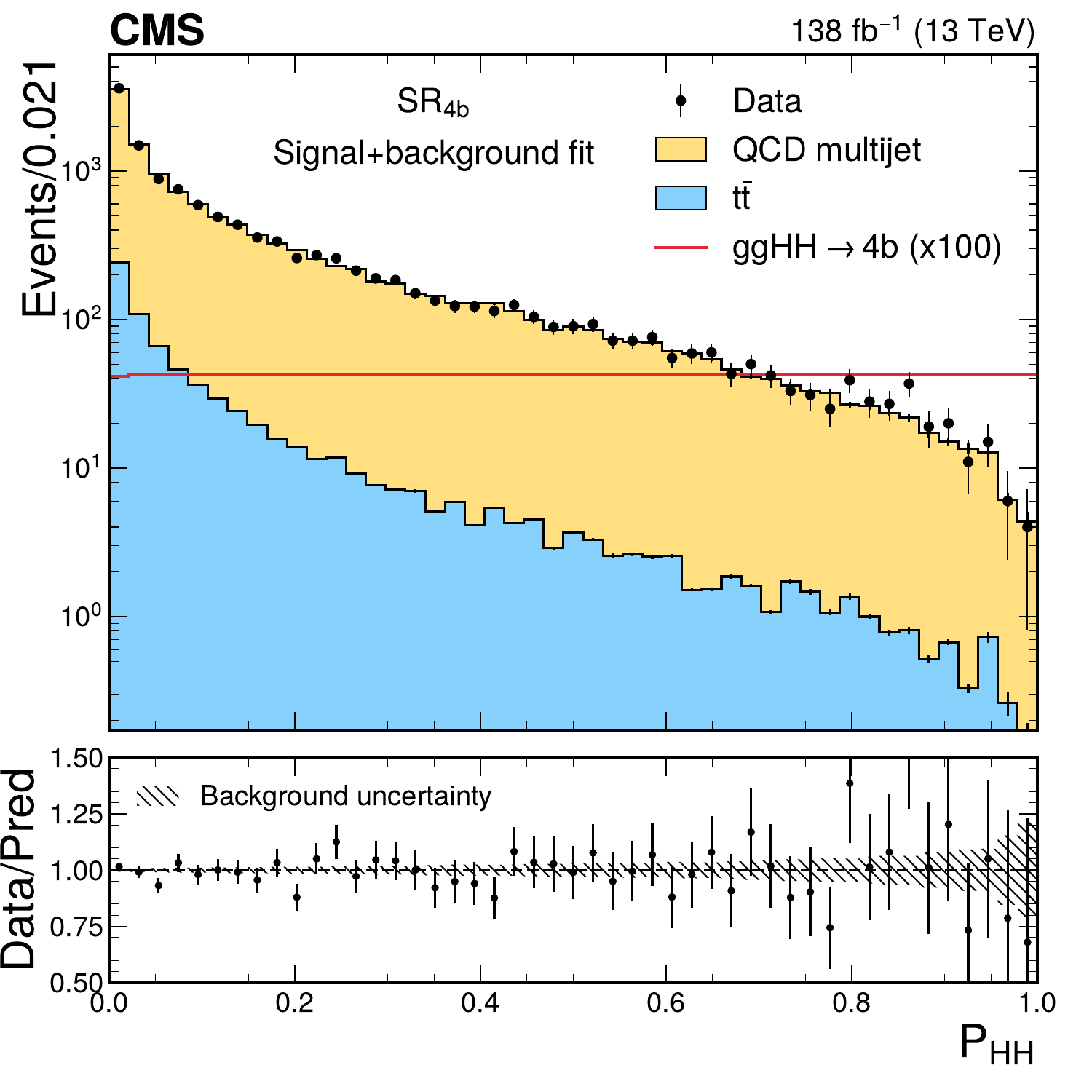}
\caption{The fitted signal+background distribution of the signal probability, \psig, in the \HH \regSRFourb. The black points show the 4\PQb events from data. The yellow and blue regions show the predictions from the QCD multijet model and the \ttbar simulation, respectively. The prediction for the SM \gghh signal distribution is given by the red histogram, multiplied by 100. The lower panel shows the data-to-background ratio, with the hatched area representing the background uncertainty.}
\label{fig:run2-resolved}
\end{figure}

The same HCR architecture is also trained to distinguish signal from background. Five classes are considered, each yielding an output probability score for the QCD multijet, \ttbar, \ZZ, \ZH, and \HH processes. The training leverages simulations for all processes, except for the QCD multijet process, which is estimated from data. Mutually exclusive signal extraction regions are defined according to which of the \HH, \ZH, and \ZZ output probabilities is the largest. The sum of the three aforementioned probabilities for events where the \HH probability is largest, \psig, is monotonically transformed such that the expected \gghh signal distribution in the \HH \regSRFourb is uniform. The \gghh signal is then measured via a maximum-likelihood fit to the transformed \psig distribution. Figure~\ref{fig:run2-resolved} shows the transformed \psig in the \HH \regSRFourb for the observed data, the QCD multijet background estimate from CRs in data, the \ttbar background from simulation, and the expected \gghh signal.

Systematic uncertainties in the QCD multijet prediction are evaluated utilizing 3\PQb data events to construct a synthetic data set. This data set is specifically designed to have the same statistical power as the 3\PQb data events, which notably exceeds that of the 4\PQb events by a factor of fifteen. Comprehensive details regarding the construction and characteristics of this data set are provided in Ref~\cite{CMS:2024tdk}. To ensure a robust assessment of these uncertainties, the QCD multijet estimation procedure is iteratively applied to fifteen distinct subsamples from this synthetic data set. The resulting differences in the signal-vs-background distributions among these fifteen QCD multijet predictions and their overall average are then parameterized by a set of orthogonal Fourier basis functions, thereby allowing for the systematic uncertainties to be assigned to the nominal background model.

\subsection{Systematic uncertainties}  \label{sec:resolved_systematics}

Systematic uncertainties can affect both the predicted number of signal events and the predicted shape of the SvsB classifier score used for the \HHbbbb signal extraction. Theoretical uncertainties arise from the limited precision in the computation of the inclusive and differential cross sections of the simulated \gghh and \qqhh processes, and their effect is fully correlated between event categories and data-taking periods. Experimental uncertainties comprise those (i) related to the imperfect simulation of the detector response, (ii) in the integrated luminosity measurement, and (iii) in the modeling of the pileup interactions. All systematic uncertainties are modeled in the signal extraction fit as nuisance parameters with log-normal or Gaussian external constraints, as described in Section~\ref{sec:resolved_results}.

Theoretical uncertainties affecting the inclusive SM \gghh cross section include the choice of the PDF set ($\pm1.5\%$), variation in the QCD $\mu_{\text{R}}$ and $\mu_{\text{F}}$ scales ($+2.1\%,-4.9\%$), the value of $\alpS$ ($\pm 1.7\%$), and the uncertainty related to the value and definition of the top quark mass ($+4\%,-18\%$). The values of these uncertainties are taken from Refs.~\cite{Grazzini:2018bsd,PhysRevD.103.056002}. The theoretical uncertainty in the \hbb branching fraction is 2.5\%~\cite{deFlorian:2016spz}. Uncertainties in the inclusive \qqhh production cross section from QCD scale variations ($+0.05\%,-0.03\%$), PDF and $\alpha_S$ ($\pm2.7\%$) are taken from Refs.~\cite{Ling:2014sne,Dreyer:2018rfu,Dreyer:2018qbw}. The dependence on \kappal of the inclusive cross section and event kinematic features for both \gghh and \qqhh production, and related uncertainties, is accounted for when interpreting the results of this analysis for non-SM hypotheses. Theoretical uncertainties in the predicted \HHbbbb signal acceptance in the SR are determined from simulated events. These include the uncertainty arising from the limited knowledge of \textsc{NNPDF} parameters, evaluated following Ref.~\cite{Butterworth:2015oua}, variations in the $\mu_{\text{R}}$ and $\mu_{\text{F}}$ scales by a factor 2 or 0.5, and variations of initial- and final-state radiation scales of the parton shower.

Experimental uncertainties with the largest impact impact on the signal prediction are those related to the JES and JER, which originate from various sources with limited correlations~\cite{CMS:2016lmd}. These uncertainties depend on the jet kinematics ($\pt$, $\eta$) and flavor composition. Variations of the jet four-momenta caused by shifts in the JES and JER have an impact on both the number of events accepted in the SR and the shape of the SvsB classifier score distribution. As described in Section~\ref{sec:ak4_btag_pt_regression}, the JES and JER closure of the post-regression jet \pt was assessed on $\zmm+\PQb$ data and residual differences between data and simulation in the post-regression JES (1--2\%) and JER (4--8\%) are included as systematic uncertainties, correlated across event categories but uncorrelated among data-taking periods. Uncertainties in the \PQb tagging efficiency also have an important impact on the \HHbbbb signal prediction and, as for JES and JER, originate from a number of independent sources~\cite{CMS-DP-2024-024,CMS-DP-2024-025}. The uncertainty sources with the largest impact are the systematic uncertainty components in the measurement of the \PQb jet tagging efficiency and the light-flavor quark or gluon jet misidentification rate. Uncertainties in the measured simulation-to-data trigger efficiency SFs, in the integrated luminosity, and associated with the modeling of pileup interactions are also considered but have a small impact on the result.

The background prediction is fully based on CRs in data, in both the \gghh and \qqhh categories and the normalization of the total background yield is fit to data, with no prior constraint, independently for pre-ParkingHH and post-ParkingHH data. The background prediction in each bin of the SvsB classifier score has an uncertainty related to the limited count of the \regSRTwob data events used in the background reweighting procedure. Random fluctuations in the background model prediction are accounted for via independent binwise uncertainties, which consider both the statistical uncertainty in the \regSRTwob data and the variance among individual $k$-fold background predictions. These uncertainties are treated as uncorrelated across the bins of the SvsB score distributions and are modeled using the Barlow--Beeston ``lite`` technique~\cite{Conway:2011in,BARLOW1993219}. In the \gghh category, shape uncertainties affecting the background prediction as a function of the SvsB score distributions are included using a set of independent Fourier basis functions, independently for pre-ParkingHH and post-ParkingHH data.  These account for the statistical variance and potential extrapolation bias in the background prediction. The methods used to quantify and account for these shape uncertainties in the background prediction are described in detail in Sections~\ref{sec:resolved_transformer} and~\ref{sec:resolved_graph} for the first and second approaches, respectively. The shape uncertainties are at the percent-level at low and intermediate SvsB classifier score, increasing to as much as 5--6\% in the highest-score bins. Because of the limited number of data events in the \qqhh category and the coarse binning used, the binwise uncertainty is sufficient to account for any variance or bias effects.

\subsection{Results} \label{sec:resolved_results}

The \HHbbbb signal, from both the \gghh and \qqhh production modes, is extracted via a binned maximum likelihood fit to the output of the SvsB classifier in the SRs. The distributions in data collected in the different data-taking periods are fitted simultaneously. The signal strength modifiers \mugghh and \muqqhh for the \gghh and \qqhh production modes, respectively, are defined as the ratio between the measured \HHbbbb rate and the expectation from the SM, while \muhh similarly corresponds to the total \HHbbbb signal rate including both production modes.
Confidence intervals on the signal strengths are estimated using a profile likelihood ratio test statistic~\cite{Cowan:2010js}, in which systematic uncertainties are modeled as nuisance parameters. 
Upper limits at 95\% \CL on the corresponding production cross sections are derived using the modified frequentist \CLs criterion~\cite{ATLAS:2011tau} with an asymptotic approximation~\cite{Junk:1999kv,Read:2002hq}. Results are determined using the CMS statistical analysis tool \textsc{Combine}~\cite{CMS:2024onh}, which is based on the \textsc{RooFit}~\cite{Verkerke:2003ir} and \textsc{RooStats}~\cite{Moneta:2010pm} frameworks. The results presented in this paper for \muqqhh are determined without any constraint on \mugghh, which is profiled in the relevant fits.

Table~\ref{tab:resolved_expected} summarizes the expected 95\% \CL upper limits and best fit signal strengths on \mugghh, \muqqhh, and \muhh from the two approaches described in Sections~\ref{sec:resolved_transformer} and \ref{sec:resolved_graph}. The first approach achieves stronger expected exclusion values compared to the second approach for the signal strength modifiers, as well as for anomalous values of \kappal and \kappaVV. Table~\ref{tab:resolved_observed} summarizes the measured signal strength modifiers and the corresponding 68\% \CL intervals, as well as the observed 95\% \CL upper limits for each signal strength modifier. No significant excess is observed over the background-only expectation, and the measured signal rates are well compatible between the two approaches.

\begin{table}[!htb]
  \centering
  \topcaption{Expected best fit values and 95\% \CL upper limits (U.L.) for \muhh, \mugghh, and \muqqhh in the \HHbbbb resolved analysis following the two approaches described in Sections~\ref{sec:resolved_transformer} and ~\ref{sec:resolved_graph}, respectively. The uncertainties given for the best fit signal strengths correspond to the 68\% \CL intervals. The best fit signal strengths are calculated with a SM \HHbbbb signal injected, while the upper limits are calculated in the absence of signal.}
  \renewcommand{\arraystretch}{1.2}
  \begin{scotch}{l c c c c c c }
     & \multicolumn{2}{c}{\muhh} & \multicolumn{2}{c}{\mugghh} & \multicolumn{2}{c}{\muqqhh} \\ [\cmsTabSkip]
    \hline
    Expected & Best fit & U.L. & Best fit & U.L. & Best fit & U.L. \\
    \hline
    First approach & $1.0^{+2.6}_{-2.4}$ & 5.3 & $1.0^{+2.6}_{-2.4}$ & 5.3 & $1^{+46}_{-39}$ & 94 \\
    Second approach & $1.0^{+2.9}_{-2.7}$ & 6.1 & $1.0^{+3.0}_{-2.7}$ & 6.2 & $1^{+58}_{-50}$ & 128 \\
  \end{scotch}
  \label{tab:resolved_expected}
\end{table}

\begin{table}[!htb]
  \centering
  \topcaption{Observed best fit values and 95\% \CL upper limits (U.L.) for \muhh, \mugghh, and \muqqhh in the \HHbbbb resolved analysis following the two approaches described in Sections~\ref{sec:resolved_transformer} and \ref{sec:resolved_graph}, respectively. The uncertainties given for the best fit signal strengths correspond to the 68\% \CL intervals.}
  \renewcommand{\arraystretch}{1.2}
  \begin{scotch}{l c c c c c c }
   & \multicolumn{2}{c}{\muhh} & \multicolumn{2}{c}{\mugghh} & \multicolumn{2}{c}{\muqqhh} \\  
    \hline
    Observed & Best fit & U.L. & Best fit & U.L. & Best fit & U.L. \\
    \hline
    First approach & $0.9^{+2.6}_{-2.4}$ & 6.3 & $0.9^{+2.6}_{-2.4}$ & 6.4 & $-3.0^{+45}_{-39}$ & 93 \\
    Second approach & $-1.3^{+2.7}_{-2.9}$ & 5.0 & $-1.1^{+2.7}_{-2.9}$ & 5.2 & $-62^{+46}_{-44}$ & 81 \\
  \end{scotch}
  \label{tab:resolved_observed}
\end{table}

The dominant uncertainty arises from the limited number of observed data events, contributing to about 80\% of the total uncertainty in the measured \muhh. The systematic uncertainty sources with the largest impact are those related to the background prediction, as described further in Section~\ref{sec:combinations}. Despite significant differences in the implementation of the major analysis components and in the assessment of the background shape uncertainties, the impacts are compatible between the two approaches. This gives further confidence in the reliability and stability of the uncertainties employed in the background model prediction. 

In addition, 95\% \CL upper limits on \muhh are computed as functions of the \kappal and \kappaVV coupling modifiers separately, with all couplings except those being fixed to the SM prediction. The observed and expected \kappal and \kappaVV intervals outside of which the theoretical prediction for the \HH production cross section is excluded at 95\% \CL are summarized in Table~\ref{tab:resolved_kappas}.

\begin{table*}[!htb]
  \centering
  \topcaption{Observed and expected, in absence of signal, intervals for \kappal and \kappaVV outside of which the theoretical prediction for the \HH cross section is excluded at 95\% \CL in the \HHbbbb resolved analysis, following the two approaches described in Sections~\ref{sec:resolved_transformer} and \ref{sec:resolved_graph}, respectively.}
  \renewcommand{\arraystretch}{1.2}
  \begin{scotch}{l c c c c}
    \multirow{2}{*}{ 95\% \CL interval}  & \multicolumn{2}{c}{\kappal} & \multicolumn{2}{c}{\kappaVV} \\ [\cmsTabSkip]
    & Observed & Expected & Observed & Expected \\
    \hline
    First approach  & $[-4.4, 11.7]$ & $[-3.8,11.0]$ & $[-0.08, 2.20]$ & $[-0.08, 2.19]$ \\ 
    Second approach & $[-3.2, 10.0]$ & $[-4.4, 11.6]$ & $[-0.05, 2.24]$ & $[-0.25, 2.4]$ \\
  \end{scotch}
  \label{tab:resolved_kappas}
\end{table*}

The sensitivity of this measurement is substantially improved compared to the previously published result based on Run~2 data~\cite{CMS:2022cpr}. Scaled to the same integrated luminosity, the improvement in expected limit is more than a factor 2. This major step forward in the CMS \HH measurement program implies that, as of today, the \HHbbbb resolved analysis is the most sensitive individual analysis in probing the \PH self-coupling for a given integrated luminosity. The large improvement arises from a combination of key factors: a fully redesigned trigger strategy, innovative new methods in \PQb jet tagging and \pt regression, and several new analysis techniques.

\subsubsection{Updated results with Run 2 data}\label{sec:results_resolved_run2}

The \HHbbbb signal is extracted via a binned maximum likelihood fit to the SvsB distribution in the \HH \regSRFourb. Systematic uncertainties are treated as nuisance parameters with either Gaussian (shape uncertainties) or log-normal (normalization uncertainties) function priors included in the likelihood function. The observed (expected) signal strength \mugghh is found to be $3.1^{+2.4}_{-1.9}$ ($1.0^{+2.1}_{-1.9}$). The 95\% \CL observed and expected upper limits on the signal strength are set to be $10.0$ and $5.9^{+3.8}_{-1.9}$, respectively. Compared to the previous measurement, the updated result achieves about a 25\% improvement in the expected upper limit. The analysis furthermore benefits from an improved background modeling and a more robust treatment of associated uncertainties. Constraints are also set on the \kappal and the observed (expected) allowed intervals outside of which the theoretical prediction for \HH production is excluded at 95\% \CL are $[-7.7, 16.2]$ ($[-3.8, 11.0]$). These new results, summarized in Table~\ref{tab:resolved_runtwo}, supersede the prior published results on the \gghh signal strength reported in Ref.~\cite{CMS:2022cpr}.

\begin{table}[!htb]
   \centering
   \topcaption{Expected and observed \mugghh and their corresponding upper limits (U.L.) at 95\% \CL in the \HHbbbb resolved channel with Run 2 data (described in Section~\ref{sec:resolved_run2}). The observed and expected intervals outside of which the theoretical prediction for the \HH cross section is excluded at 95\% \CL for \kappal are also reported.}
   \renewcommand{\arraystretch}{1.2}
   \begin{scotch}{l l l c}
     \mugghh & \multicolumn{1}{c}{Best fit} & \multicolumn{1}{c}{U.L.} & \kappal \\ [\cmsTabSkip]
     \hline
     Expected & $1.0^{+2.1}_{-1.9}$ & $5.9^{+3.8}_{-1.9}$ & $[-3.8, 11.0]$ \\
     Observed & $3.1^{+2.4}_{-1.9}$ & 10.0 & $[-7.7, 16.2]$ \\
   \end{scotch}
   \label{tab:resolved_runtwo}
\end{table}

\section{The merged topology}\label{sec:merged_analysis}

\subsection{Overview} \label{sec:merged_common}

The merged category targets \HH events in which both Higgs bosons are highly Lorentz boosted. In this regime, the \PQb quarks from each \PH decay are emitted in a collimated configuration and their hadronization products can be reconstructed as a single AK8 jet. For \PH with ${\pt \gtrsim 300\GeV}$, this approach is more effective than using a pair of AK4 jets. Although the QCD multijet background remains substantial, its differential cross section falls more steeply with increasing event energy compared to the \HH signal, improving signal-vs-background discrimination at high \pt. As a result, despite representing a small fraction of the inclusive \HH signal, merged topologies offer enhanced sensitivity to rare processes, as demonstrated in the previous Run~2 search~\cite{CMS:2022fxs}.

The Run~3 data set is analyzed in the merged topology by leveraging the dedicated trigger paths and improved \bbbar-identification introduced during Run~3. Events passing the trigger requirements must contain at least two AK8 jets with ${\pt > 300}$ and 250\GeV, respectively and with a soft-drop mass ${\mSD > 50\GeV}$ for each jet. The two AK8 jets with the highest and second-highest \bbbar-tagging scores, according to the \GloParT algorithm described in Section~\ref{sec:glopart}, are selected as the primary (\Hone) and secondary (\Htwo) \PH candidates, respectively. The leading jet is also required to have a \bbbar-tagging score exceeding the threshold applied at the trigger level.

In this region, the dominant background arises from QCD multijet events, particularly those with one or more AK8 jets containing ${\Pg\to\bbbar}$, which can mimic the \hbb signature. This process constitutes about 75\% of the total background in the preselected region. The \ttbar process contributes most of the remaining background, about 24\%. The selected AK8 jets in \ttbar events originate from a variety of sources, such as partially merged hadronic top quark decays, merged hadronic \PW boson decays, ${\Pg\to\bbbar}$, or the fragmentation of single \PQb quarks. The remaining 1\% of the background arises from \Vjets, \VV, and \VH processes. The primary observables to reject these backgrounds are the \GloParT regressed masses and \bbbar tagging scores of the two \PH candidate AK8 jets.

The merged topology is particularly sensitive to \qqhh production for certain non-SM coupling hypotheses ${\kappaVV \neq 1}$ that significantly increase the \qqhh cross section, especially for highly Lorentz-boosted \PH pairs~\cite{Dolan:2013rja,Dolan:2015zja,Bishara:2016kjn,Arganda:2018ftn}. To enhance the sensitivity to the \qqhh production mode, events are further categorized using additional AK4 jets that do not overlap with the AK8 jets. These jets are required to have ${\pt > 25\GeV}$ and ${\abs{\eta} < 4.7}$. Additional selections on the dijet system (jj), specifically \mjj and \detajj, are also applied either explicitly or through a multivariate algorithm. All preselected events not assigned to the \qqhh category are considered in the \gghh categories.

The merged analysis is performed following two complementary approaches, both of which use ML to separate the signal from background via an SvsB classifier. The first approach introduces new methods for the merged analysis in both the background estimation and the signal extraction, employing a background prediction method similar to that used in the resolved analysis (Section~\ref{sec:resolved_transformer}) and training the SvsB classifier with reweighted data sideband events. The second approach models the dominant QCD multijet background via a parametric extrapolation from a highly populated data sideband and uses MC simulated events to train the SvsB classifier, following the strategy adopted in the previous analysis of Run 2 data~\cite{CMS:2022gjd}.

\subsection{The first approach: background estimation using multidimensional DNN-based reweighting} \label{sec:merged_184}

In this section, we present a new approach to the measurement of \HHbbbb in the merged topology. At its core is a background estimation strategy derived from the methods used in the resolved analysis described in Section~\ref{sec:resolved_analysis}. This approach eliminates any dependence on simulated QCD multijet background events or the use of parametric functions for the background model, enabling a more performant and more flexible signal extraction strategy than was possible with previous methods. In addition to the selections mentioned in Section~\ref{sec:merged_common}, the regressed masses of both AK8 jet \hbb candidates, as estimated by the \GloParT algorithm, are required to be between 50 and 200\GeV. The \TXbb score for both jets is also required to exceed a value of 0.65. In the SR, the $\mreg({\Hone})$ is additionally required to be between 100 and 150\GeV. The expected signal fraction after applying these selections is less than 0.1\%. 

In order to better isolate the SM \HHbbbb process, a feedforward DNN with four hidden layers was trained to separate the \HHbbbb signal from the backgrounds based on: a) the \pt and $\eta$ of the two \PH candidates, b) the \TXbb score and \mreg of the \hbb candidate jet with highest \TXbb score ($\Hone$), c) the kinematics (\pt, $\eta$, \mhh) of the $\PH\PH$ system, and d) the $\Delta R$ and mass of each \PH candidate and the nearest nonoverlapping AK4 jet. The dominant features used by the DNN to suppress the QCD multijet background are the $\TXbb(\Hone)$ and $\mreg(\Hone)$, while \ttbar is further suppressed by using variables that aim to reconstruct the full set of top quark decay products by combining the \hbb candidate AK8 jet with nearby AK4 jets. The \TXbb score is discretized according to the calibration regions described in Section~\ref{sec:zbb}. 

The prediction for the dominant QCD multijet background in the SvsB DNN training is obtained by reweighting data events in a \bbbar tag score sideband region, as described later in this section. The SvsB DNN achieves an additional background rejection of more than one order of magnitude while preserving a signal efficiency of about 60\%. Further background rejection is achieved by tightening the $\TXbb(\Htwo)$ selections. Events in the preselected region are partitioned into three categories based on the SvsB DNN classifier score, denoted as \psig, and the value of $\TXbb(\Htwo)$:
\begin{itemize}
    \item{High-purity SR (HPSR):} ${\psig>0.9}$ and ${\TXbb(\Htwo) > 0.95}$; 
    \item{Low-purity SR (LPSR):} ${\psig>0.9}$ and ${0.85 < \TXbb(\Htwo) < 0.95}$; 
    \item{Validation region (VR):} ${\psig>0.7}$ and ${\TXbb(\Htwo) > 0.7}$. Veto events selected in either the HPSR or LPSR.
\end{itemize}

The HPSR region corresponds to about 90\% of the measurement sensitivity in the merged analysis, while the LPSR and VR regions serve to validate the background model. 

\begin{figure}[!htb]
\centering
\includegraphics[width=0.425\textwidth]{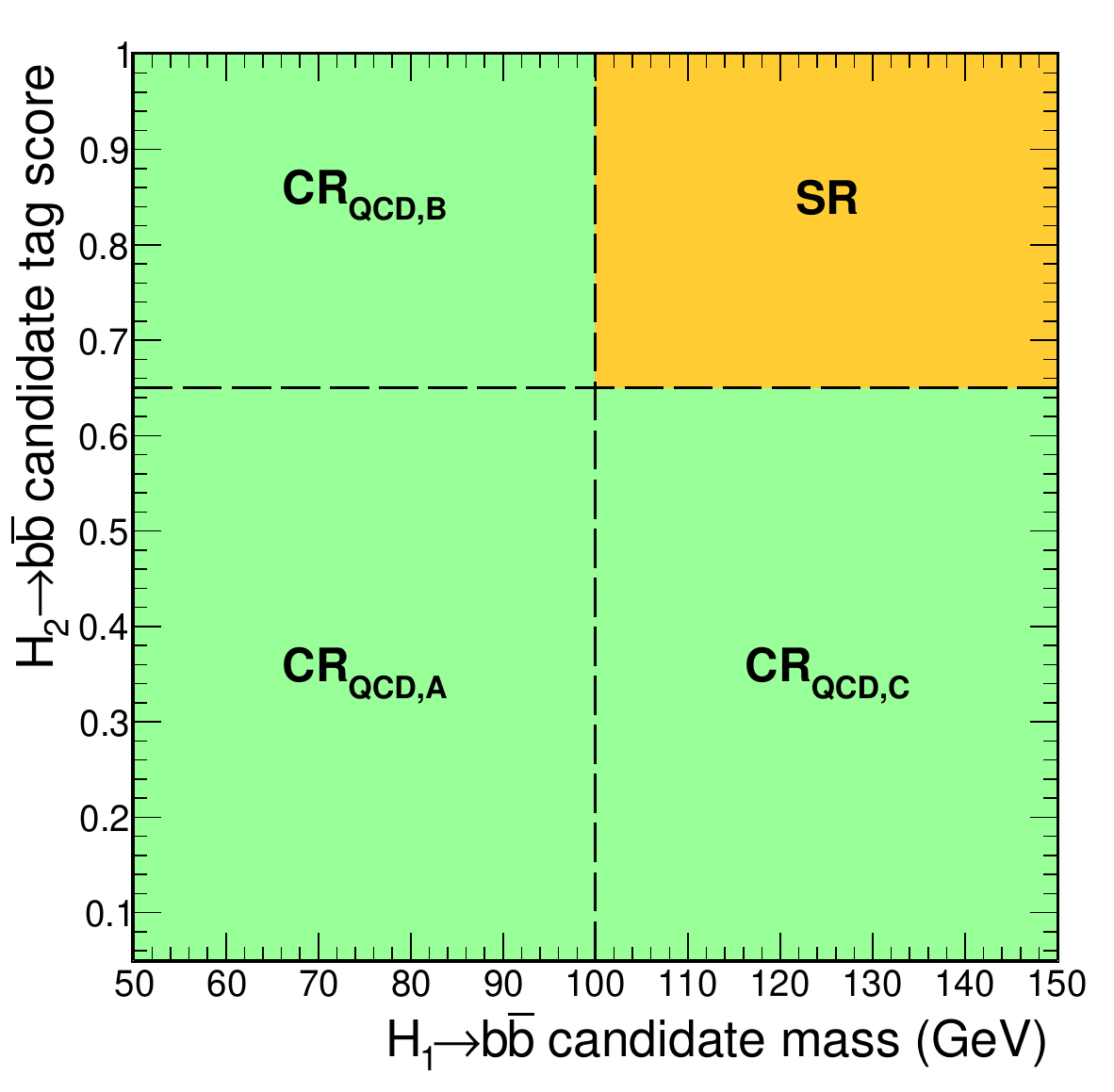}
\includegraphics[width=0.425\textwidth]{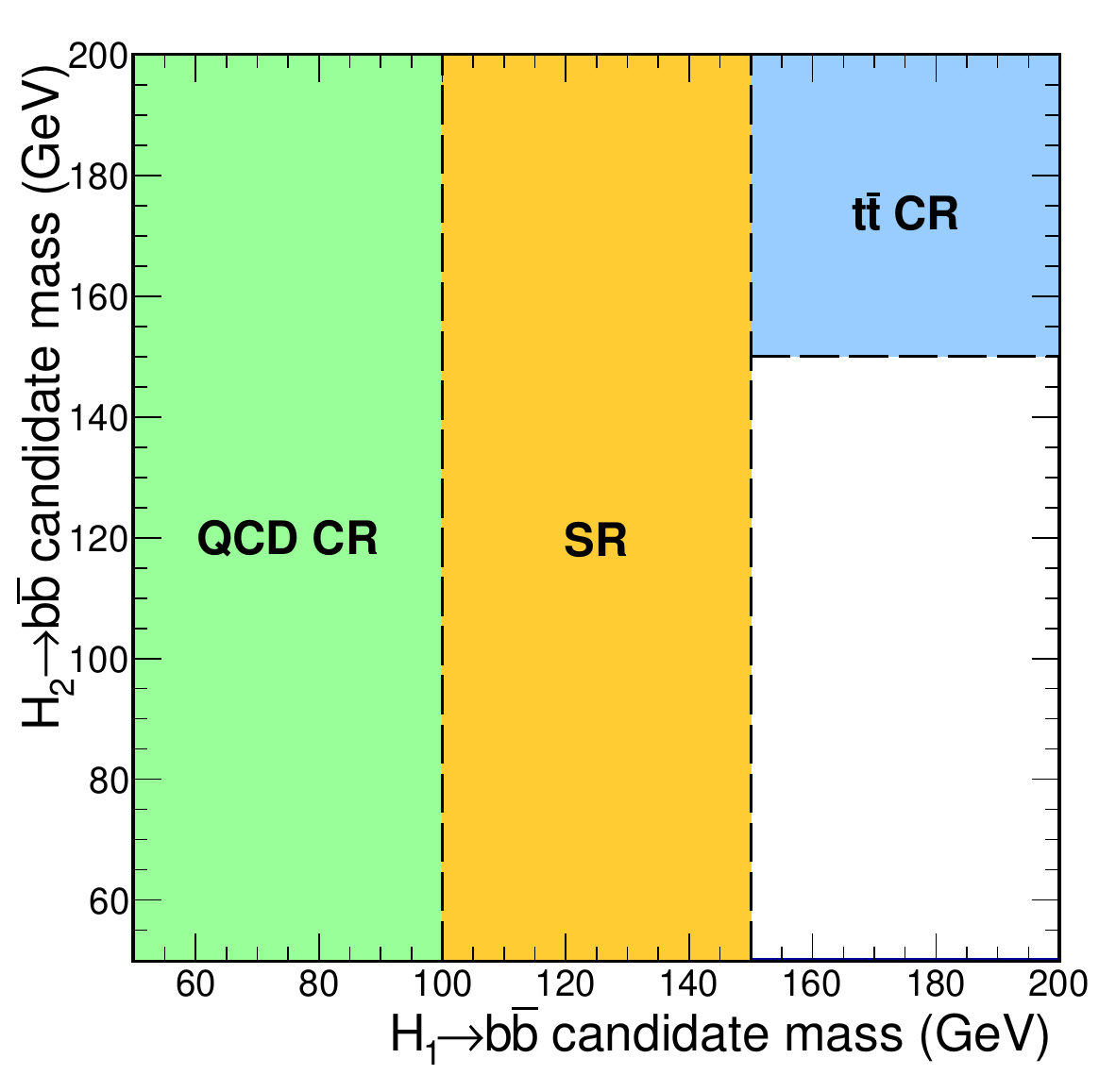}
\caption{\cmsLLeft: schematic diagram showing the regions used for the background estimation strategy used in the merged channel. The purity in QCD multijet events is reported in each of the background-enriched regions (\regCRQA, \regCRQB, and \regCRQC). \cmsRRight: schematic diagram of the signal (orange area) and CRs (QCD in green, \ttbar in azure) in the plane defined by the $\mreg({\Hone})$ and $\mreg({\Htwo})$. }
\label{fig:schematic_boosted_cartoon_def}
\end{figure}

The background model employs a DNN to perform a multidimensional reweighting of data events, learning from events in a mass sideband (${50 < \mreg({\Hone}) < 100 \GeV}$) the mapping of observables from a region, \regCRQA, with a loosened $\TXbb(\Htwo)$ requirement ${0.05 < \TXbb(\Htwo) < 0.65}$, to another a region, \regCRQB, with the same ${\TXbb(\Htwo)>0.65}$ selection as applied in the SR. The DNN output is then used to apply a weight to data events with ${100 < \mreg({\Hone}) < 150 \GeV}$ and ${0.05 < \TXbb(\Htwo) < 0.65}$, denoted \regCRQC, to obtain the background prediction in the SR. The reweighting DNN was trained from input features in common with the SvsB DNN described above. A schematic of the regions used in this background estimation technique is shown in Fig.~\ref{fig:schematic_boosted_cartoon_def} (\cmsLeft). The requirement of ${\TXbb(\Htwo) > 0.05}$ is applied in all regions to ensure that the AK8 jets selected in background events have properties similar to those of the \hbb signal, minimizing the extrapolation of the background model.

The background model is derived from an ensemble of $k=20$ reweighting DNNs, following exactly the same methodology as described for the resolved channel in Section~\ref{sec:resolved_transformer}. For each of the trainings, 75\% of the data from the \regCRQA and \regCRQB regions are randomly selected for training and the initial weights of the DNN are also randomly sampled. The background prediction is then derived by applying the median per-event weight among the $k$-fold background predictions to \regCRQC data events. The spread, defined in Section~\ref{sec:merged_systematics}, in the background predictions among the different $k$-folds is assigned as a systematic uncertainty in the background prediction for each bin of the fitted observable. The subdominant \ttbar background constitutes up to 20\% of the total background in some regions. In order to explicitly separate the \ttbar contribution in data, the reweighting DNN is trained to differentiate simulated \ttbar events from data events in \regCRQA and \regCRQB. The data events in \regCRQC are then assigned a weight of 
\begin{equation}\label{eq:merged_rwt}
\mathrm{weight} = \frac{\mathcal{P}(\regCRDaB)-\mathcal{P}(\regCRtB)}{\mathcal{P}(\regCRDaA)-\mathcal{P}(\regCRtA)},
\end{equation}
where $\mathcal{P}(\regCRDaB)$ and $\mathcal{P}(\regCRDaA)$ estimate the likelihood that an event corresponds to data in \regCRQB or \regCRQA, respectively, while $\mathcal{P}(\regCRtB)$ and $\mathcal{P}(\regCRtA)$ estimate the likelihood that an event derives from the \ttbar process in a given region. This procedure effectively performs a multidimensional subtraction of the expected \ttbar contribution in simulation from the background prediction based on CRs in data. The total background prediction in the SR is then obtained from the sum of the reweighted \regCRQC data events and the simulated \ttbar prediction in the SR. This procedure makes it possible to differentiate the predicted QCD multijet and \ttbar background contributions and to explicitly constrain the normalization of each process separately in the fit.

The normalization of the \ttbar prediction in each SR category (HPSR, LPSR, and VR) is constrained in situ by data in a \ttbar-enriched CR (HP-TT-CR, LP-TT-CR, and TT-VR), defined with exactly the same selections applied as in the SR except that the $\mreg(\Hone)$ and $\mreg(\Htwo)$ are required to be between 150 and 200\GeV, as shown in Fig.~\ref{fig:schematic_boosted_cartoon_def} (\cmsRight). This procedure reduces the dependence of the analysis on the \ttbar background simulation. The SR category definitions include selections on \psig, the output score of the SvsB DNN, which is trained using $\mreg(\Hone)$ as an input feature. With the aim of aligning as much as possible the characteristics of the selected events in the \ttbar CRs to those of \ttbar events in the corresponding SR categories, the \psig score is evaluated in the \ttbar CRs with the $\mreg(\Hone)$ value shifted to match the range in the SR, ${100 < \mreg(\Hone) < 150 \GeV}$. The \ttbar process is the dominant contribution in these regions and is expected to be roughly 80\% of the total background in the HP-TT-CR and LP-TT-CR regions. The event yields in these \ttbar-enriched regions are included in the signal extraction fit such that the \ttbar normalization in each SR category (HPSR, LPSR, VR) is fit individually to data without any prior constraint. This effectively removes any dependence on the modeling of the \psig or \TXbb score distributions in \ttbar simulation. As shown in Fig.~\ref{fig:schematic_boosted_cartoon_def} (\cmsLeft), a low-mass QCD CR is also defined per SR category with all the selections of the corresponding SR category applied except for a shift in the $\mreg(\Hone)$ requirement, $50 < \mreg(\Hone) < 100\GeV$, following similar logic to that employed for the \ttbar CRs. The event yield in each QCD CR is included in the signal extraction fit to improve the constraint on the normalization of the QCD multijet prediction.

The new background estimation method employed in the merged analysis, based on a reweighting of \regCRQC data events (as shown in Fig.~\ref{fig:schematic_boosted_cartoon_def} \cmsRight), yields a background prediction for multiple observables, providing flexibility in the signal extraction. Two complementary signal extraction methods are presented, each based on a fit to a different observable. The ``mass fit'' method performs a fit to the $\mreg(\Htwo)$ distribution in each SR category, as was done in previous searches in this channel~\cite{CMS:2022fxs}, while the ``DNN fit'' method extracts the signal via a fit to the output of an SvsB classifier DNN, \psig, following an approach similar to that used in the resolved analysis described in Section~\ref{sec:resolved_analysis}. With the DNN fit method, the SR categories are redefined with respect to those described earlier, still partitioning events based on the same $\TXbb(\Htwo)$ criteria but without any \psig selections applied so that the full \psig distribution can be included in the signal extraction fit.

The mass fit method has the advantage of fitting a physical observable in which the expected \HHbbbb signal as well as some of the background processes (\VH, \VV) manifest as resonances in the $\mreg(\Htwo)$ distribution. The primary systematic uncertainty following the mass fit method is the limited constraint achievable for the QCD multijet background normalization from CRs in data. The DNN fit method, on the other hand, has the advantage of including more data events in the fit and therefore better constraining the QCD multijet background predictions. It also enables the SvsB classifier DNN to leverage the full event information, including correlations between the $\mreg(\Htwo)$ and the other observables. The dominant systematic uncertainties with the mass fit and DNN fit methods differ significantly--by pursuing both methods and obtaining consistent results, we build further confidence in the procedures used throughout. 

\subsubsection{The mass fit method}\label{sec:massfit_merged}

\begin{figure*}[!htb]
  \centering
    \includegraphics[width=0.32\textwidth]{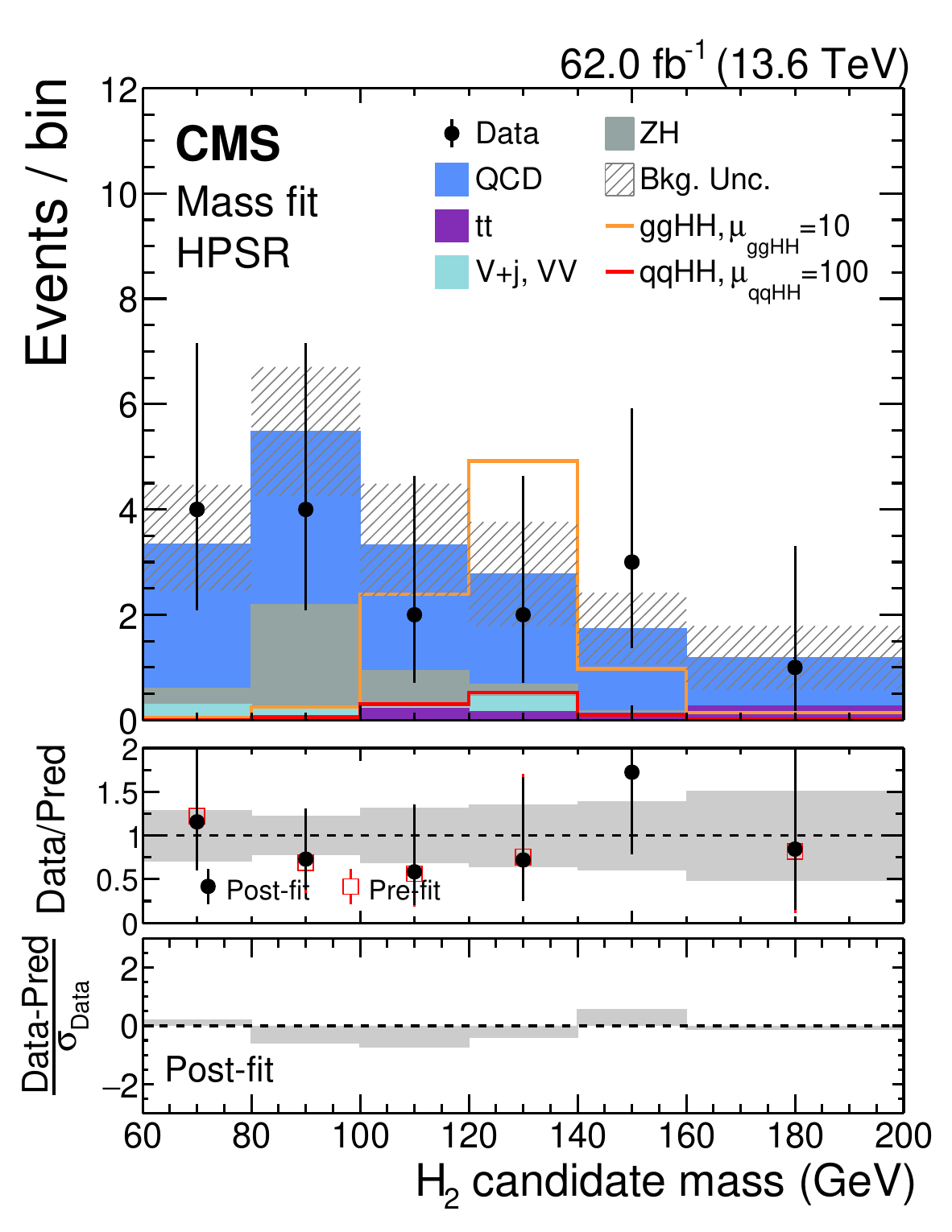}
    \includegraphics[width=0.32\textwidth]{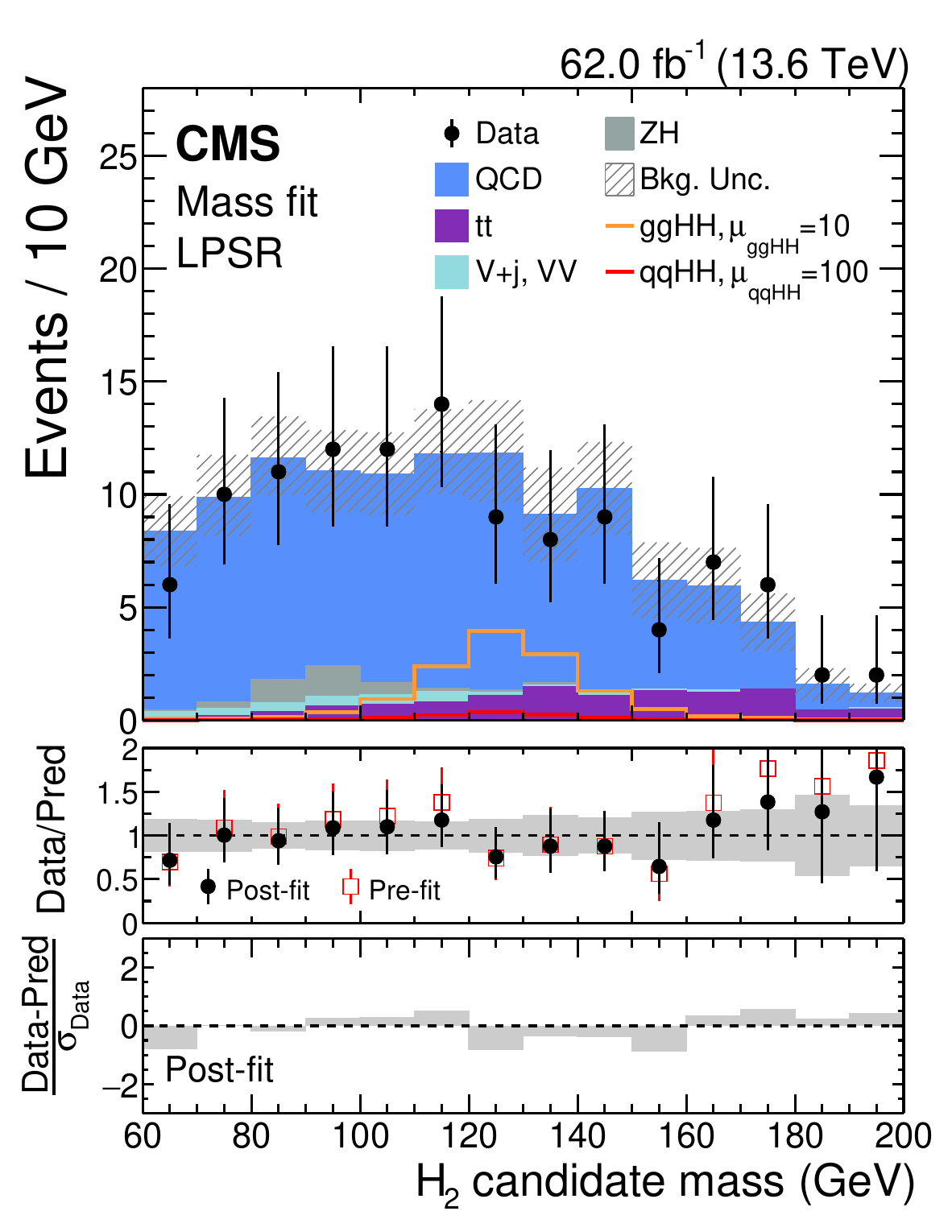}
    \includegraphics[width=0.32\textwidth]{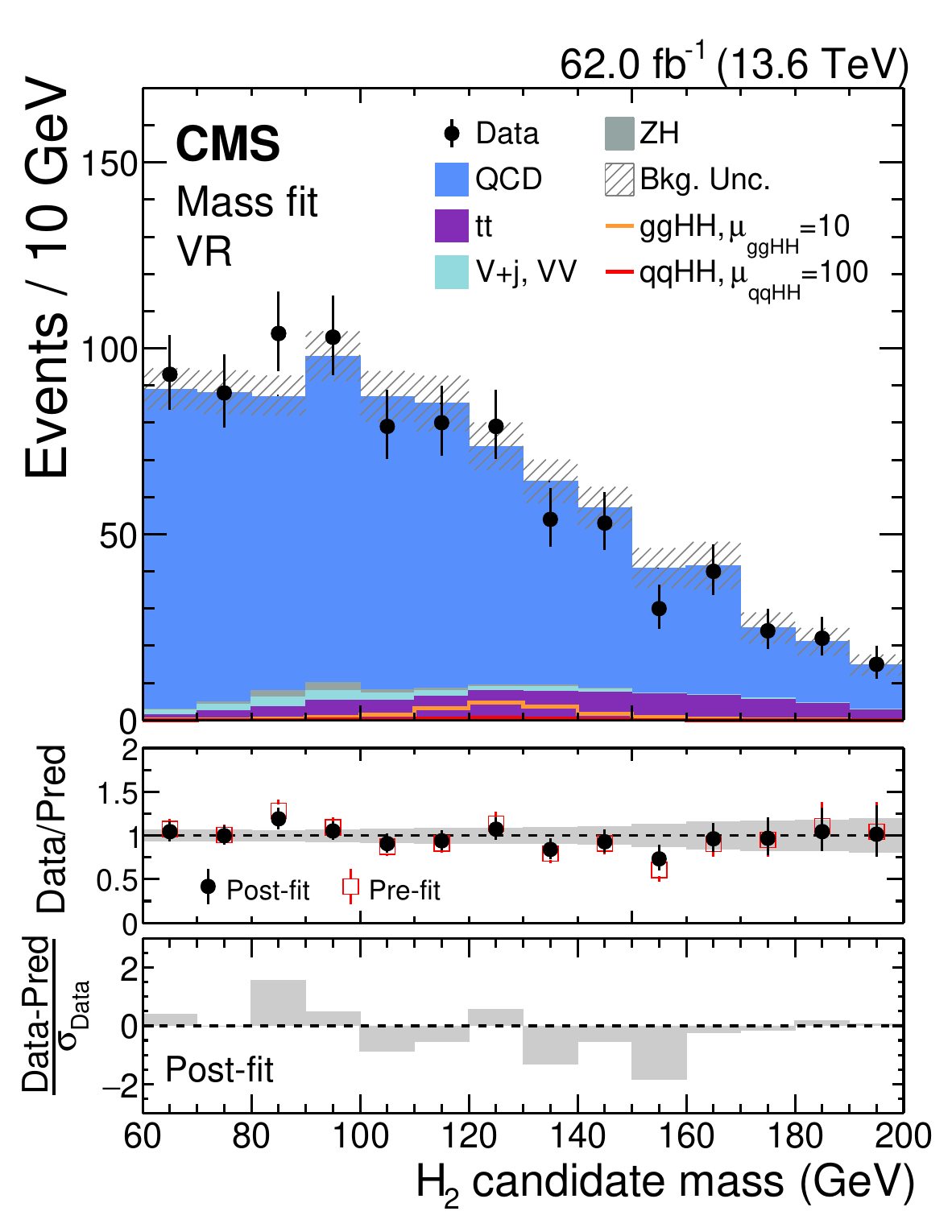}
 \caption{Pre-fit and post-fit distributions of the fitted observables for the \gghh-inclusive HPSR (left), LPSR (center), and VR (right) categories of the merged analysis following the mass fit method. Distributions are shown for data (black points) and the different background contributions from QCD multijet, \ttbar, \Vjets, \VV, and \ZH productions. The expected distributions for SM \gghh (orange) and \qqhh (red) signals are overlaid and scaled by a multiplicative factor to improve their visibility. Notations are as in Fig.~\ref{fig:dnn_3b_control_region_fitted}.}
    \label{fig:merged_massfit_postfit}

\end{figure*}

In the mass fit method, the signal is extracted via a binned maximum likelihood fit to the $\mreg(\Htwo)$ distribution in each category, in addition to the event yields in the \ttbar- and QCD-enriched CRs. Figure~\ref{fig:merged_massfit_postfit} shows these distributions in the HPSR (left), LPSR (center), and VR (right) categories. The overlaid signal contribution is the a priori prediction scaled by a multiplicative factor to enhance its visibility. A relatively coarse bin width, about twice as wide as the expected mass resolution, was chosen for the HPSR so that each bin in the \regCRQC sideband data distribution used to derive the background prediction contains at least 10 events. This binning configuration provides a comparable sensitivity, within about 5\%, to finer binning choices, while also minimizing the statistical variance in the background prediction. In the LPSR and VR regions, which have many more events, more granular bins were chosen--this has a minimal impact on the merged analysis sensitivity, but enables a more stringent test of the background model.

In both the HPSR and LPSR, the spread among the $k$-fold background predictions in most bins of the $m_{\Htwo}$ distribution, including those where the expected signal lies, is comparable to the per-bin statistical uncertainty in the reweighted \regCRQC data events (5--20\%). The $k$-fold variance, included in the bin-by-bin uncertainties, increases substantially, up to 40--50\%, for the rightmost bin(s) with ${m_{\Htwo}>160\GeV}$. In this region, the \ttbar relative contribution to the total background is larger and the increased spread in the $k$-fold predictions reflects a larger uncertainty in the per-event \ttbar subtraction performed by the reweighting (Eq.~(\ref{eq:merged_rwt})). The background prediction models the observed data well in all regions including the VR, which contains an order of magnitude more data events than the LPSR and therefore validates the background model with significantly higher precision than relied upon in the fit to the HPSR. 

\subsubsection{The DNN fit method}\label{sec:dnnfit_merged}

\begin{figure*}[!htb]
  \centering
    \includegraphics[width=0.32\textwidth]{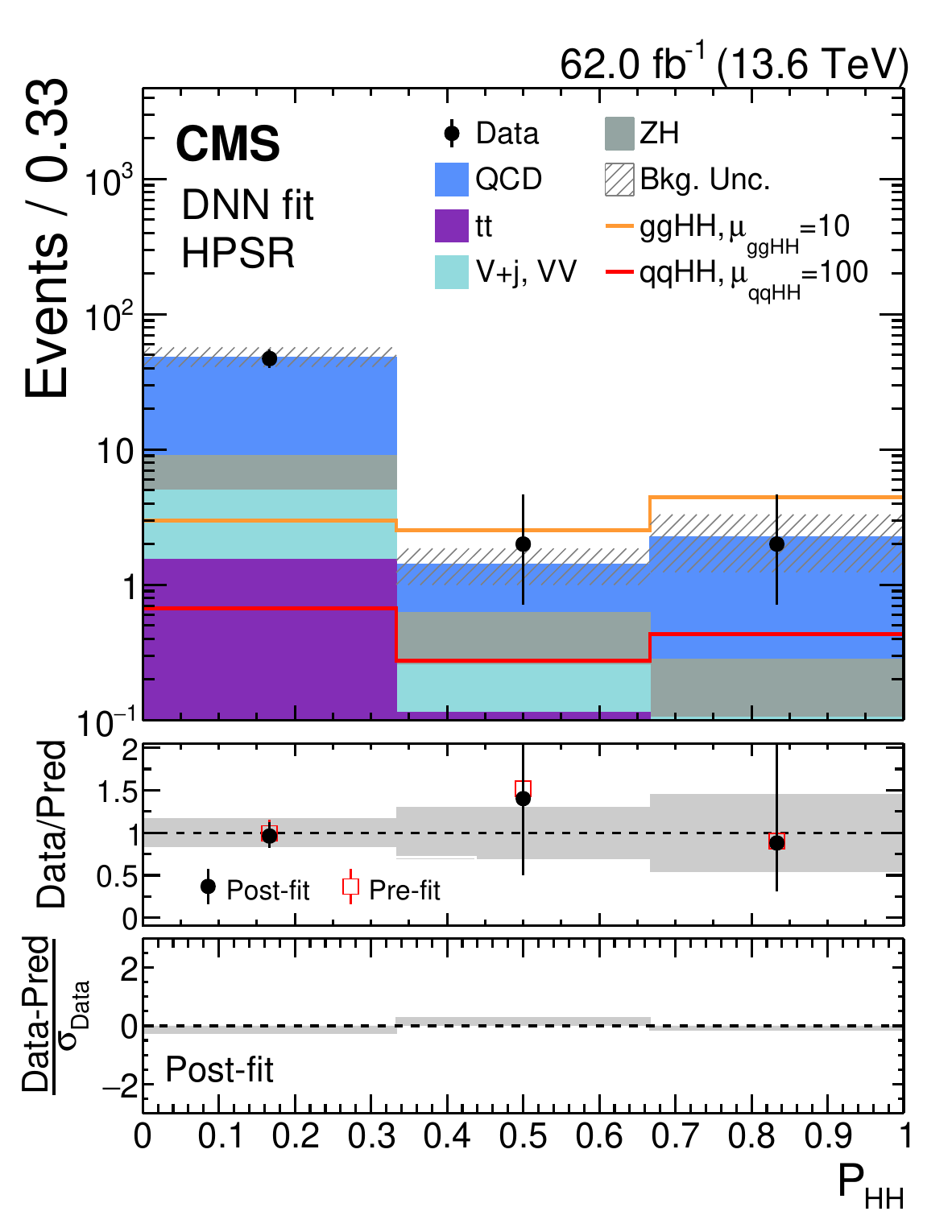}
    \includegraphics[width=0.32\textwidth]{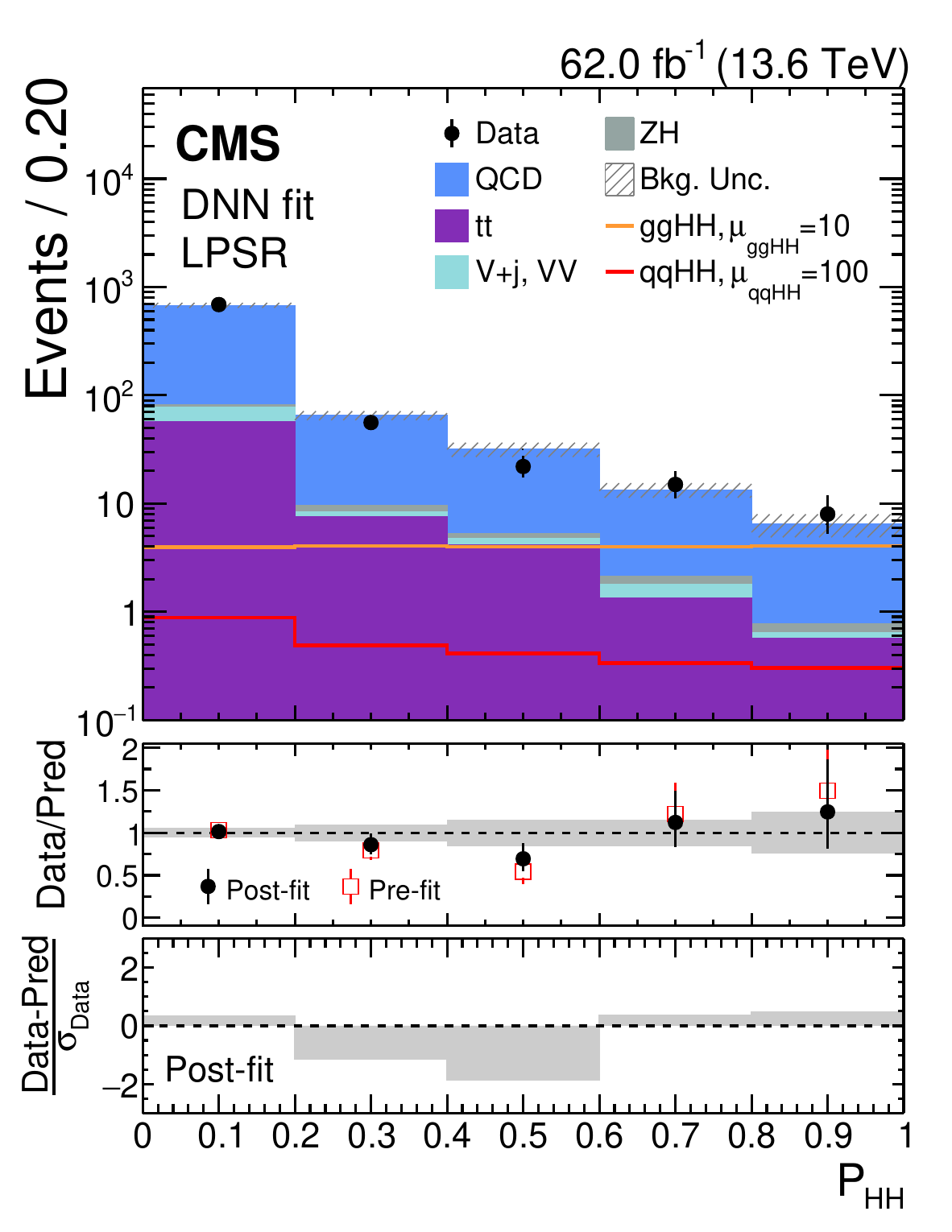}
    \includegraphics[width=0.32\textwidth]{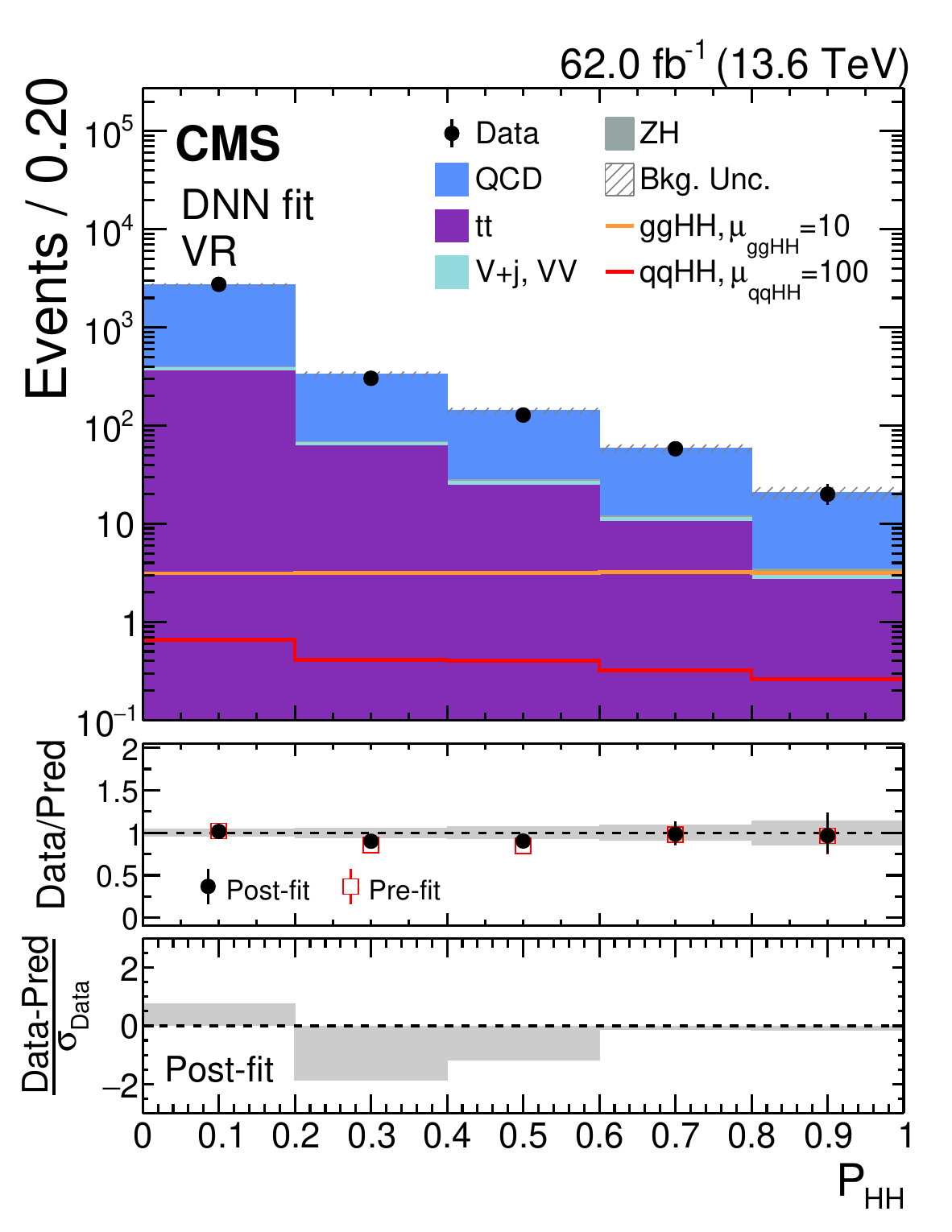}
 \caption{Pre-fit and post-fit distributions of the fitted observables for the \gghh-inclusive HPSR (left), LPSR (center), and VR (right) categories of the merged analysis following the DNN fit method. Distributions are shown for data (black points) and the different background contributions from QCD multijet, \ttbar, \Vjets, \VV, and \ZH productions. The expected distributions for SM \gghh (orange) and \qqhh (red) signals are overlaid and scaled by a multiplicative factor to improve their visibility. Notations are as in Fig.~\ref{fig:dnn_3b_control_region_fitted}.}
    \label{fig:merged_dnnfit_postfit}

\end{figure*}

The SvsB DNN developed for the DNN fit method has the same architecture and input features as used for the SvsB DNN employed in the mass fit analysis, with the addition of $\mreg(\Htwo)$ as an input feature. The SR categories are redefined with respect to those used for the mass fit method, removing any selections on \psig. Figure~\ref{fig:merged_dnnfit_postfit} shows the \psig distributions in the HPSR (left), LPSR (center), and VR (right) categories of the DNN fit analysis. A minimal coarse binning was chosen for the HPSR in order to simplify the fit and to ensure that any variance in the background model correlated across bins is negligible. In the LPSR, instead, a more granular binning was chosen in order to more precisely test the background model. The variance in the background model prediction, evaluated from the spread among $k$-fold background predictions, is 1\% or less in the background-dominated bins and increases to 10--20\% in the rightmost high-score bins where the expected signal purity is highest. The observed data in the VR, which has an order of magnitude more events than the LPSR, are consistent with the background prediction, within uncertainties.

\subsubsection{The VBF \texorpdfstring{\HHbbbb}{HHbbbb} category}\label{sec:vbf_merged}

Similar to the resolved \qqhh analysis described in Section~\ref{sec:vbf_hh4b_resolved}, events with at least two additional AK4 jets with ${\pt > 25\GeV}$ and ${\abs{\eta} < 4.7}$ are considered in the merged \qqhh category if the maximum \mjj among additional jet pairs exceeds 400\GeV and their $\eta$ separation satisfies ${\abs{\detajj}>3.5}$. Since measurements of \qqhh at high \pt are particularly sensitive to deviations in \kappaVV, the merged analysis specifically targets sensitivity to an anomalous $\kappaVV=0$ \qqhh signal, a representative example of the non-SM \qqhh signals that could be present in this search.

In the \qqhh merged analysis, the same background model for the QCD multijet background was employed, using the same region definitions reported in Fig.~\ref{fig:schematic_boosted_cartoon_def}. A feedforward SvsB DNN classifier was trained using the same input features used in the \gghh merged analysis described in Section~\ref{sec:resolved_transformer}, as well as: a) the AK4 dijet invariant mass \mjj and $\abs{\detajj}$, b) the four-vectors of the two VBF jets, c) the centrality of each \PH candidate. The DNN aims to separate the $\kappaVV=0$ \qqhh signal from the QCD multijet and \ttbar backgrounds. All events in the \qqhh merged analysis are required to pass the same minimum threshold on the \qqhh category SvsB DNN output score, $\psigvbf>0.9$. This requirement has an efficiency of about 86\% for the $\kappaVV=0$ \qqhh signal and reduces the background by nearly two orders of magnitude after the \qqhh category selections described above. All events that fail this selection are included in the merged analysis \gghh production category. Following a similar methodology to that used in the inclusive \gghh category, the \qqhh merged analysis SR was then partitioned based on the $\TXbb({\Htwo})$ into three regions with different levels of signal purity--VBF-HPSR, VBF-LPSR, and VBF-VR.

\begin{figure*}[!htb]
  \centering
    \includegraphics[width=0.425\textwidth]{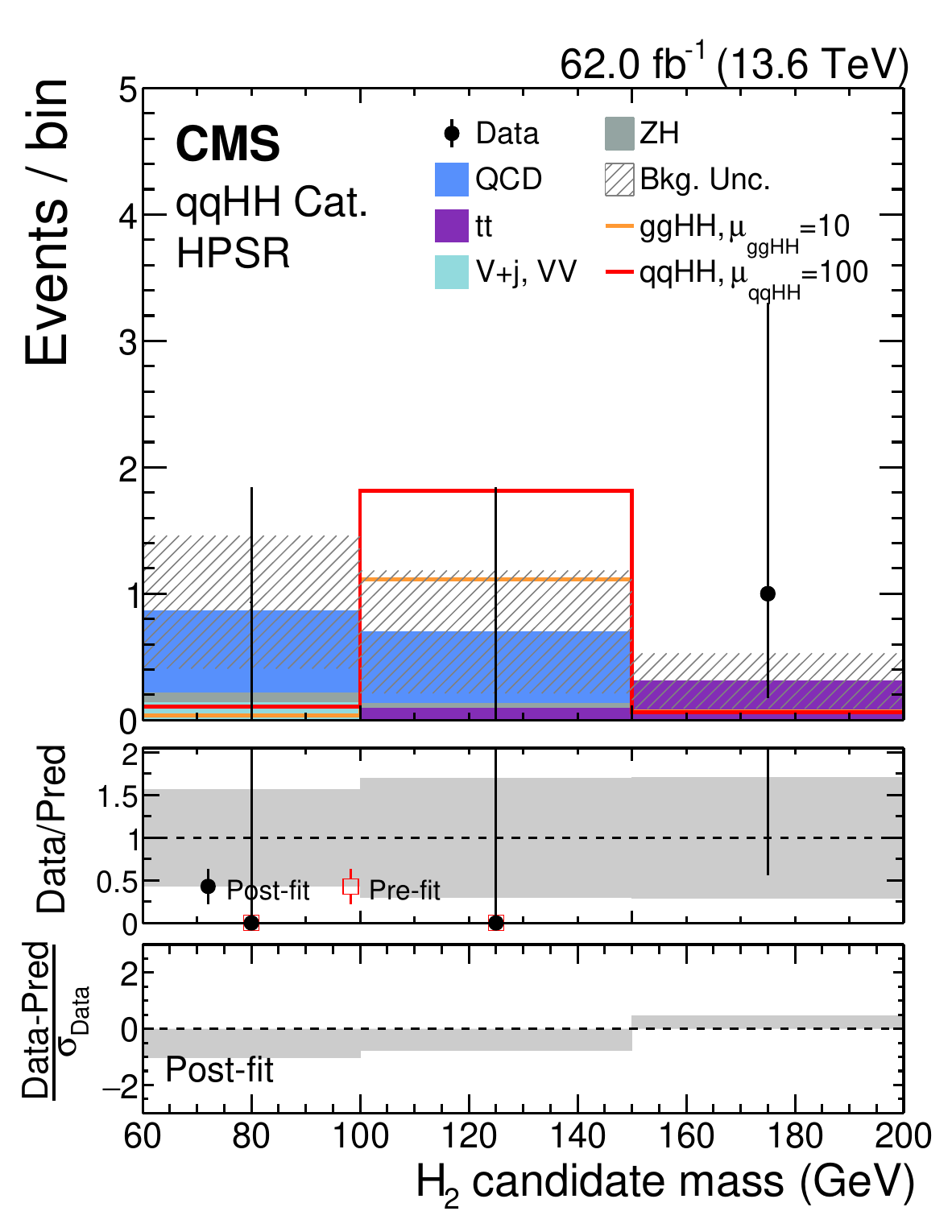}
    \includegraphics[width=0.425\textwidth]{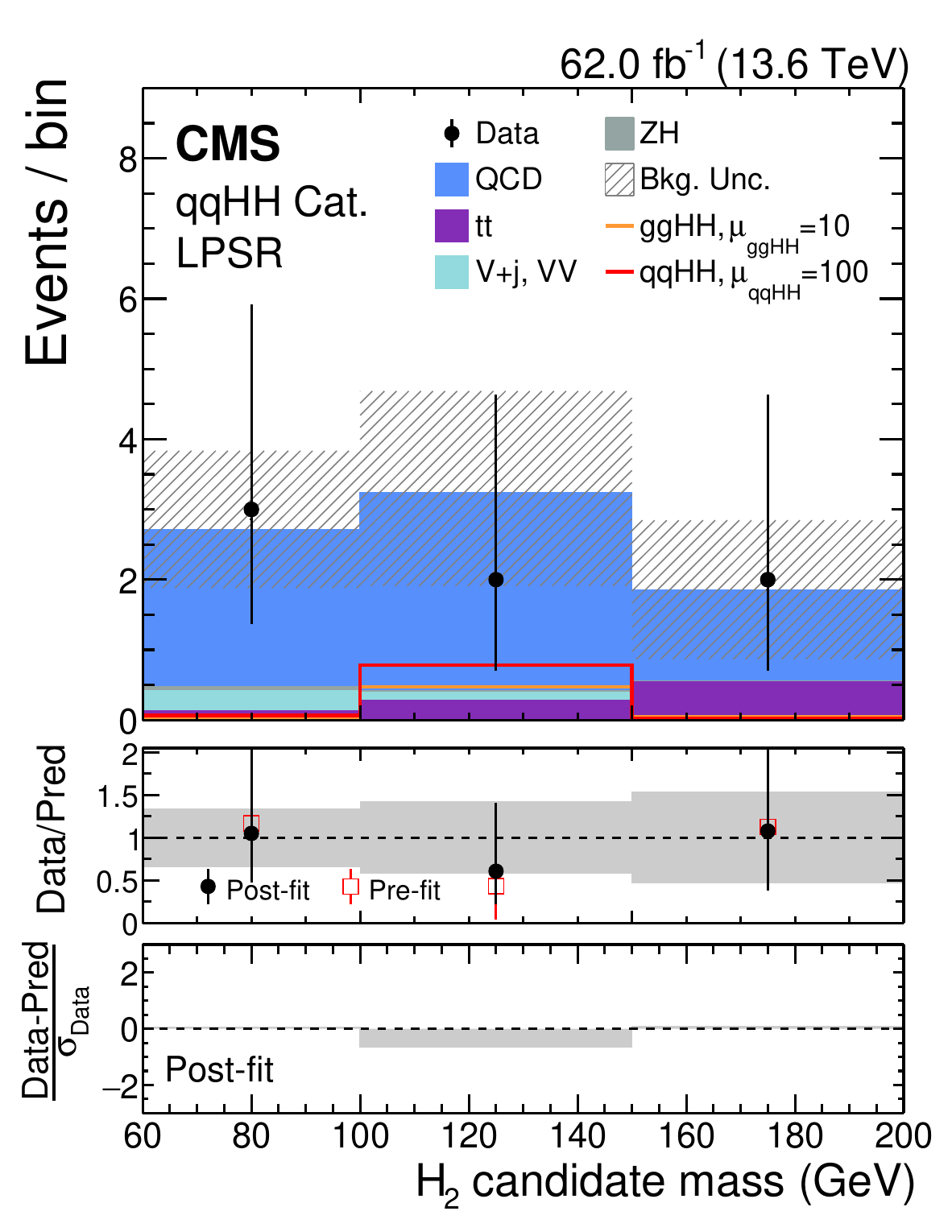}
    \caption{Pre-fit and post-fit distributions of the fitted observables for the \qqhh HPSR (left) and LPSR (right) categories of the merged analysis. Distributions are shown for data (black points) and the different background contributions from QCD multijet, \ttbar, \Vjets, \VV, and \ZH productions. The expected distributions for SM \gghh (orange) and \qqhh (red) signals are overlaid and scaled by a multiplicative factor to improve their visibility. Notations are as in Fig.~\ref{fig:dnn_3b_control_region_fitted}.}
    \label{fig:merged_vbf_postfit}

\end{figure*}

The signal in the VBF merged analysis is extracted via a fit to the $\mreg(\Htwo)$ distribution in each category, including the event yield in dedicated QCD- and \ttbar-enriched CRs per category, as depicted in Fig.~\ref{fig:schematic_boosted_cartoon_def} (\cmsRight). The measurement uncertainty in the VBF category is dominated by statistical uncertainties, so the fit is simplified and the mass fit method is used. Figure~\ref{fig:merged_vbf_postfit} shows the fitted distributions in the VBF-HPSR (left) and VBF-LPSR (right) categories. The number of events in the VBF SR categories is very small, with approximately one event expected in the VBF-HPSR. The binning of the $\mreg({\Htwo})$ distribution in the fit is therefore very coarse. The sensitivity is driven by the middle bin with $\mreg({\Htwo})$ between 100 and 150\GeV, while the other bins effectively serve for further background model validation. The variance in the background model, evaluated from the spread in $k$-fold background predictions, ranges from 10--15\% for ${\mreg({\Htwo})< 150\GeV}$ and from 30--60\% for ${\mreg({\Htwo}) > 150\GeV}$, where the \ttbar contribution is the most significant.

\subsection{The second approach: background estimation using a parametric alphabet method}\label{sec:merged_parametric}

A different approach is explored in the merged topology following the strategy adopted in the analysis of Run~2 data~\cite{CMS:2022gjd}. In this approach, a boosted decision tree (BDT) is trained to discriminate the \gghh and \qqhh process from the QCD multijet and \ttbar backgrounds. The input features of the BDT are similar to the ones used in the approach described in Section~\ref{sec:merged_184} and include kinematic properties of the \PH and VBF jet candidates, the \HH system, the $N$-subjettiness ($\tau_{32}$)~\cite{Thaler:2010tr} of the two \PH candidates, the \TXbb score of the \PH candidate with the leading \TXbb score, angular separation between the final state objects and event-level observables. The \TXbb score and the regressed mass of the subleading in \TXbb \PH candidate jet, $\mreg({\Htwo})$, are excluded from the input features as these variables are used to define the event categories of the signal extraction.
The \TXbb score is discretized following the working points calibrated for the \GloParT algorithm. The BDT yields four output probabilities: $\Prob(\gghh)$, $\Prob(\qqhh)$, $\Prob(\text{QCD multijet})$ and $\Prob(\ttbar)$, which are combined to define two discriminant scores targeting the \gghh and \qqhh categories:

\begin{equation*}
\begin{aligned}
\mathcal{D}(\text{\gghh-vs-bkg}) &= \frac{\Prob(\gghh)}{\Prob(\gghh) + \Prob(\text{QCD}) + \Prob(\ttbar)}, \\
\mathcal{D}(\text{\qqhh-vs-bkg}) &= \frac{\Prob(\qqhh)}{\Prob(\qqhh) + \Prob(\text{QCD}) + \Prob(\ttbar)}.
\end{aligned}
\end{equation*}

\begin{figure*} [!hbt] 
\centering
\includegraphics[width=0.75\textwidth]{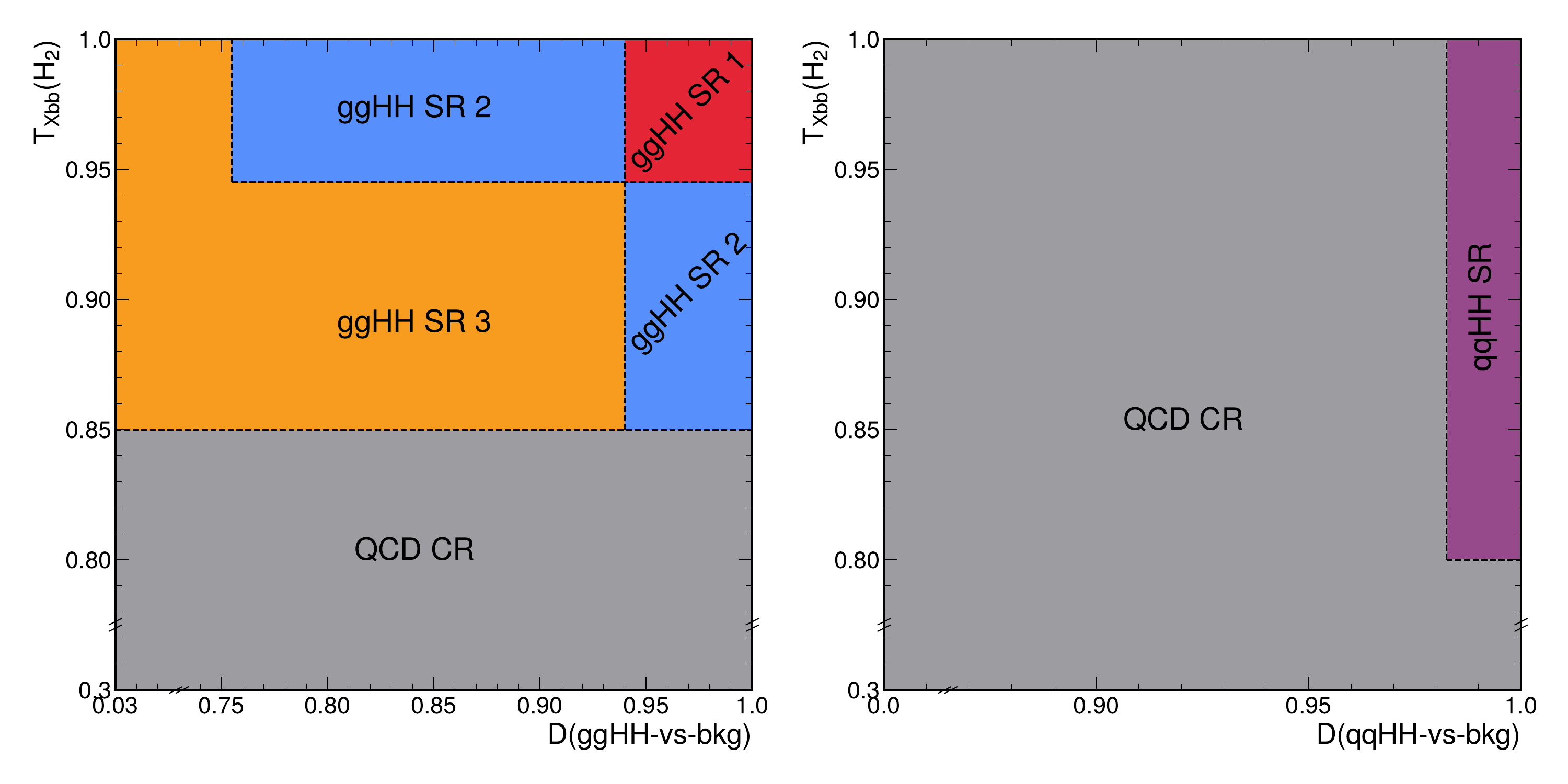}
\caption{Schematic diagrams showing the SRs and QCD CR (gray) used in the merged analysis described in Section~\ref{sec:merged_parametric}. Left: the \gghh SR 1 (red), 2 (blue), and 3 (orange) are defined based on successively lower selections on the \TXbb score of the \Htwo candidate and the $\mathcal{D}(\text{\gghh-vs-bkg})$ score. Right: the \qqhh SR (purple) is defined based on a tight selection on $\mathcal{D}(\text{\qqhh-vs-bkg})$ and a loose selection on \TXbb of the \Htwo candidate.}
\label{fig:sketch-boosted-mass-fit}
\end{figure*}

Four nonoverlapping SRs are constructed based on a two-dimensional (2D) selection applied on the $\TXbb(\Htwo)$ score and the respective $\mathcal{D}(\text{\gghh-vs-bkg})$ and $\mathcal{D}(\text{\qqhh-vs-bkg})$ scores, as illustrated in Fig.~\ref{fig:sketch-boosted-mass-fit}. Three of these SRs are sensitive to the \gghh production (\gghh SR 1--3) and one is sensitive to the \qqhh production (\qqhh SR). The highest signal purity SR, \gghh SR 1, is defined by requiring events to satisfy tight WPs for both $\mathcal{D}(\text{\gghh-vs-bkg})$ and $\TXbb(\Htwo)$. Events not in the \gghh SR 1 are assigned to the \qqhh SR only if they satisfy a tight $D(\text{\qqhh-vs-bkg})$ WP and a loose $\TXbb(\Htwo)$ WP. Remaining events are grouped into \gghh SR 2 and 3, defined by successively lower selections on the $\mathcal{D}(\text{\gghh-vs-bkg})$ and $\TXbb(\Htwo)$ WPs, as shown in Fig.~\ref{fig:sketch-boosted-mass-fit}. The categorization selections are chosen to maximize the sensitivity to the \gghh process in the \gghh SRs and to the $\kappaVV=0$ \qqhh process in the \qqhh SR. The events failing the SR conditions, but satisfy a very loose selection on $\mathcal{D}(\text{\gghh-vs-bkg})$ and $\TXbb(\Htwo)$, are enriched in QCD multijet events and used for the background modeling (QCD CR). 

The two dominant backgrounds after the SR selection are the QCD multijet process and the \ttbar process. The QCD multijet background is estimated from data in the QCD CR using the parametric alphabet method, following the strategy employed in the Run~2 analysis in the merged topology~\cite{CMS:2022gjd}, as well as previous searches~\cite{CMS:2020zge,CMS:2017bcq,CMS:2017dcz,CMS:2017nuu}. The estimation is performed as a function of $\mreg(\Htwo)$, after subtracting \ttbar (8\%) and other minor contributions (2\%). In the absence of correlations between $\mreg(\Htwo)$ and the \TXbb score or BDT discriminator, the QCD multijet background is expected to have the same $\mreg(\Htwo)$ distribution in the QCD CR as in the SRs. Therefore, since the \GloParT \bbbar tagger is trained to be independent of the \PH mass assumption and the BDT inputs are chosen to avoid correlations with $\mreg(\Htwo)$, the QCD CR can be used to predict the shape of the QCD multijet background in the SRs. The background model is defined as the product of the $\mreg(\Htwo)$ shape obtained from the QCD CR and an $n^{\text{th}}$-order Bernstein polynomial transfer function, with the parameters determined from the fit, to account for any residual correlations between the $\mreg$ and \TXbb score or BDT discriminator. A Fisher F-test and a goodness-of-fit test are performed in the QCD CR and each SR to determine the order of the polynomial. A zero-order polynomial is found to be sufficient in all four SRs. The overall normalization is considered as a systematic uncertainty and a floating normalization is assigned and determined from the sidebands.

The \ttbar background is estimated using simulated events with corrections extracted in dedicated CRs in data, which will be described in the following paragraphs. While the overall kinematic distributions are generally well modeled, several features relevant to this analysis require these corrections. The \TXbb score is calibrated using a CR enriched in \ttbar events including one semileptonic top quark decay. This region is defined using single-lepton triggers and requires exactly one isolated lepton with ${\pt > 50\GeV}$, ${\ptmiss > 50\GeV}$, and one high-\pt AK8 jet with ${\pt > 300}$ (250)\GeV for 2022 (2023) and ${\mSD > 50\GeV}$.  Events with additional leptons or AK8 jets are rejected. A \PQb-tagged AK4 jet, well separated from the lepton and the jet, is required to enhance \ttbar purity. Finally, a requirement on ${\tau_{32} < 0.5}$ is applied on the AK8 jet. The correction factors are found to be in the range of 10--20\% in the highest \TXbb score region. Simulation-to-data correction factors are also derived as a function of $\tau_{32}$ after subtracting the estimated contributions from non-\ttbar processes.

The recoil of the \ttbar system is corrected using a CR enriched in fully hadronic \ttbar events. Events are selected by requiring two large-radius jets, each with ${\pt > 450\GeV}$, ${\abs{\eta} < 2.5}$, ${\mSD > 50\GeV}$, ${\TXbb > 0.1}$, and ${\tau_{32} < 0.46}$. Correction factors are derived by comparing the distribution of the \pt of the \ttbar system in data and simulation in bins of dijet \pt ($\pt(\mathrm{jj})$), with typical corrections of 1--2\% across most of the recoil momentum.

The distribution of the BDT discriminant is validated in the fully hadronic \ttbar CR for both the \gghh and \qqhh SRs. Events are required to have two large-radius jets with ${\pt>300\GeV}$, ${\mSD > 40\GeV}$, and ${\TXbb>0.1}$. The \Hone candidate jet must have ${150 < \mreg < 200\GeV}$ and ${\tau_{32} < 0.6}$, while the \Htwo candidate jet must have ${60 <\mreg < 250\GeV}$. The ratio between background subtracted data and \ttbar simulation prediction is taken as the correction factor as a function of $\mathcal{D}(\text{\gghh-vs-bkg})$ for the \gghh SR 1--3, or $\mathcal{D}(\text{\qqhh-vs-bkg})$ for the \qqhh SR. In the highest BDT score bins, corrections of approximately 20--50\% are derived to account for discrepancies between data and simulation.

\begin{figure*}[!htb]
    \centering
    \includegraphics[width=0.425\textwidth]{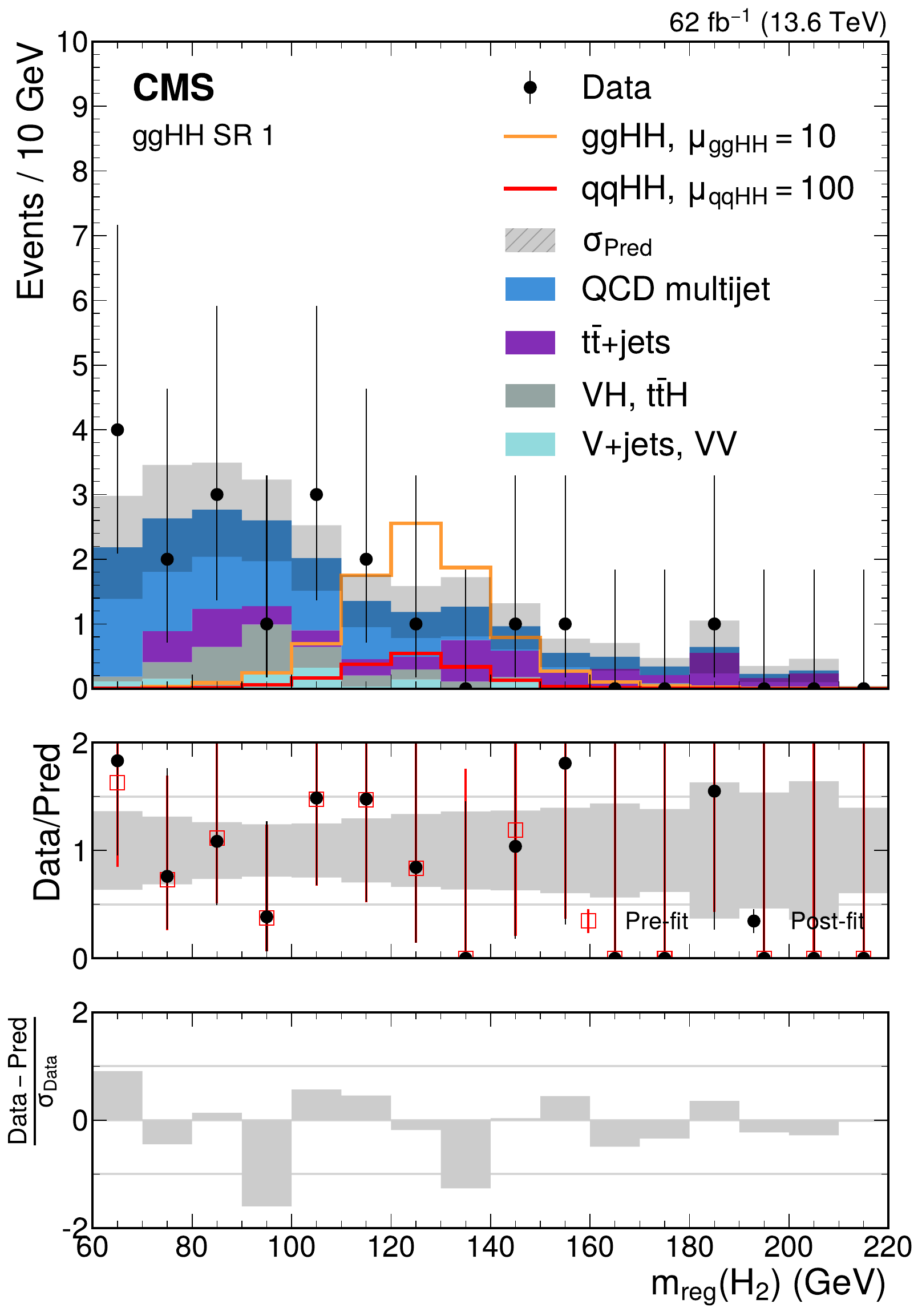}
    \includegraphics[width=0.425\textwidth]{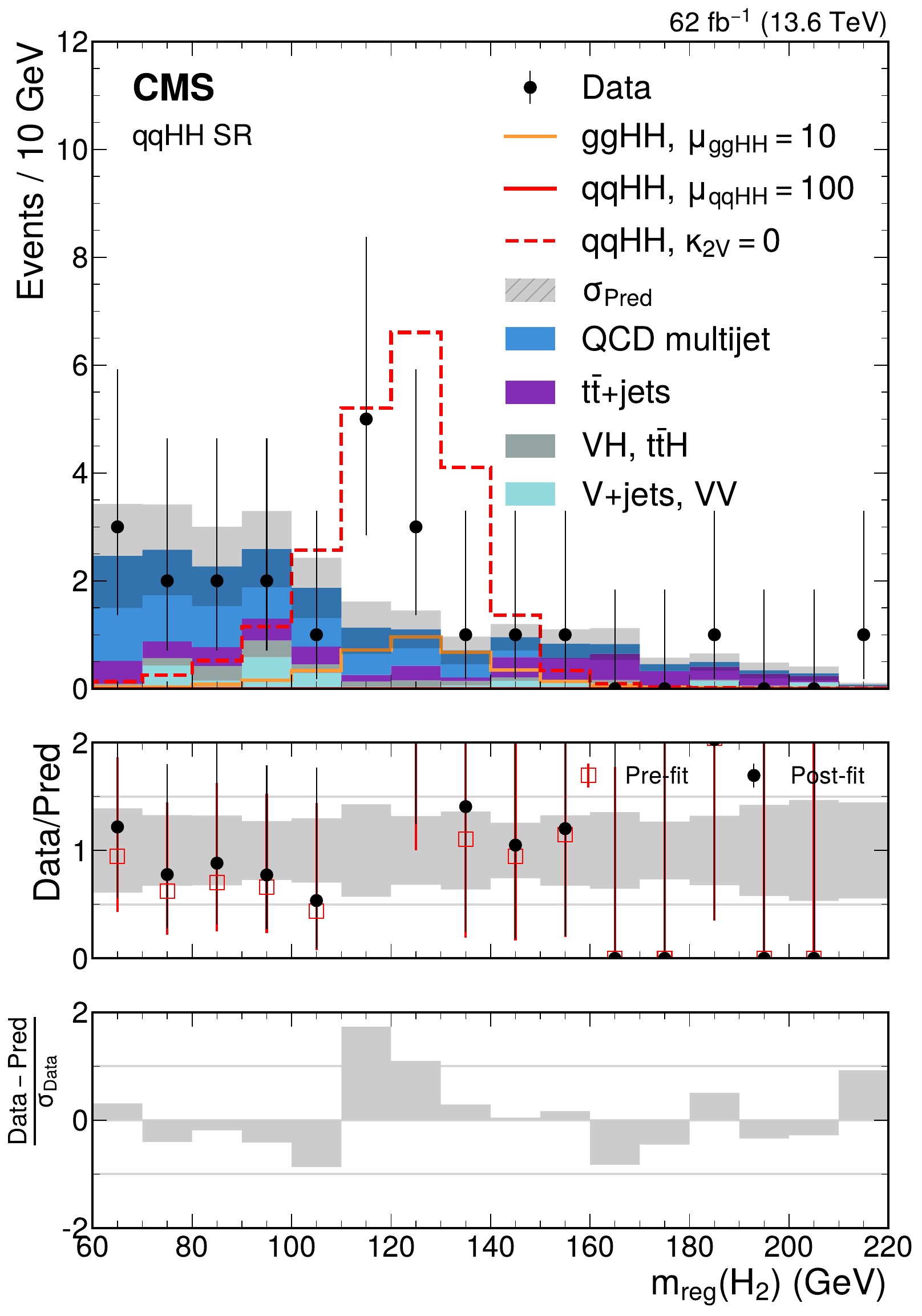}\\
    \includegraphics[width=0.425\textwidth]{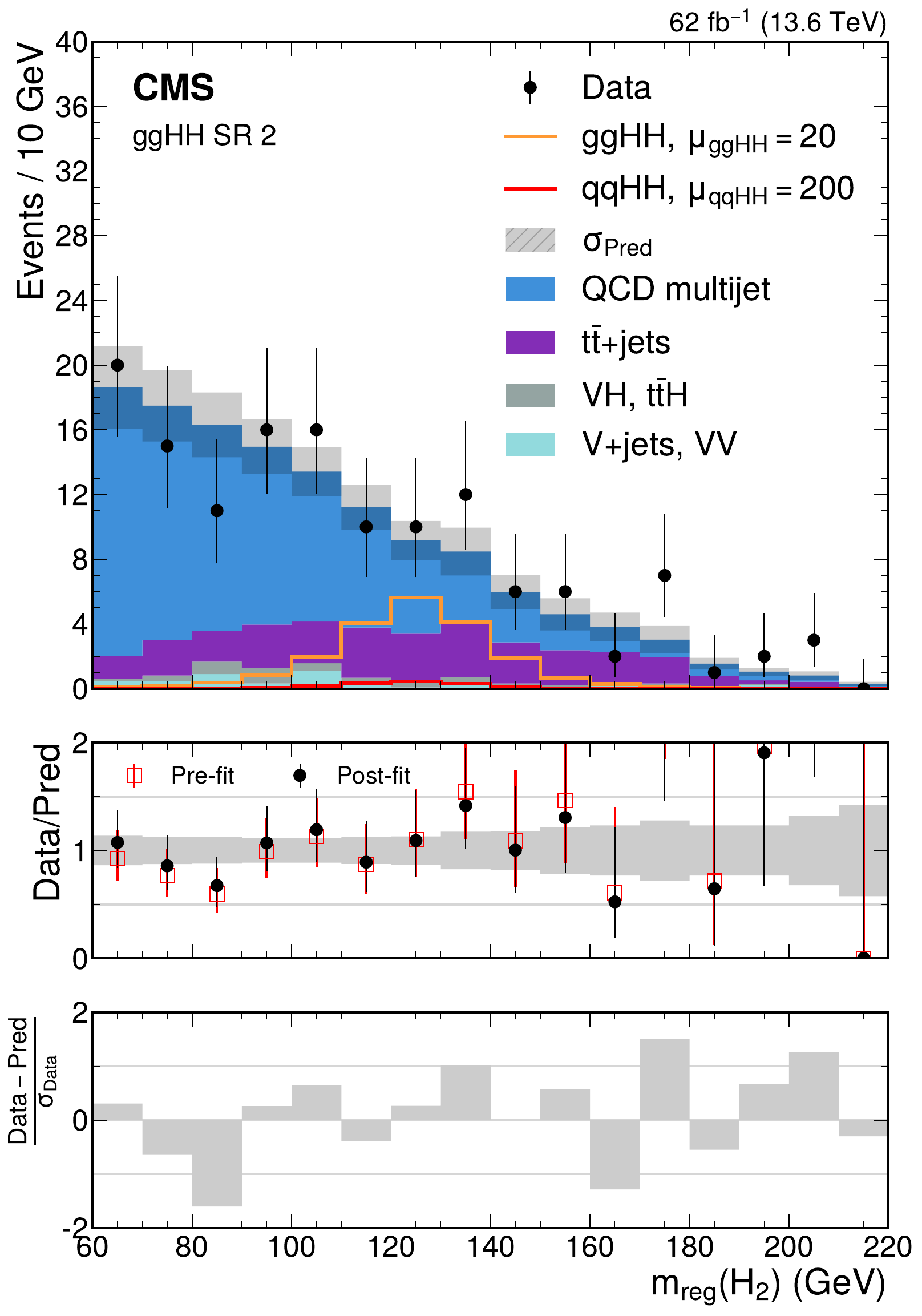}
    \includegraphics[width=0.425\textwidth]{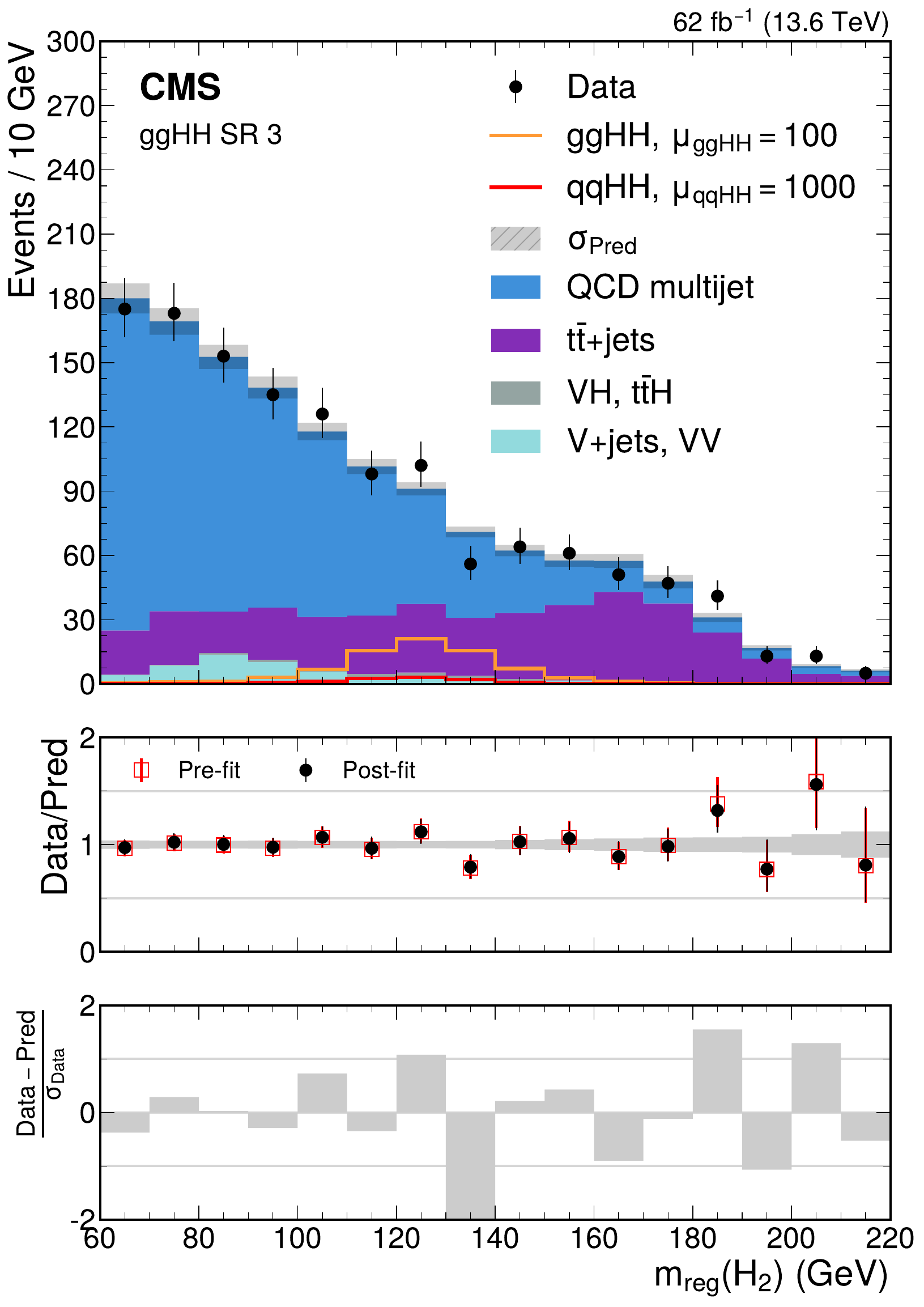}
    \caption{The background-only fit distributions of the regressed mass of the subleading \PH candidate $\mreg(\Htwo)$ in \gghh SR 1 (upper left), \qqhh SR (upper right), \gghh SR 2 (lower left), and \gghh SR 3 (lower right). The SM \gghh and \qqhh signal are overlaid in the \gghh SRs scaled by different factors. The targeted $\kappaVV=0$ \qqhh signal is also overlaid in the \qqhh SR. Notations are as in Fig.~\ref{fig:dnn_3b_control_region_fitted}.}
    \label{fig:parametric-mass-fit-data-mc}
\end{figure*}

Figure~\ref{fig:parametric-mass-fit-data-mc} shows the $\mreg(\Htwo)$ distributions for all categories. Signal is extracted via a fit to these distributions. The \ttbar background, as well as the minor backgrounds from \Vjets, \VV, and single \PH production, are obtained from the MC templates with corrections based on CRs in data.

\subsection{Systematic uncertainties}\label{sec:merged_systematics}

Many of the theoretical and experimental uncertainties affecting either the \HHbbbb signal prediction or that of background processes modeled via simulation are estimated following the same procedures as outlined for the resolved analysis in Section~\ref{sec:resolved_systematics}. Uncertainties that are exclusive to the merged analysis are those affecting the \bbbar tagging efficiency, the jet mass scale (JMS) and resolution (JMR) for the two \hbb-candidate jets, and the uncertainties in the QCD multijet and \ttbar background predictions.

The simulated predictions for the \HHbbbb signal, \ZH, and \VV processes are corrected with the \bbbar tagging efficiency SFs reported in Section~\ref{sec:zbb}. The corresponding uncertainties, which range 10--20\%, are included in the fit as independent nuisance parameters for each WP (VHP, HP, and MP) and data-taking period. These uncertainties are uncorrelated between AK8 jets with ${\pt>450\GeV}$, for which efficiency corrections are measured using high-\pt \zbb decays, and AK8 jets with ${\pt<450\GeV}$, for which no explicit calibration was made. The \bbbar tagging efficiency for jets with \TXbb score less than the range of the MP, which contribute only to the regions with small signal purity, is assigned a relative uncertainty twice as large than the closest calibrated WP. Differences in the JMS and JMR between data and simulation, measured with \zbb decays for the same \TXbb WPs, are used to correct predictions from the simulation, and the corresponding uncertainties are propagated as systematic uncertainties in the simulated predictions. 

The systematic uncertainties with the largest impact in the merged analysis are those affecting the QCD multijet background prediction, followed by uncertainties in the \ttbar background prediction. 
In the first approach, the QCD multijet background prediction is derived from a reweighting of data events in a \bbbar tag sideband region (\regCRQC). Because of the close synergies with the background prediction method used in the resolved analysis, the uncertainty in the QCD multijet prediction is evaluated following a very similar procedure to that described in Section~\ref{sec:resolved_systematics}. The normalization of the QCD multijet prediction is fit to data with no prior constraint, independently per SR category (HPSR, LPSR, and VR). The spread, defined as the half-difference between 84\% and 16\% quantiles, among each of the $k=20$ individual background predictions is added in quadrature to the binwise statistical uncertainty from the \regCRQC data. These uncertainties are uncorrelated across bins of the fitted distributions. The normalization of the \ttbar background prediction is fit to data without any prior, independently per SR category, and binwise uncertainties take into account the limited statistical precision of the \ttbar simulation.

For the second approach, the uncertainty in the QCD multijet background prediction is propagated through profiling of the unconstrained transfer factor parameters in the signal extraction fit.
For the \ttbar background, we propagate uncertainties in the corrections for \TXbb, $\tau_{32}$, and the recoil $\pt(\text{jj})$ distribution. These systematic uncertainties are implemented as lognormal constraints because the variations from the uncertainties are not dependent on $\mreg(\Htwo)$. The bin-by-bin statistical uncertainty in the correction on the $\mathcal{D}(\text{\gghh-vs-bkg})$ and $\mathcal{D}(\text{\qqhh-vs-bkg})$ shapes for \ttbar, derived in the all-hadronic \ttbar CR, is propagated through additional nuisance parameters to the signal extraction fit.

\subsection{Results} \label{sec:merged_results}

Table~\ref{tab:merged_limits_exp} summarizes the expected upper limits at the 95\% \CL on \muhh, \mugghh, and \muqqhh from the two approaches described in Sections~\ref{sec:merged_184} and \ref{sec:merged_parametric}. The first approach achieves stricter expected upper limits compared to the second approach.

\begin{table*}[!htb]
\centering
\topcaption{The expected best fit signal strengths and 95\% \CL upper limits (U.L.) on the inclusive, \gghh, and \qqhh signal strengths for the SM scenario for the two approaches described in Sections~\ref{sec:merged_184}~and~\ref{sec:merged_parametric}, respectively. The expected best fit signal strength and 95\% CL upper limit on the \qqhh signal strength for the non-SM $\kappaVV=0.5$ scenario are also reported. The uncertainties given for the best fit signal strengths correspond to the 68\% \CL intervals. The best fit signal strengths are calculated with a SM \HHbbbb signal injected, while the upper limits are calculated in the absence of signal.}   
\renewcommand{\arraystretch}{1.2}
\cmsTable{
  \begin{scotch}{l c c c c c c c c }
  Scenario  & \multicolumn{6}{c}{SM}  & \multicolumn{2}{c}{$\kappaVV=0.5$} \\
  \hline
  & \multicolumn{2}{c}{\muhh} & \multicolumn{2}{c}{\mugghh} & \multicolumn{2}{c}{\muqqhh} & \multicolumn{2}{c}{\muqqhh} \\ [\cmsTabSkip]
  \hline
  Expected & Best fit & U.L. & Best fit & U.L. & Best fit & U.L. & Best fit & U.L. \\
  \hline
  First approach (mass fit) & $1.0^{+3.4}_{-2.9}$ & 7.1 & $1.0^{+3.5}_{-3.0}$ & 7.3 & $1^{+83}_{-67}$   & 173 & $1.0^{+0.8}_{-0.6}$    & 0.78  \\ 
  First approach (DNN fit)  & $1.0^{+3.1}_{-2.6}$ & 6.6 & $1.0^{+3.2}_{-2.7}$ & 6.7 & $1^{+82}_{-66}$   & 185 & $1.0^{+0.8}_{-0.6}$    & 0.87  \\ 
  Second approach           & $1.0^{+3.7}_{-3.1}$ & 8.1 & $1.0^{+3.8}_{-3.2}$ & 8.3 & $1^{+224}_{-194}$ & 518 & $1.0^{+1.0}_{-0.8}$    & 1.91  \\ 
  \end{scotch}
}
\label{tab:merged_limits_exp}
\end{table*}

Table~\ref{tab:merged_limits_obs} summarizes the measured signal strength modifiers and the corresponding 68\% \CL intervals for each signal strength modifier.  No significant excess is observed over the background-only expectation, and the measured signal rates are compatible between the two approaches.

\begin{table*}[!htb]
\centering
\topcaption{The observed best fit signal strengths and 95\% \CL upper limits (U.L.) on the inclusive, \gghh, and \qqhh signal strengths for the SM scenario for the two approaches described in Sections~\ref{sec:merged_184}~and~\ref{sec:merged_parametric}, respectively. The best fit and observed 95\% CL upper limits on the \HH signal strength for the non-SM $\kappaVV=0.5$ are also reported. The uncertainties given for the best fit signal strengths correspond to the 68\% \CL intervals.}   
\renewcommand{\arraystretch}{1.2}
\cmsTable{
\begin{scotch}{l c c c c c c c c }
 Scenario & \multicolumn{6}{c}{SM}  & \multicolumn{2}{c}{$\kappaVV=0.5$} \\
 \hline
 & \multicolumn{2}{c}{\muhh} & \multicolumn{2}{c}{\mugghh} & \multicolumn{2}{c}{\muqqhh} & \multicolumn{2}{c}{\muqqhh} \\ [\cmsTabSkip]
 \hline
 Observed & Best fit & U.L. & Best fit & U.L. & Best fit & U.L. & Best fit & U.L. \\
 \hline
 First approach (mass fit) & $-2.6^{+2.9}_{-3.2}$ & 5.3 & $-2.5^{+2.9}_{-3.2}$ & 5.5 & $-56^{+45}_{-65}$ & 131 & $-0.24^{+0.20}_{-0.29}$ & 0.59 \\ 
 First approach (DNN fit) & $-0.8^{+2.7}_{-2.5}$ & 5.8 & $-0.8^{+2.7}_{-2.6}$ & 6.1 & $-68^{+45}_{-65}$ & 131 & $-0.30^{+0.20}_{-0.30}$ & 0.64 \\ 
 Second approach & $2.1^{+4.5}_{-3.7}$ & 11.1 & $2.0^{+4.6}_{-3.8}$ & 11.2 & $408^{+188}_{-230}$ & 1060 & ${1.37}^{+0.89}_{-0.80}$ & 2.86 \\
\end{scotch}
}
\label{tab:merged_limits_obs}
\end{table*}

In addition, 95\% \CL upper limits on \muhh  are computed as a function of the \kappal and \kappaVV coupling modifiers separately, with all couplings except for the one being varied fixed to the SM prediction. The observed and expected intervals outside of which the theoretical prediction for the \HH production rate is excluded at 95\% \CL for \kappal and \kappaVV are summarized in Table~\ref{tab:merged_couplings}.

\begin{table*}[!htb]
    \centering
    \topcaption{The observed and expected intervals for which the predicted \HH production cross section is excluded at 95\% CL, for \kappal and \kappaVV, in the merged analysis for the two approaches described in Sections~\ref{sec:merged_184}~and~\ref{sec:merged_parametric}, respectively.}   
    \label{tab:merged_couplings}
    \renewcommand{\arraystretch}{1.2}
    \begin{scotch}{l c c c c}
     \multirow{2}{*}{ 95\% \CL interval}  & \multicolumn{2}{c}{\kappal} & \multicolumn{2}{c}{\kappaVV}\\ [\cmsTabSkip] 
     & Observed & Expected & Observed & Expected \\\hline
     First approach (mass fit) & $[-5.1, 10.9]$ & $[-6.8, 12.8]$ & $[0.63, 1.42]$ & $[0.54, 1.51]$ \\ 
     First approach (DNN fit) & $[-6.5, 13.4]$ & $[-7.7, 14.9]$ & $[0.64, 1.41]$ & $[0.56, 1.49]$\\ 
     Second approach & $[-11.0, 19.2]$ & $[-10.3, 18.4]$ & $[0.30, 1.74]$ & $[0.47, 1.58]$ \\
 \end{scotch}
\end{table*}

\section{Combined results}\label{sec:combinations}

The analyses of the resolved and merged topologies are combined, using the approach with the best expected sensitivity for each topology. This corresponds to the resolved analysis ``first approach'' described in Section~\ref{sec:resolved_transformer} and the merged analysis ``first approach'' described in Section~\ref{sec:merged_184} using the mass-fit method. Although the high-\pt events selected in the merged analysis are a small fraction of the events accepted in the resolved analysis, the resolved analysis SvsB classifier learns that events with high-\pt jets have significantly better S/B and therefore tends to assign high \psig values to these events. It is therefore important to address this event overlap between topologies, making sure that no events are considered twice and ensuring that the combination of analyses yields the best possible expected sensitivity. 

Of all the event categories considered, the merged analysis HPSR category achieves the highest expected S/B. It is therefore natural to assign priority to merged HPSR events, removing them from the resolved analysis before performing a combined signal extraction fit including both analyses. In the combined fit, the merged HPSR categories are included following the mass fit method (Section~\ref{sec:massfit_merged}), as this minimizes the scale of the event overlap with the resolved analysis and because the mass-fit method achieves the most stringent expected constraints on \kappal. This removes 2--10\% of the expected signal events from the resolved analysis at high SvsB classifier score and has a very small effect on the background, removing a few events from the resolved \regSRFourb and affecting the background distribution by much less than 1\%. The overlap removal degrades by about 5\% the expected 95\% \CL upper limit on \muhh from the resolved analysis for all \kappal hypotheses, compared to the results summarized in Section~\ref{sec:resolved_results}. Omitting the VR and LPSR regions from the merged analysis degrades the merged analysis-only expected limit by about 14\%. Alternative combination configurations were tested, such as the inclusion of the merged analysis LPSR regions or using the merged analysis DNN fit method (Section~\ref{sec:dnnfit_merged}) rather than the mass fit. In all cases, the combination approach described above achieved comparable to or better sensitivity than the alternative options.

Systematic uncertainties affecting the \HHbbbb signal prediction from theoretical sources are fully correlated across the \gghh and \qqhh resolved and merged analyses, as well as across data-taking periods. Experimental uncertainties in the signal prediction from pileup, integrated luminosity, JES, and JER are correlated among all event categories but uncorrelated across data-taking periods. The remaining experimental uncertainties, such as the trigger efficiency, the \PQb tagging efficiencies, the \bbbar tagging efficiency, and the AK8 jet mass scale and resolution are uncorrelated across event categories and data-taking periods. The uncertainties in the QCD multijet background predictions of the two analyses, affecting either the rate or the shape of the backgrounds, are also uncorrelated. The dominant systematic uncertainties impacting the combined fit are the uncertainty in the normalization of the QCD multijet background in the merged HPSR as well as the shape and normalization uncertainties in the background prediction of the resolved analysis. The remaining uncertainties have a more marginal impact on the combined result, with the exception of the uncertainty in the theoretical prediction for the inclusive \gghh cross section related to the scheme and scale choice of the virtual top mass $m_{\text{top}}$, which has a significant relative impact for \muhh values near the exclusion limits.

\begin{figure*}[!htb]
  \centering
    \includegraphics[width=0.45\textwidth]{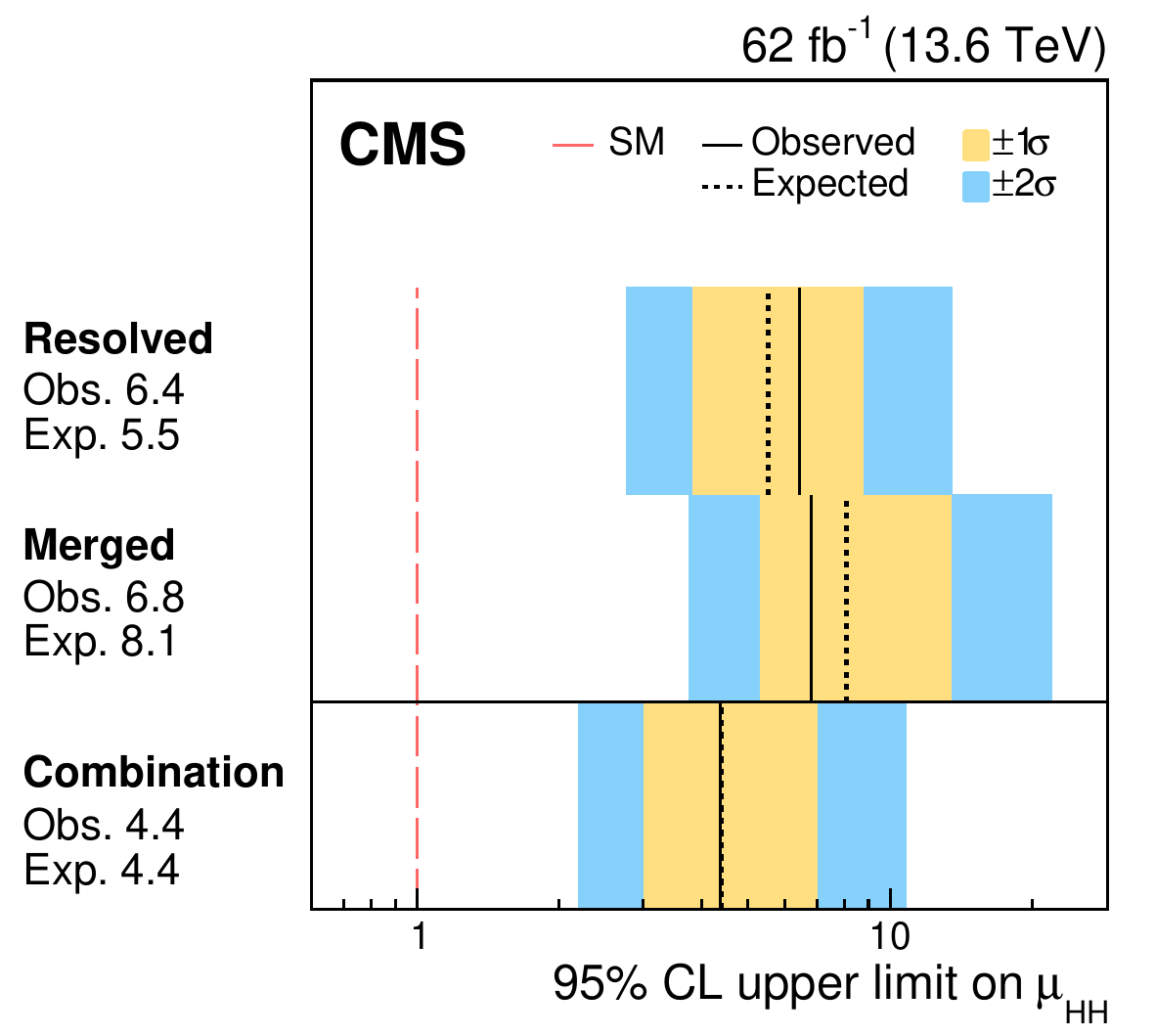}
    \includegraphics[width=0.45\textwidth]{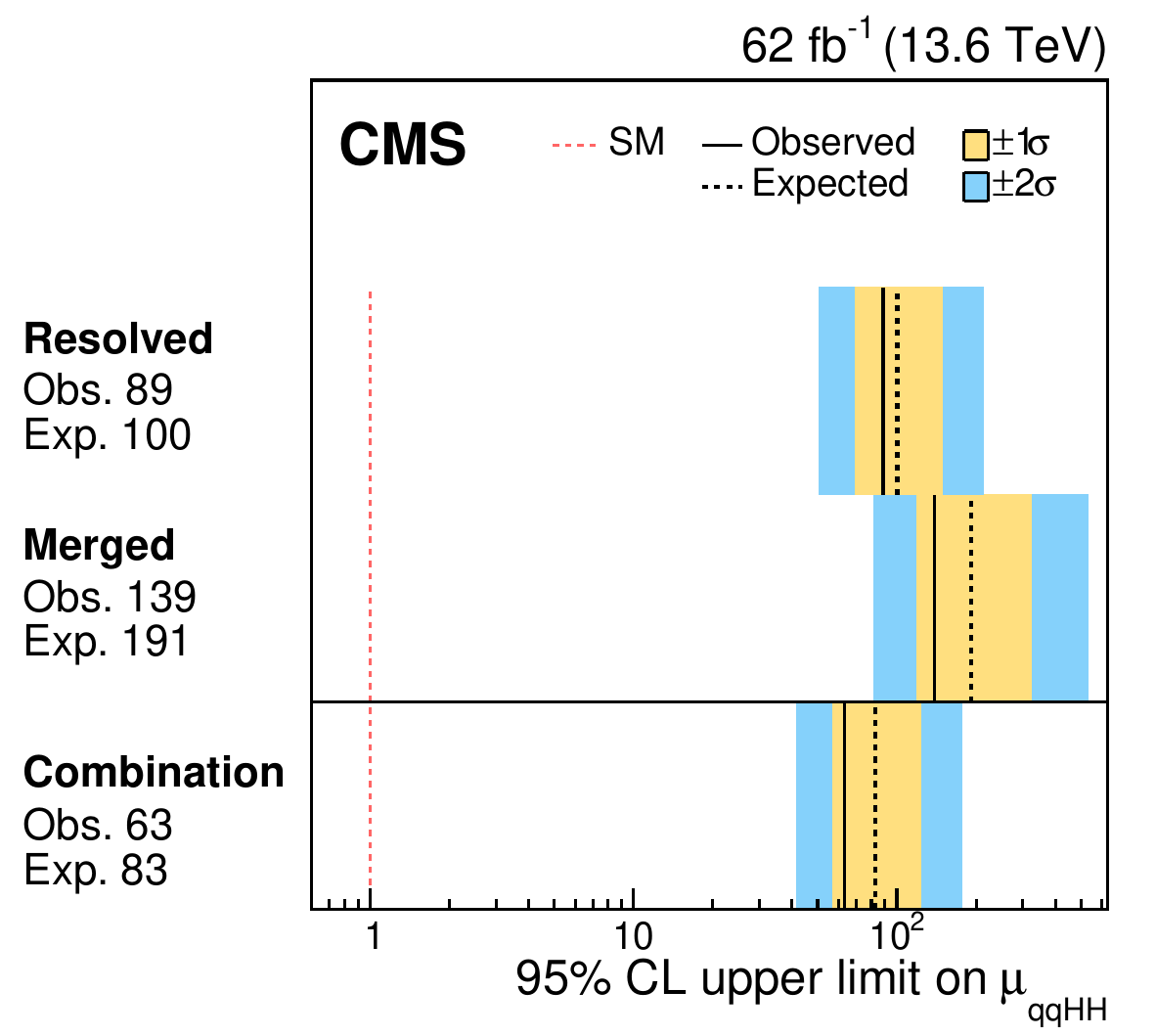}
    \caption{Left: the observed (solid) and expected (dashed) 95\% \CL upper limits on the signal strength of \HH production from the resolved analysis with overlap removed, the merged HPSR mass fit category, and their combination. The orange and blue bands represent, respectively, the 68 and 95\% \CL intervals around the expected limit. Right: the same breakdown of 95\% \CL upper limits on the signal strength of \qqhh production.}
    \label{fig:combination_results}

\end{figure*}

The observed data in the combined result are compatible with the best fit signal-plus-background prediction, with a $p$-value of 72\% from a goodness-of-fit test. Figure~\ref{fig:combination_results} (left) shows 95\% \CL upper limits on the \HH signal strength \muhh, assuming that relative contributions of the \gghh and \qqhh production modes are fixed to the SM prediction within their uncertainties. Upper limits are presented separately for the resolved \HHbbbb analysis, after overlap removal, for the merged HPSR mass fit category, and for their combination. Upper limits are obtained by fixing the signal event kinematic features to those predicted for the SM \PH couplings, $\kappal=\kappaVV=1$. The observed (expected) upper limit on \muhh from the combined fit is 4.38~(4.39). Figure ~\ref{fig:combination_results} (right) shows, instead, the 95\% \CL upper limits on the \qqhh signal strength, \muqqhh, calculated with the \gghh signal strength profiled.

\begin{figure*}[!htb]
  \centering
    \includegraphics[width=0.45\textwidth]{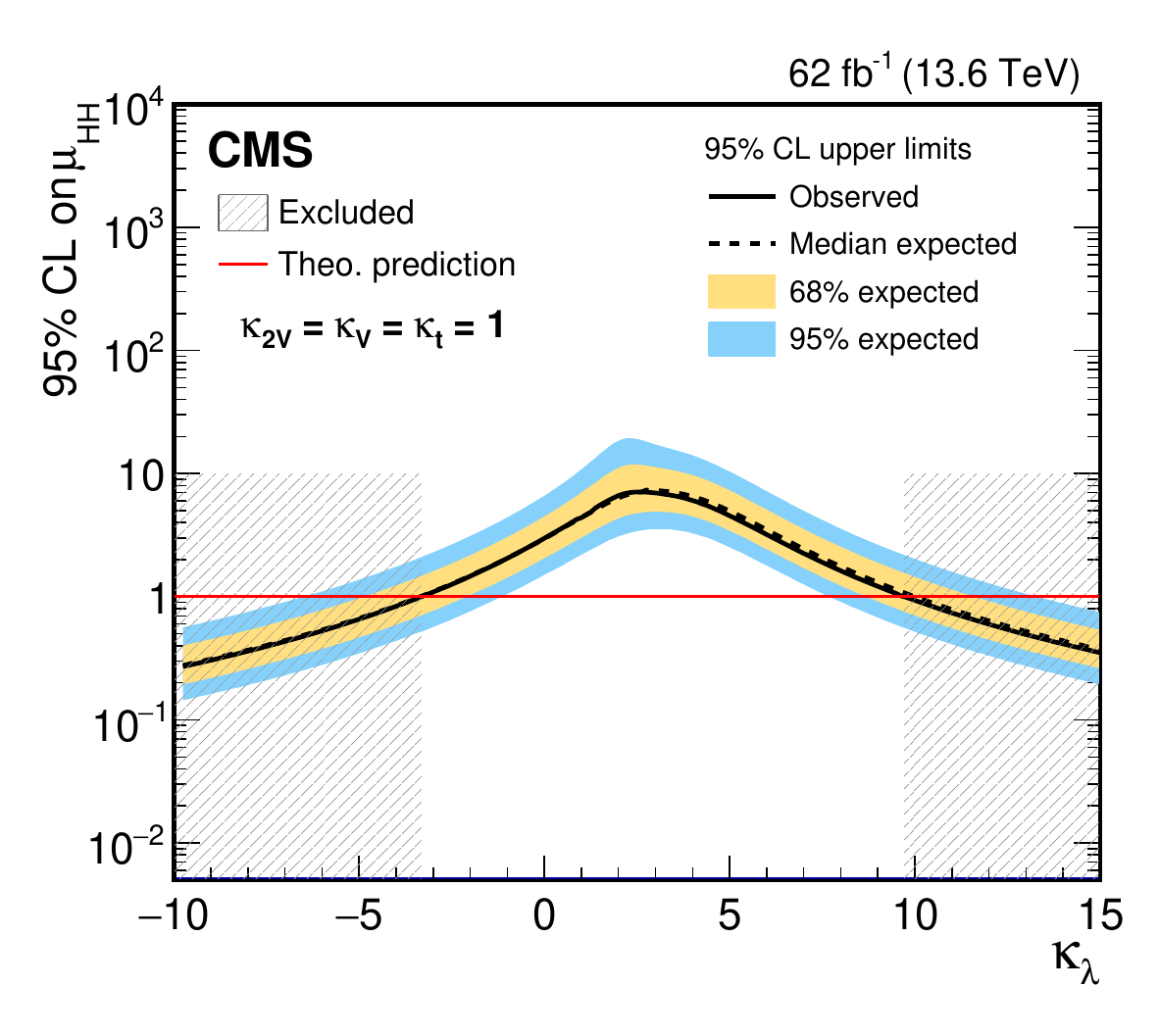}
    \includegraphics[width=0.45\textwidth]{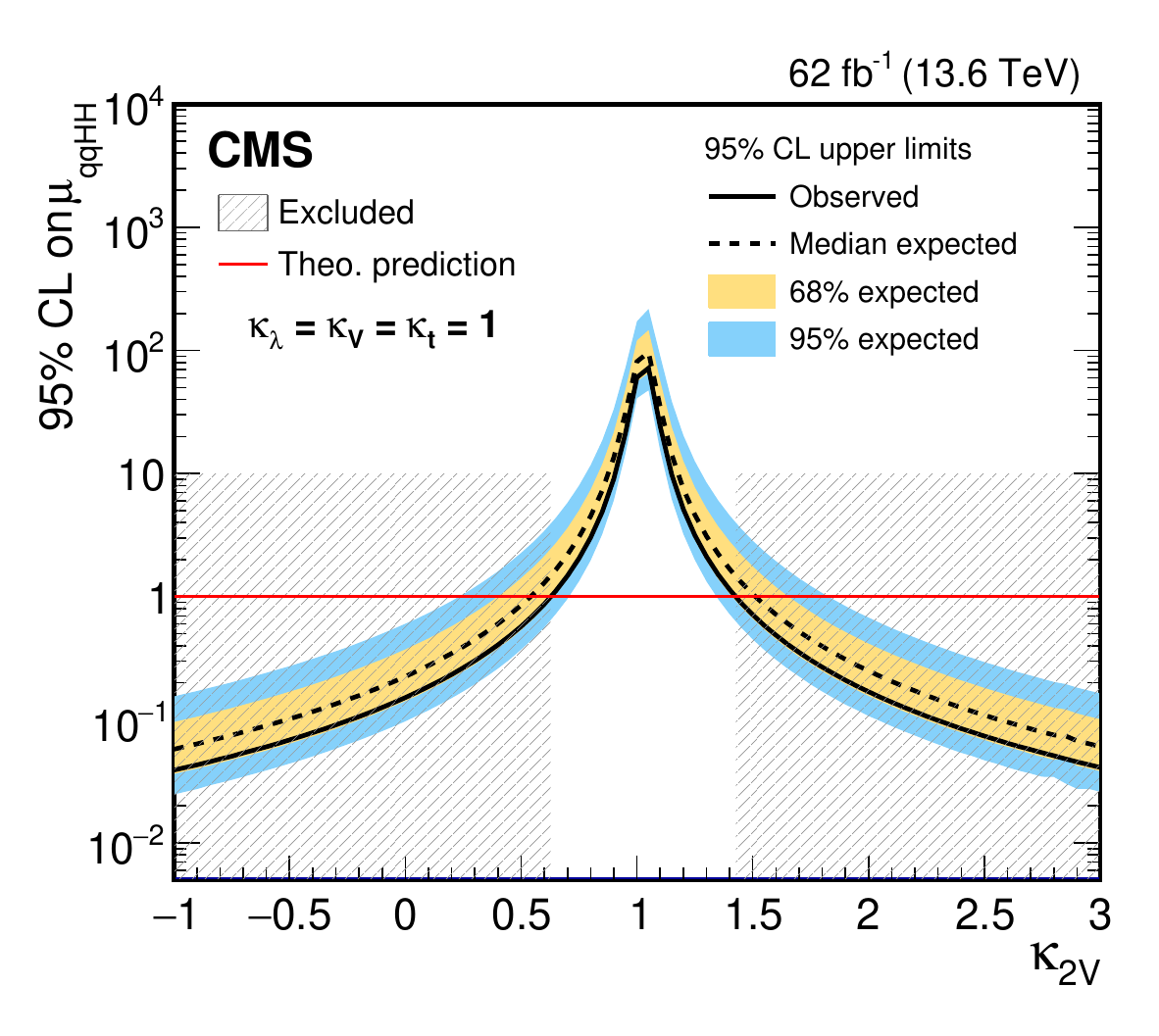}
    \caption{The observed (solid) and expected (dashed) 95\% \CL upper limits on the signal strength of the \HH production (\muhh) obtained as functions of \kappal (left) and \kappaVV (right) for the combined fit of resolved and merged \HHbbbb analyses. The orange and blue bands represent, respectively, the 68 and 95\% \CL intervals around the expected limit. The horizontal red lines indicate the SM prediction.}
    \label{fig:combination_scan_ul_results}

\end{figure*}

In addition, observed and expected upper limits at 95\% \CL on the \HH signal strength are reported in Fig.~\ref{fig:combination_scan_ul_results} as functions of \kappal (left) and \kappaVV (right). For values of \kappal outside the observed (expected, in absence of a signal) range of $[-3.3,9.7]~([-3.4,10.0])$, the theoretical prediction for \HH production is excluded at 95\% \CL. Similarly, the prediction for \HH production is excluded at 95\% \CL for \kappaVV values outside the observed (expected) range of $[0.63,1.43]~([0.54,1.51])$. 

\begin{figure*}[!htb]
  \centering
    \includegraphics[width=0.45\textwidth]{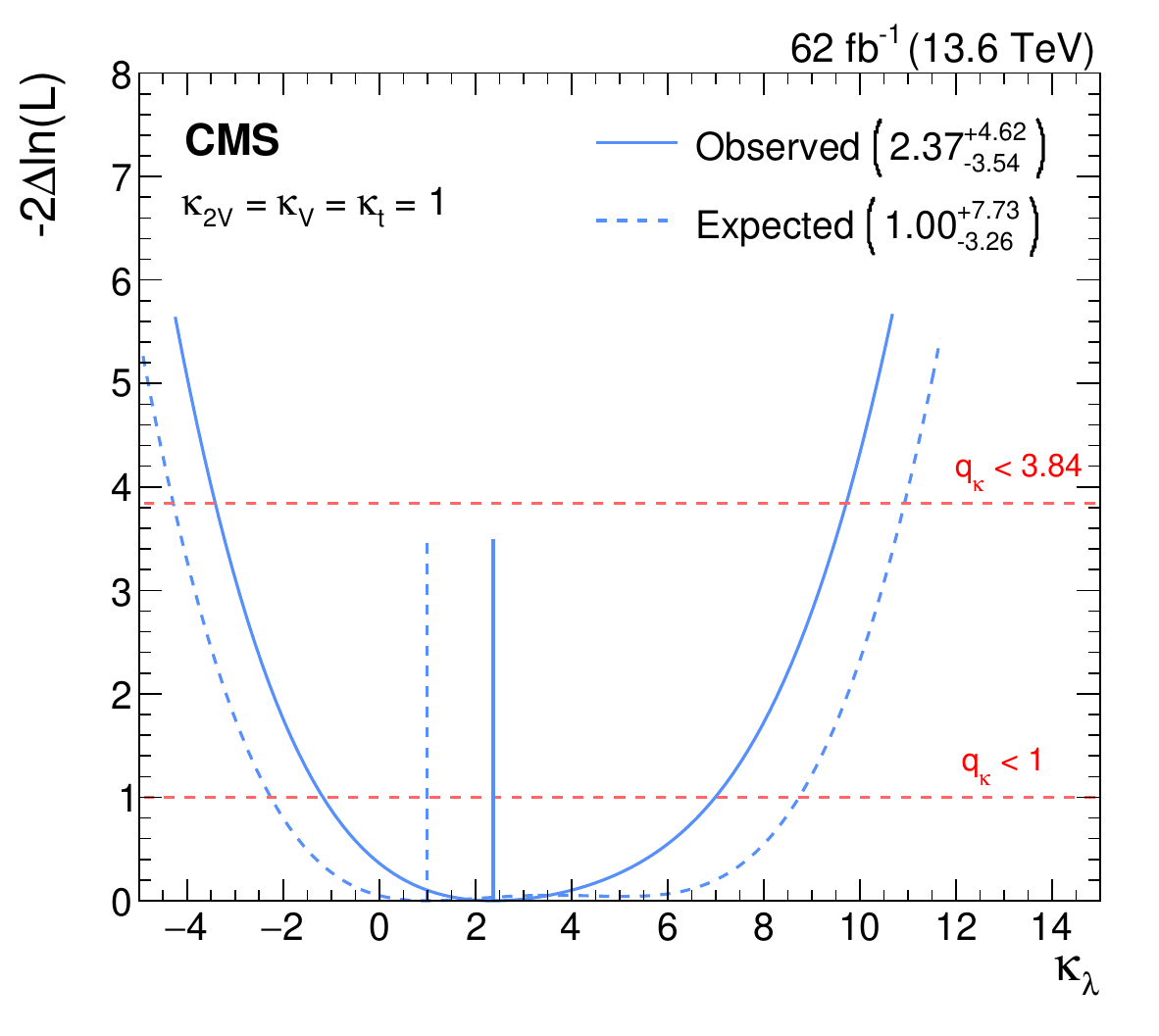}
    \includegraphics[width=0.45\textwidth]{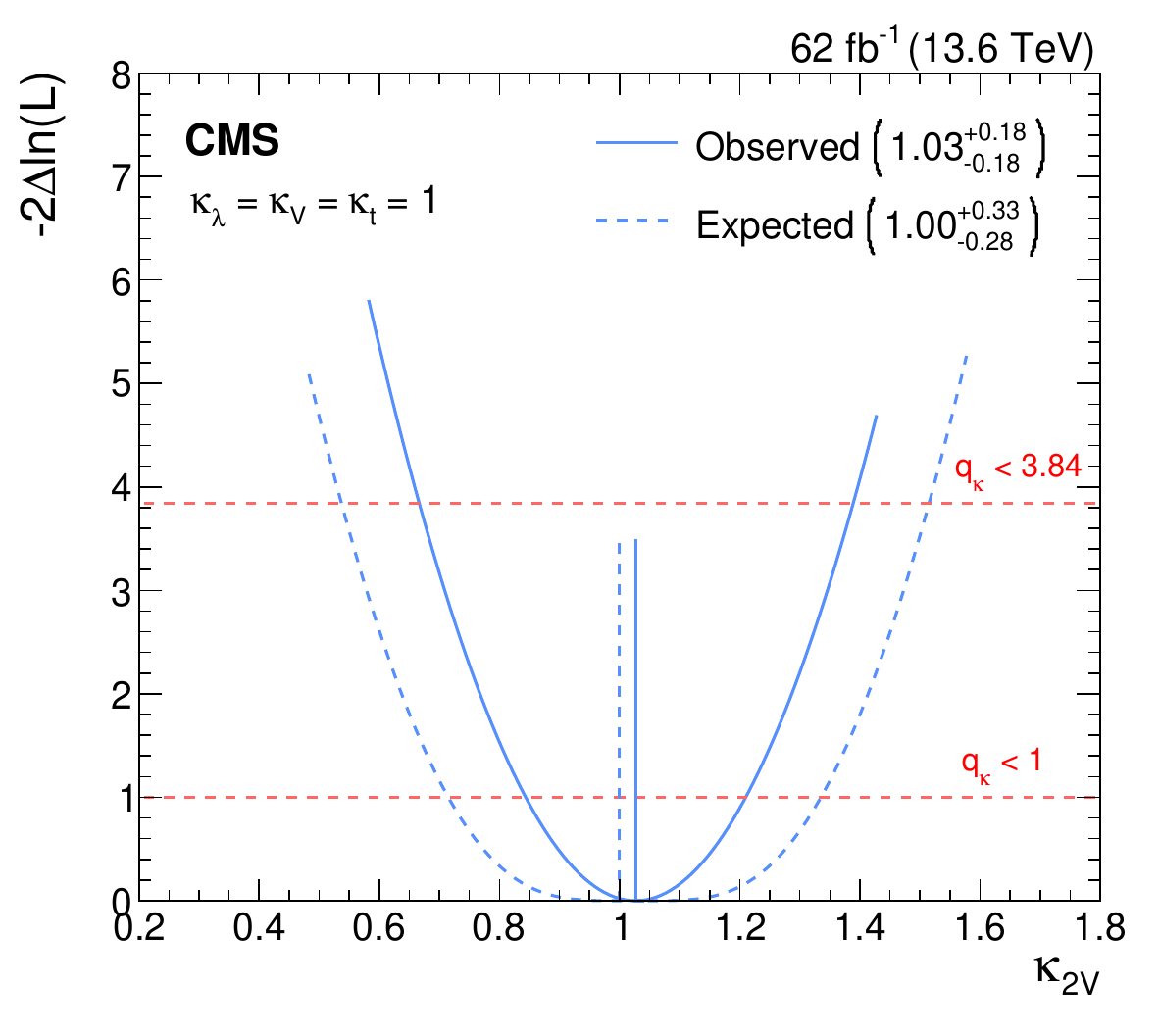}
    \caption{The observed (solid) and expected (dashed) profile likelihood ratios as function of \kappal (left) and \kappaVV (right) for the combined fit of resolved and merged \HHbbbb analyses, where the expected is obtained from an Asimov dataset~\cite{Cowan:2010js} defined by fixing the nuisance parameters to values obtained from a fit to data with ${\muhh=1}$. The ${q_{\kappa}<1}$ and ${q_{\kappa}<3.84}$ levels are indicated with the dashed red lines.}
    \label{fig:combination_scan_profile_results}

\end{figure*}

Figure~\ref{fig:combination_scan_profile_results} shows the observed and expected (for SM) profiled likelihood ratios as a function of \kappal (left) and \kappaVV (right). The \kappal and \kappaVV intervals presented in the latest \HH combinations by the ATLAS~\cite{PhysRevLett.133.101801} and CMS Collaborations~\cite{CMS:2022dwd,CMS:2025ngq} were obtained by applying Wilks' theorem~\cite{Wilks} in the asymptotic regime, such that the intervals for which the profiled-likelihood test statistic, ${q_{\kappa} = -2(\ln \mathcal{L}_{\kappa} - \ln \mathcal{L}_{\hat{k}})}$, where $\hat{\kappa}$ refers to the best fit value of $\kappa$, falls below 1 and 3.84 can be interpreted as 68\% and 95\% \CL intervals, respectively. Although the asymptotic regime applies to signal strength parameters, it is not necessarily valid for the coupling modifiers, \kappal and \kappaVV, which modify the predicted cross section via both linear and quadratic terms. In particular, the quadratic dependence of the expected yields on the values of the coupling modifiers violates the conditions for asymptotic convergence~\cite{Bernlochner:2022oiw}. Therefore, the likelihood ratio intervals predicted by the application of Wilks' theorem may undercover or overcover. 

For the results presented in this paper, the coverage predicted by Wilks' theorem was compared with that obtained following a Feldman-Cousins (FC) procedure~\cite{Feldman:1997qc}. The bounds on $q_{\kappa}$ corresponding to the 68\% and 95\% \CL intervals obtained via the FC procedure were observed to be more stringent than those predicted by Wilks' theorem, with differences ranging from a few percent to as much as 30\% in regions of the coupling space where post-fit uncertainties in the measured signal cross section are largest. The \kappal and \kappaVV intervals for which ${q_{\kappa}<1}$ and ${q_{\kappa}<3.84}$ are nonetheless reported in this paper, in order to directly compare with previous results and because the calculation of the FC intervals is computationally expensive, particularly for 2D intervals. For the ${q_{\kappa}<3.84}$ intervals on \kappal presented in this paper, the 95\% \CL interval obtained from the FC procedure is within a few percent of the ${q_{\kappa}<3.84}$ interval that provides 95\% \CL under Wilks' theorem.

\begin{figure*}[!htb]
  \centering
    \includegraphics[width=0.45\textwidth]{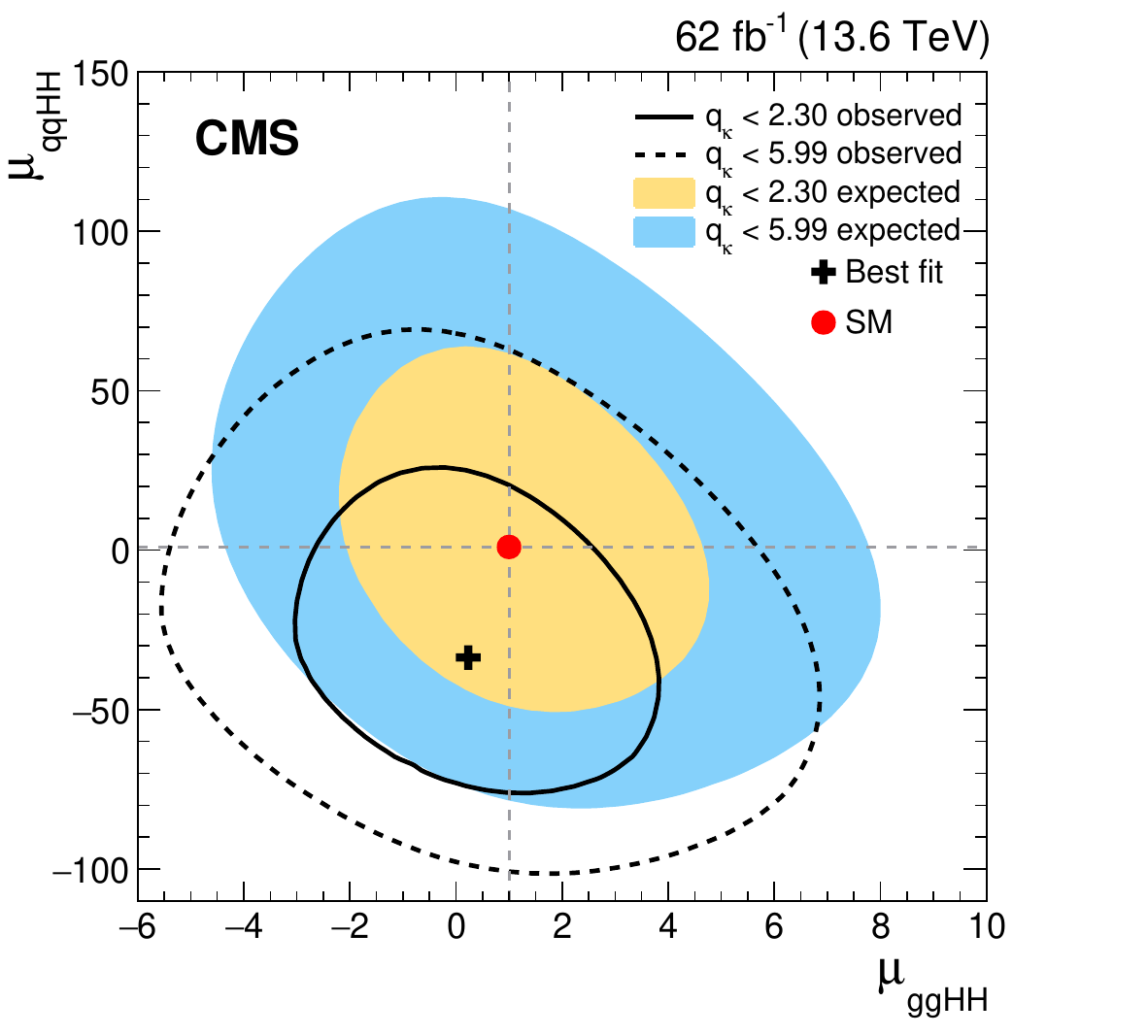}
    \includegraphics[width=0.45\textwidth]{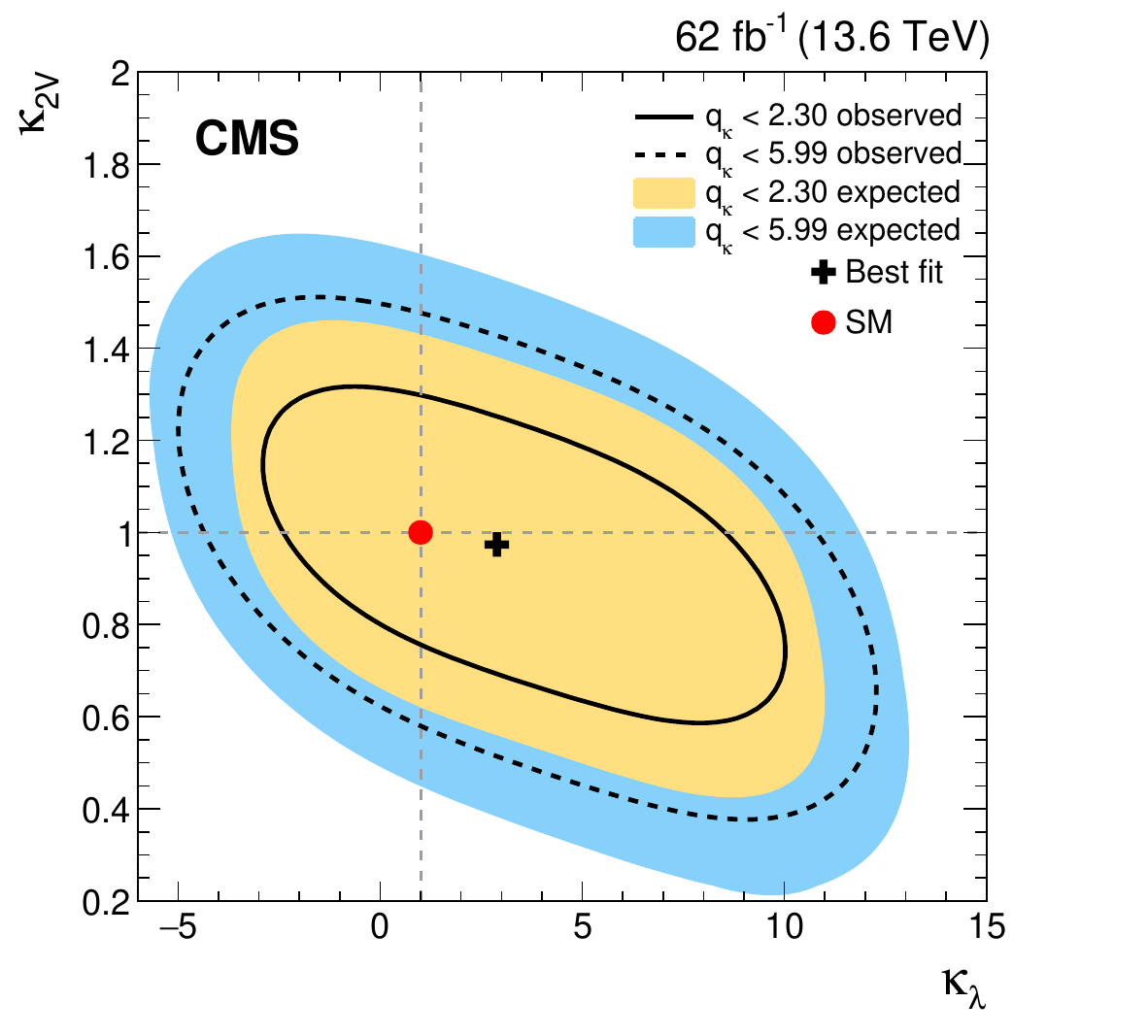}
    \caption{The observed and expected 2D exclusion range for (\mugghh, \muqqhh) (left) and (\kappal,\kappaVV) (right) for the combined fit of resolved and merged \HHbbbb analyses. The black solid and dashed lines show the observed exclusion contours for ${q_{\kappa}<2.30}$ and ${q_{\kappa}<5.99}$, respectively, while the orange and blue shaded regions show the expected exclusion contours. The red circle indicates the SM prediction, while the black cross shows the best fit result.}
    \label{fig:combination_scan_kappa2D_results}

\end{figure*}

Table~\ref{tab:combination_bestfit} summarizes the observed and expected best fit values for the signal strengths, \kappal, and \kappaVV as well as their corresponding uncertainties, defined as the interval in which the test statistic $q_{\kappa}$ is less than one. The observed (expected) allowed range for \kappal for which ${q_{\kappa}<3.84}$ is ${[-3.4, 9.7]}\,{([-4.3, 10.9])}$, while for \kappaVV the allowed $q_{\kappa}<3.84$ range is ${[0.68, 1.38]}\,{([0.54, 1.50])}$.In both figures, all couplings other than the one being varied are fixed to the SM values. Figure~\ref{fig:combination_scan_kappa2D_results} left shows the ${q_{\kappa}<2.30}$ and ${q_{\kappa}<5.99}$ exclusion contours in the 2D (\mugghh, \muqqhh) plane. Figure~\ref{fig:combination_scan_kappa2D_results} right shows the exclusion contours in the 2D (\kappal, \kappaVV) plane, obtained via a fit with all other couplings fixed to the SM values. The best fit points in the (\mugghh, \muqqhh) and (\kappal, \kappaVV) planes are observed to be consistent with the SM predictions, within uncertainties. 

\begin{table}[!htb]
  \centering
  \topcaption{Expected and observed best fit values for the signal strengths, \kappal, and \kappaVV from the combined fit of the \HHbbbb resolved and merged analyses. The uncertainties given correspond to the ${q_{\kappa}<1}$ intervals. The expected results are calculated with a SM \HHbbbb signal injected.}
  \renewcommand{\arraystretch}{1.2}
  \begin{scotch}{l c c c c c}
    & \muhh & \mugghh &  \muqqhh & \kappal  & \kappaVV \\
   \hline
   Expected & $1.0^{+2.2}_{-2.0}$ & $1.0^{+2.2}_{-2.0}$ & $1^{+40}_{-35}$ & $1.0^{+7.7}_{-3.3}$ & $1.00^{+0.33}_{-0.28}$ \\
   Observed & $-0.2^{+2.1}_{-2.0}$ & $-0.1^{+2.2}_{-2.1}$ & $-34^{+38}_{-29}$ & $2.4^{+4.6}_{-3.5}$ & $1.03^{+0.18}_{-0.18}$ \\
   \end{scotch}
  \label{tab:combination_bestfit}
\end{table}

Table~\ref{tab:syst_breakdown} provides a breakdown of the major uncertainty sources in the measurement of \muhh. The largest contribution to the overall uncertainty, denoted as $\Delta \muhh$, is the ``statistical'' uncertainty, arising from the finite number of data events in the distributions fitted for the signal extraction. The systematic uncertainty is dominated by uncertainties in the normalization and shape, as well as bin-by-bin fluctuations, in the background prediction. The overall systematic uncertainty in the signal prediction is dominated by the theoretical uncertainty in the \gghh cross section prediction, primarily the uncertainty related to $m_{\text{top}}$. The method used for the background prediction and assessment of its uncertainty components explicitly differentiates between ``variance'' uncertainties, related to finite data, and ``bias'' uncertainties, related to potential limitations in the method itself. This separation makes future projections of this result onto larger data samples easier and more accurate. Given the largely statistical nature of the major uncertainty sources, it is expected that the precision in measuring \muhh will improve substantially with more data.

\begin{table*}[!htb]
    \centering
    \topcaption{Major sources of uncertainty in the measurement of the \muhh signal strength for the combined fit of the resolved and merged \HHbbbb analyses. The post-fit uncertainty on \muhh is separated into overall statistical and systematic components. The systematic component is further divided into uncertainties affecting the background prediction and those affecting the $\PH\PH$ signal prediction.}
    \renewcommand{\arraystretch}{1.2}
    \begin{scotch}{lc} 
    Uncertainty source & Relative contribution to $\Delta \muhh$ \\
    \hline
    Overall statistical uncertainty & 79\% \\
    Overall systematic uncertainty & 61\% \\
    \hline
    \multicolumn{2}{c}{Uncertainties in the background prediction} \\
    \hline
    Normalization & 30\% \\
    Bin-by-bin variations & 23\% \\
    Shape uncertainties: variance & 41\% \\
    Shape uncertainties: bias     & 14\% \\
    \hline
    \multicolumn{2}{c}{Uncertainties in the signal prediction} \\
    \hline
    Theoretical uncertainties & 19\% \\
    Jet energy scale and resolution uncertainties & 5\% \\
    \PQb-tagging uncertainties & 6\% \\
    Other sources & 2\% \\
    \end{scotch}
    \label{tab:syst_breakdown}
\end{table*}

\section{Combination with Run 2 data}\label{sec:combination_run23}

The Run 3 analyses considered in Section~\ref{sec:combinations} are combined with the updated resolved analysis with Run~2 data presented in Section~\ref{sec:resolved_run2} as well as the published merged analyses with Run~2 data targeting separately \gghh and \qqhh production~\cite{CMS:2022gjd}. Priority among Run~2 events is assigned to the merged analysis, removing overlapping events in the resolved analysis. This overlap removal degrades the expected \muhh upper limit from the Run~2 resolved analysis by about 12\% compared to the results described in Section~\ref{sec:resolved_run2}. Uncertainties affecting the background predictions, which account for the majority of the total background uncertainty in the analyses considered, are uncorrelated across both data-taking periods and analysis topologies. All theoretical uncertainties in the signal predictions are correlated between Run~2 and Run~3, including uncertainties in the \gghh inclusive cross section from the QCD scale variations, PDFs, and top quark mass, uncertainties in the inclusive \qqhh cross section, and acceptance uncertainties related to the QCD scale, parton shower, and PDF parameters.

Common experimental uncertainties related to the JES and JER, integrated luminosity, and pileup are correlated across Run~2 analyses. The JES and JER uncertainties are uncorrelated across Run~2 and Run~3 analyses because of the different methods in jet \pt regression and calibration. Several \PQb tagging systematic uncertainty sources are correlated across Run~2 and Run~3 analyses, while the statistical uncertainty components are uncorrelated. The trigger efficiency and \bbbar tagging uncertainties are uncorrelated between Run~2 and Run~3 analyses because of the different algorithms and calibration methods used. The remaining experimental uncertainties, which have a minor impact, are uncorrelated across Run~2 and Run~3 analyses.

\begin{figure*}[!htb]
  \centering
    \includegraphics[width=0.45\textwidth]{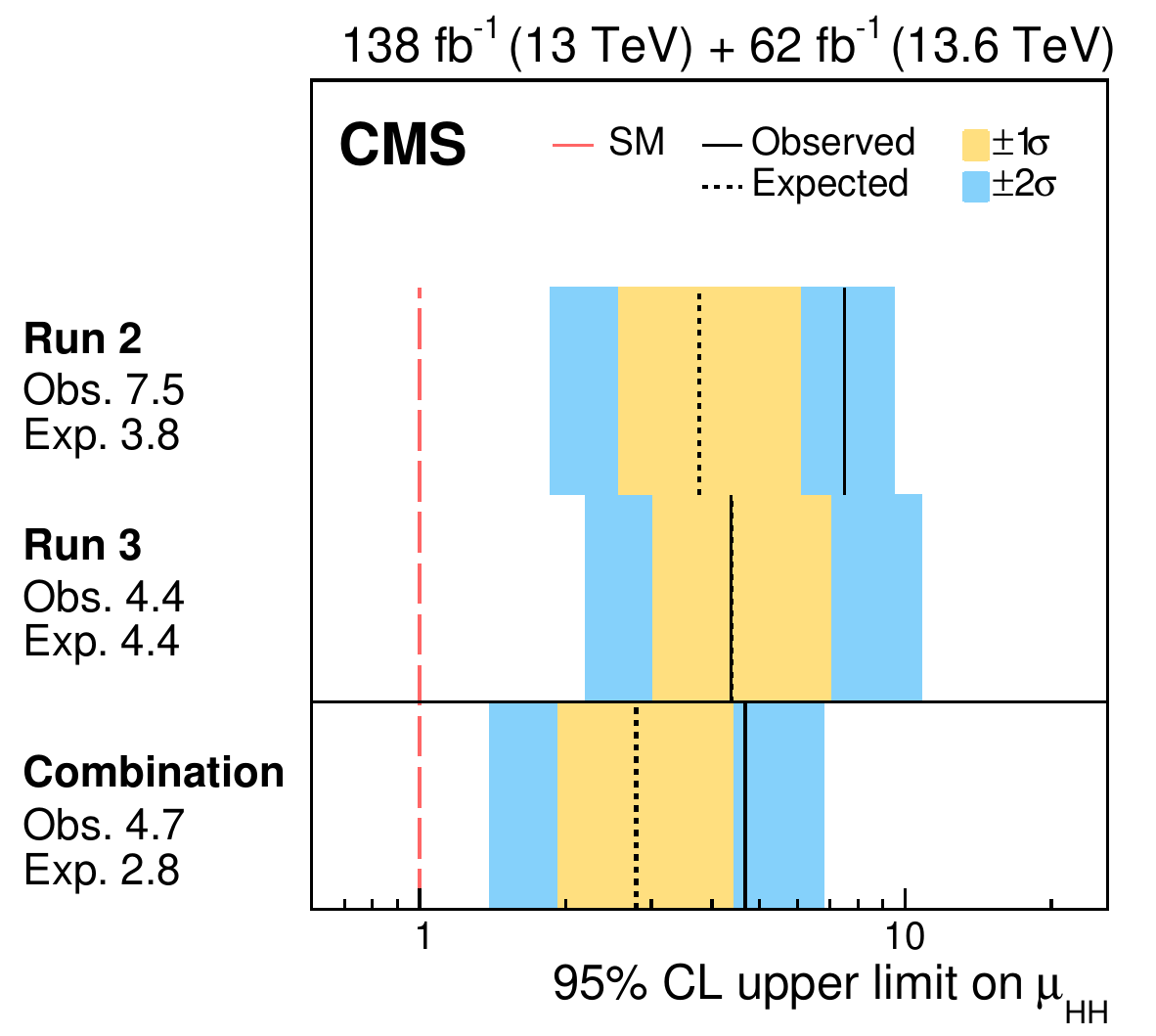}
    \includegraphics[width=0.45\textwidth]{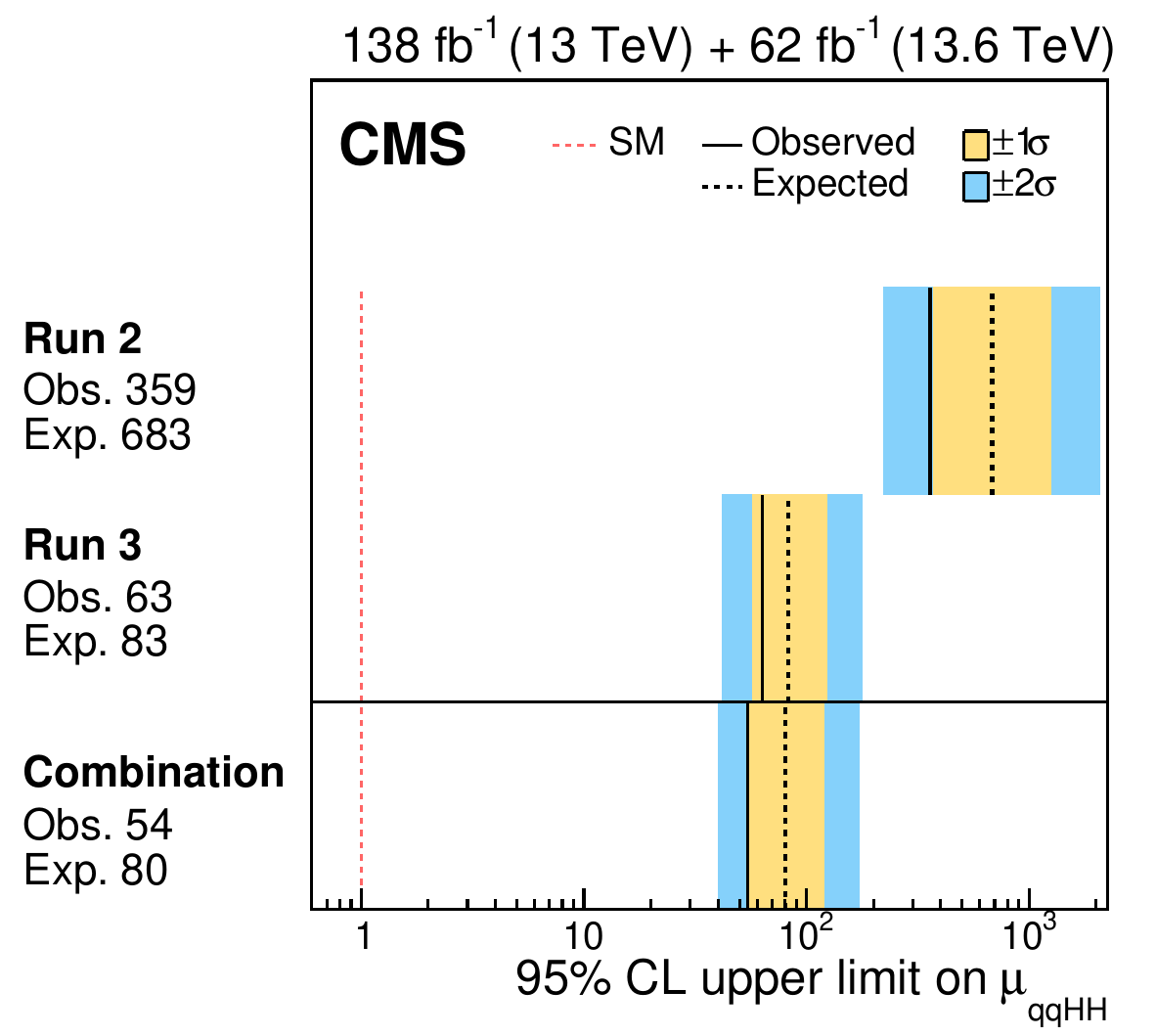}
  \caption{Left: the observed (solid) and expected (dashed) 95\% \CL upper limits on the signal strength of \HH production from the combination of Run~2 and Run~3 analyses and for each data set separately. The orange and blue bands represent, respectively, the 68 and 95\% \CL intervals around the expected limit. Right: the same breakdown of 95\% \CL upper limits on the signal strength of \qqhh production.}
  \label{fig:combination_run23_limits}
\end{figure*}

\begin{figure*}[!htb]
  \centering
    \includegraphics[width=0.45\textwidth]{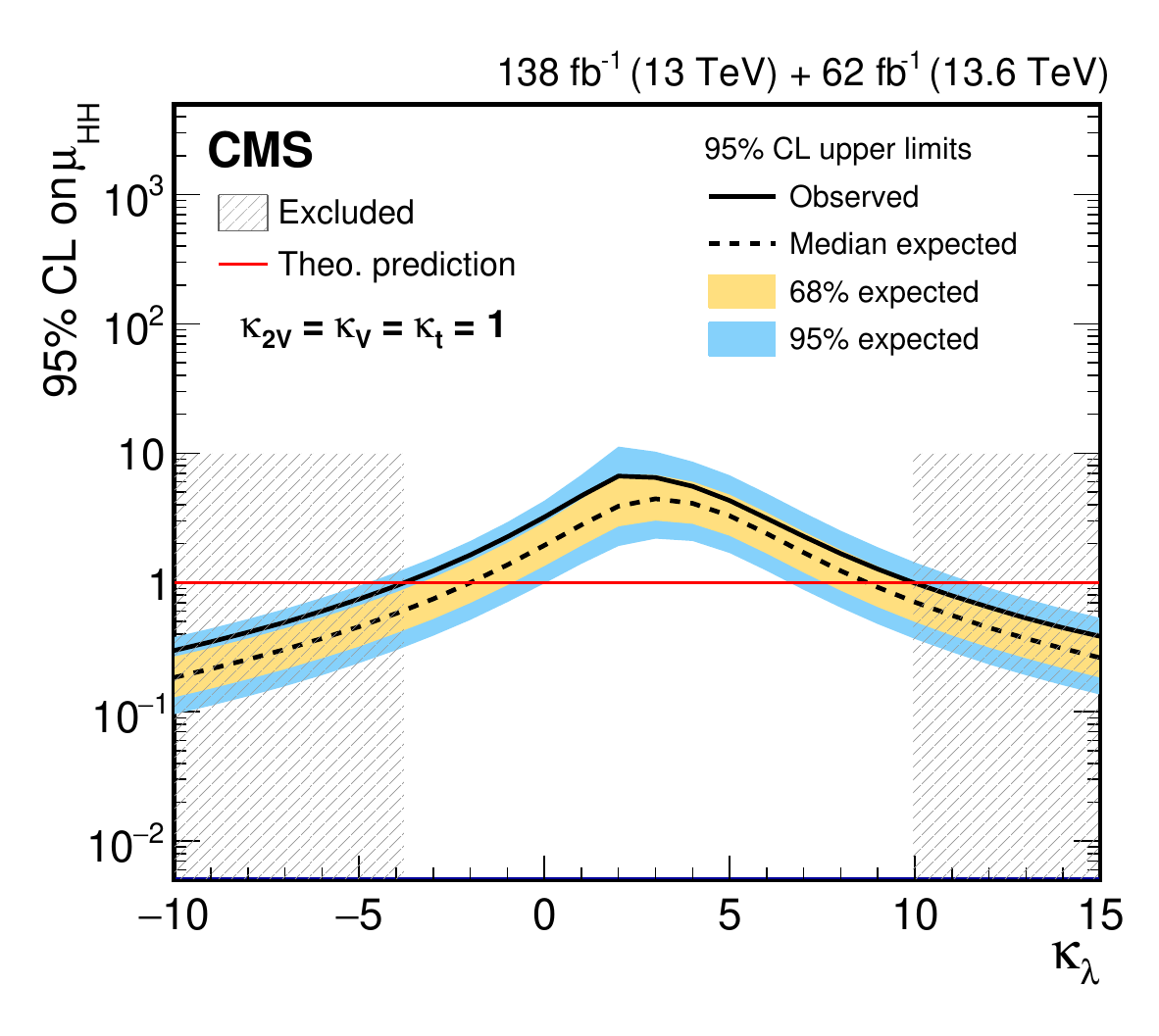}
    \includegraphics[width=0.45\textwidth]{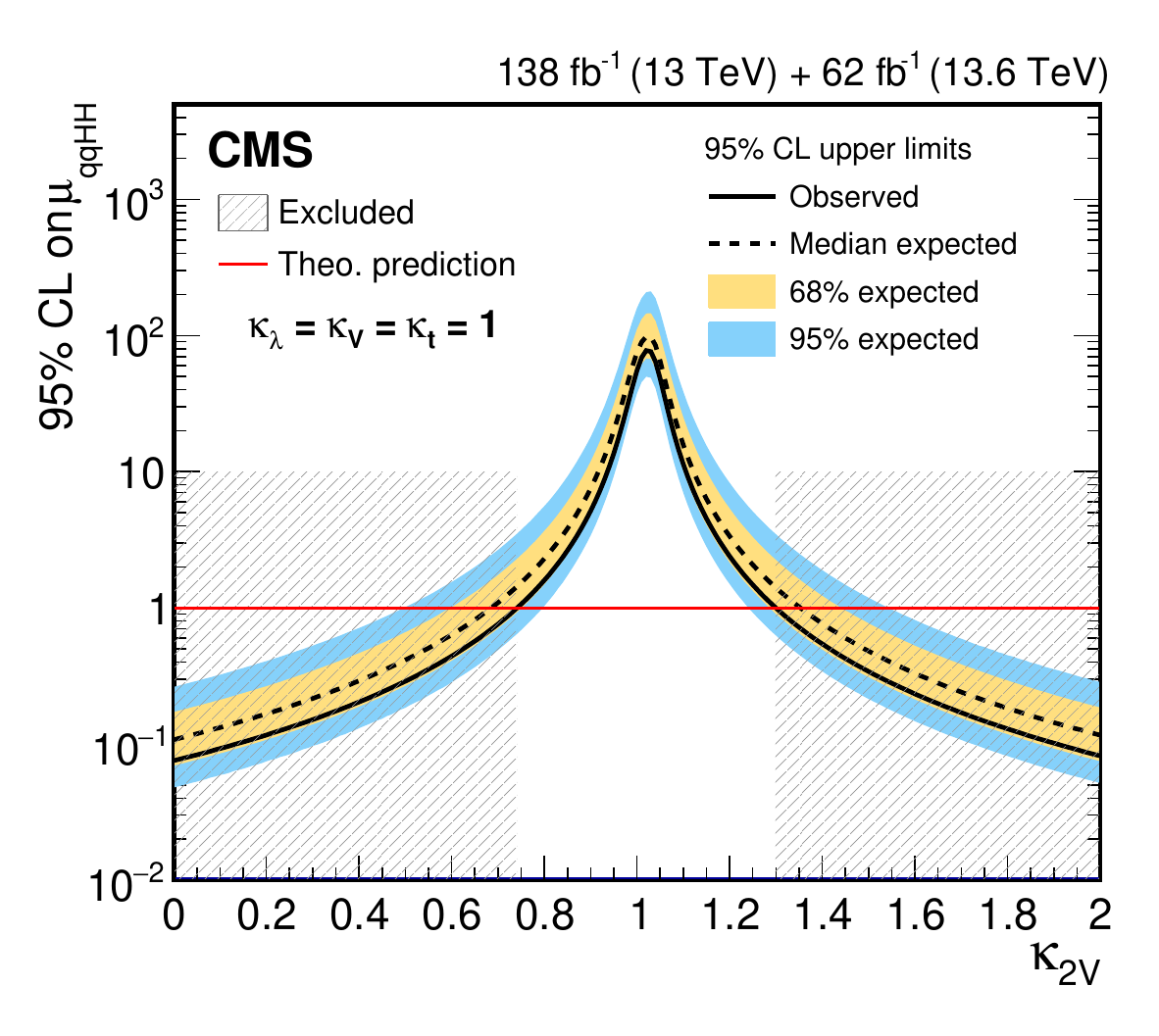}
    \caption{The observed (solid) and expected (dashed) 95\% \CL upper limits on the signal strength of the \HH production (\muhh) obtained as function of \kappal (left) and \kappaVV (right) for the combined fit of Run~2 and Run~3 \HHbbbb analyses. The orange and blue bands represent, respectively, the 68 and 95\% \CL intervals around the expected limit. The horizontal red lines indicate the SM prediction.}
    \label{fig:combination_run23_scan_ul_results}

\end{figure*}

The combined Run~2 and Run~3 observed data are compatible with the best fit signal-plus-background prediction, with a $p$-value of 90\% from the goodness-of-fit test. Figure~\ref{fig:combination_run23_limits} shows 95\% \CL upper limits on \muhh (left) and \muqqhh (right) for the Run~2 and Run~3 data separately, as well as for the combined fit. The observed (expected) upper limit on \muhh from the combined fit is 4.7~(2.8). The observed and expected upper limits at 95\% \CL on the \HH signal strength are shown in Fig.~\ref{fig:combination_run23_scan_ul_results} as function of \kappal (left) and \kappaVV (right). For values of \kappal outside the observed (expected, in absence of a signal) range of ${[-3.8, 10.0]}\,{([-2.0, 8.7])}$, the theoretical prediction for \HH production is excluded at 95\% \CL. Similarly, the prediction for \HH production is excluded at 95\% \CL for \kappaVV values outside the observed (expected) range of ${[0.74, 1.30]}\,{([0.69, 1.35])}$. 

\begin{figure*}[!htb]
  \centering
    \includegraphics[width=0.45\textwidth]{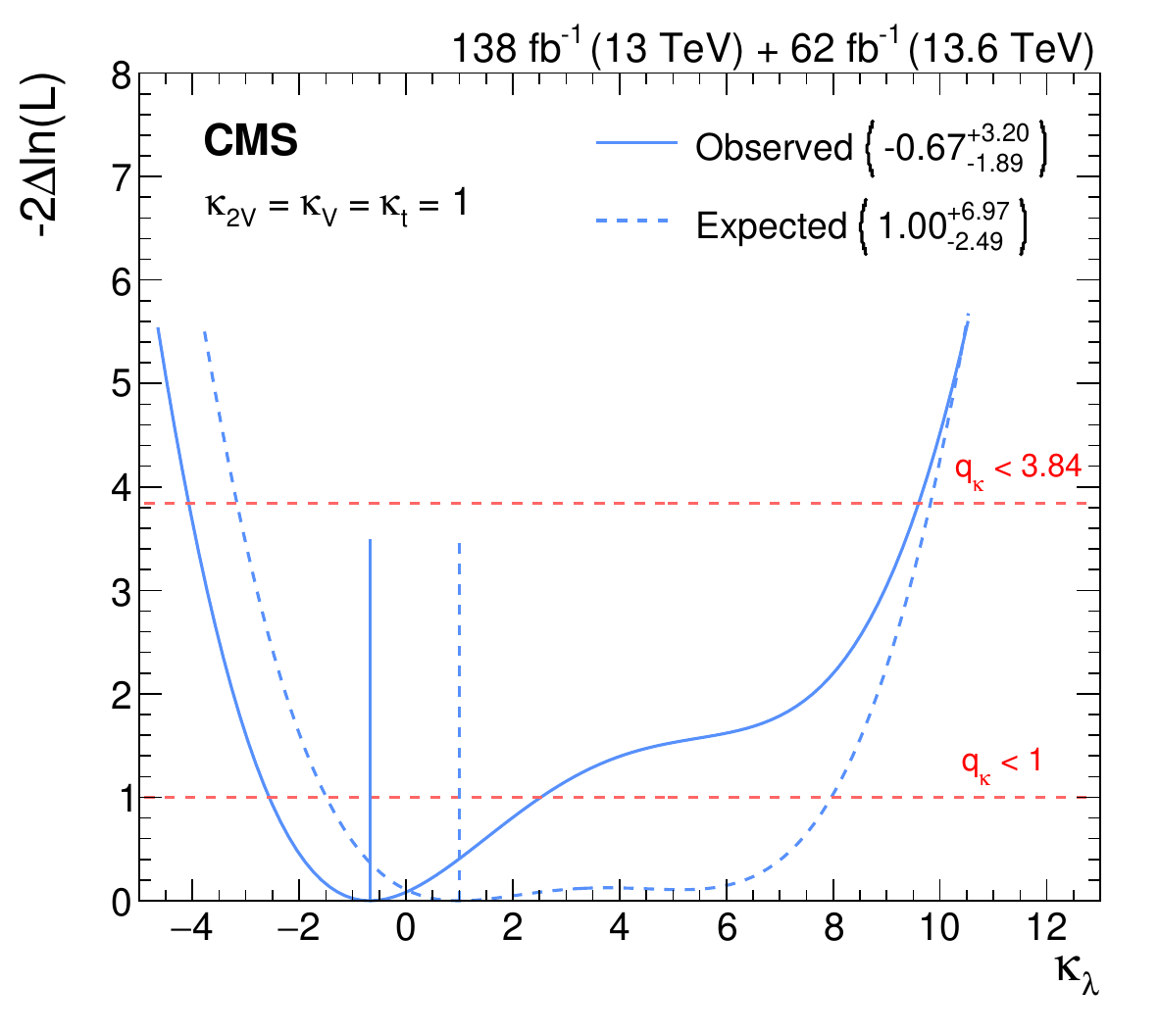}
    \includegraphics[width=0.45\textwidth]{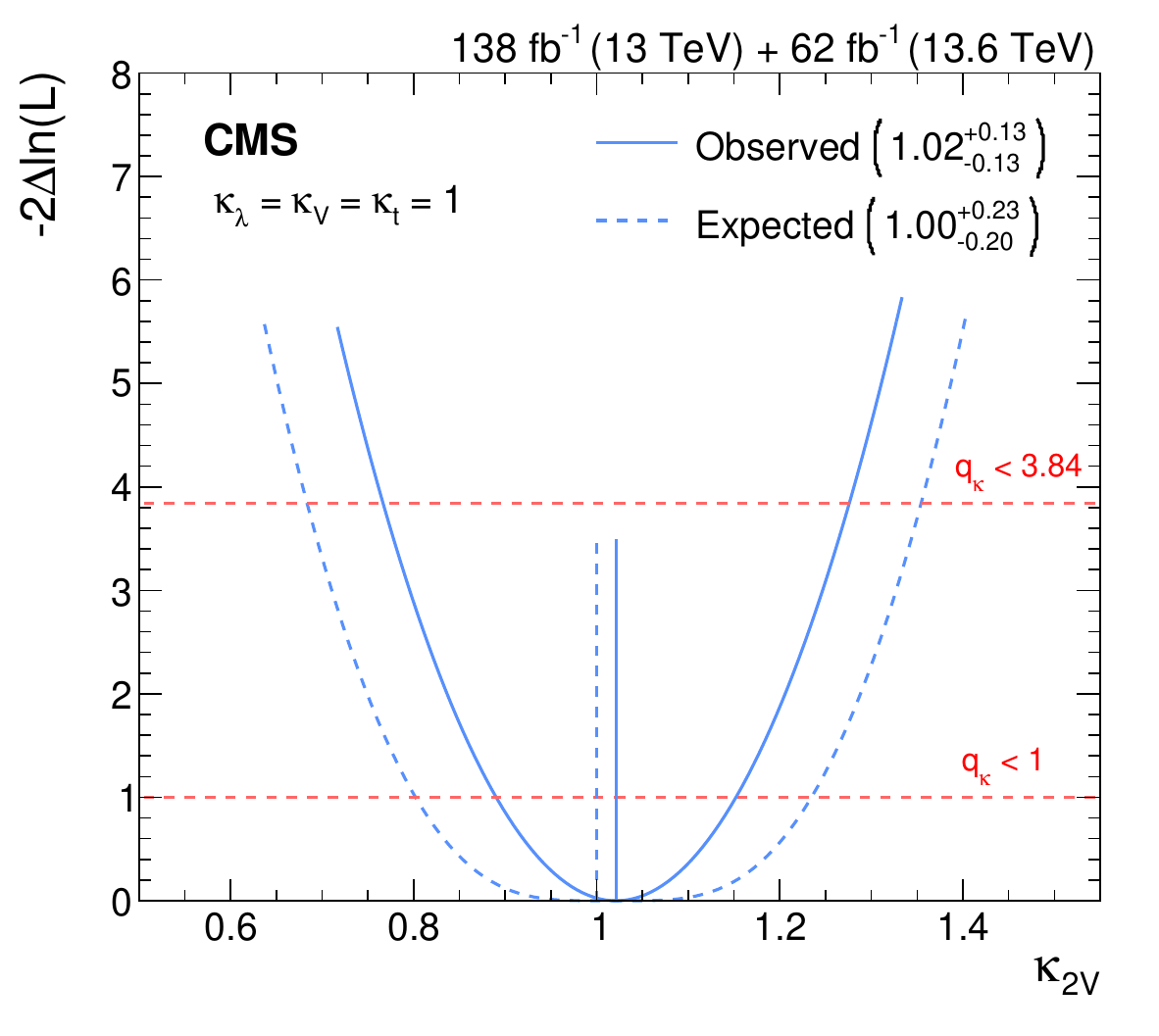}
    \caption{The observed (solid) and expected (dashed) profile likelihood ratios as function of \kappal (left) and \kappaVV (right) for the combined fit of Run~2 and Run~3 \HHbbbb analyses, where the expected is obtained from an Asimov dataset~\cite{Cowan:2010js} defined by fixing the nuisance parameters to values obtained from a fit to data with ${\muhh=1}$. The ${q_{\kappa}<1}$ and ${q_{\kappa}<3.84}$ levels are indicated with the dashed red lines.}
    \label{fig:combination_run23_scan_profile_results}

\end{figure*}

\begin{figure*}[!htb]
  \centering
    \includegraphics[width=0.45\textwidth]{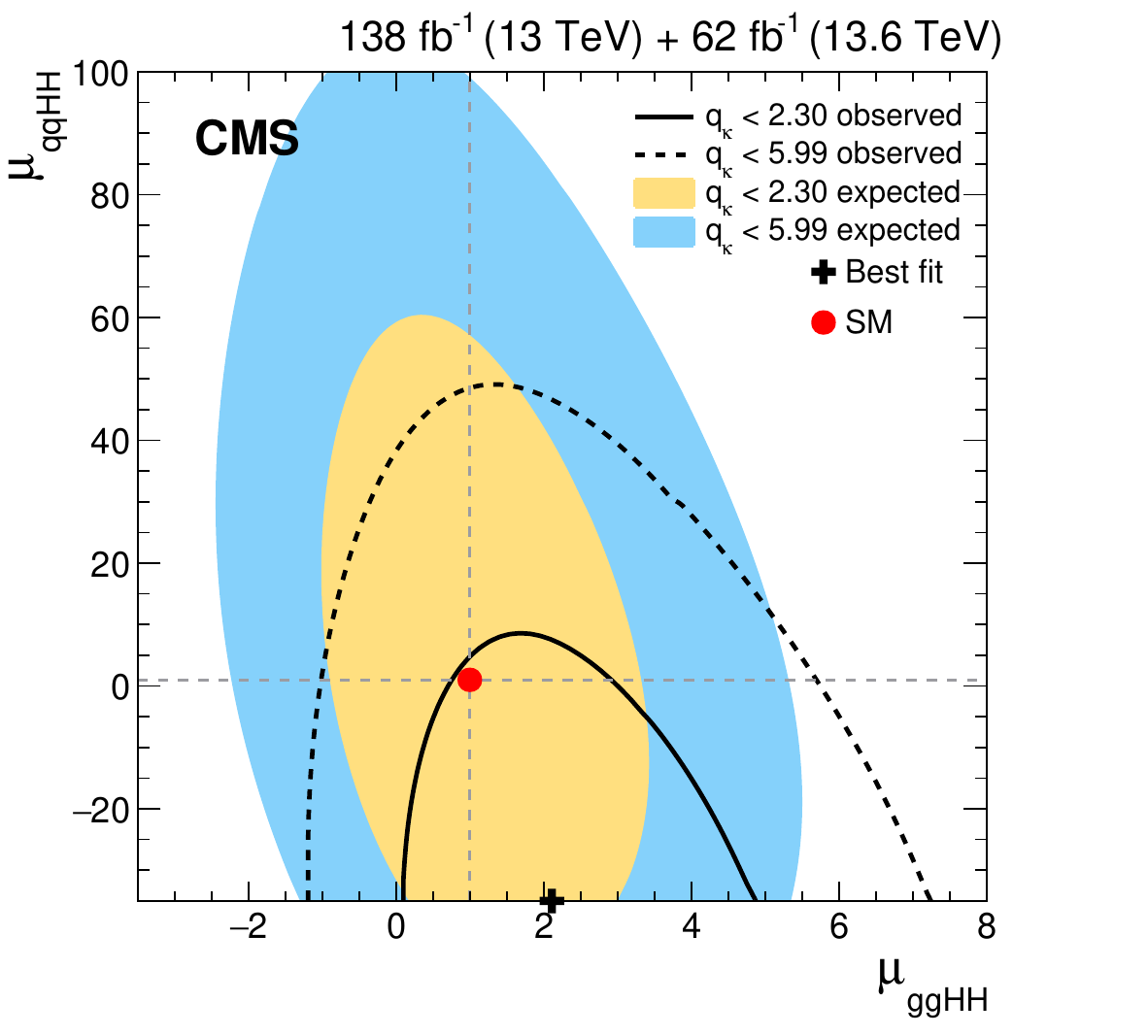}
    \includegraphics[width=0.45\textwidth]{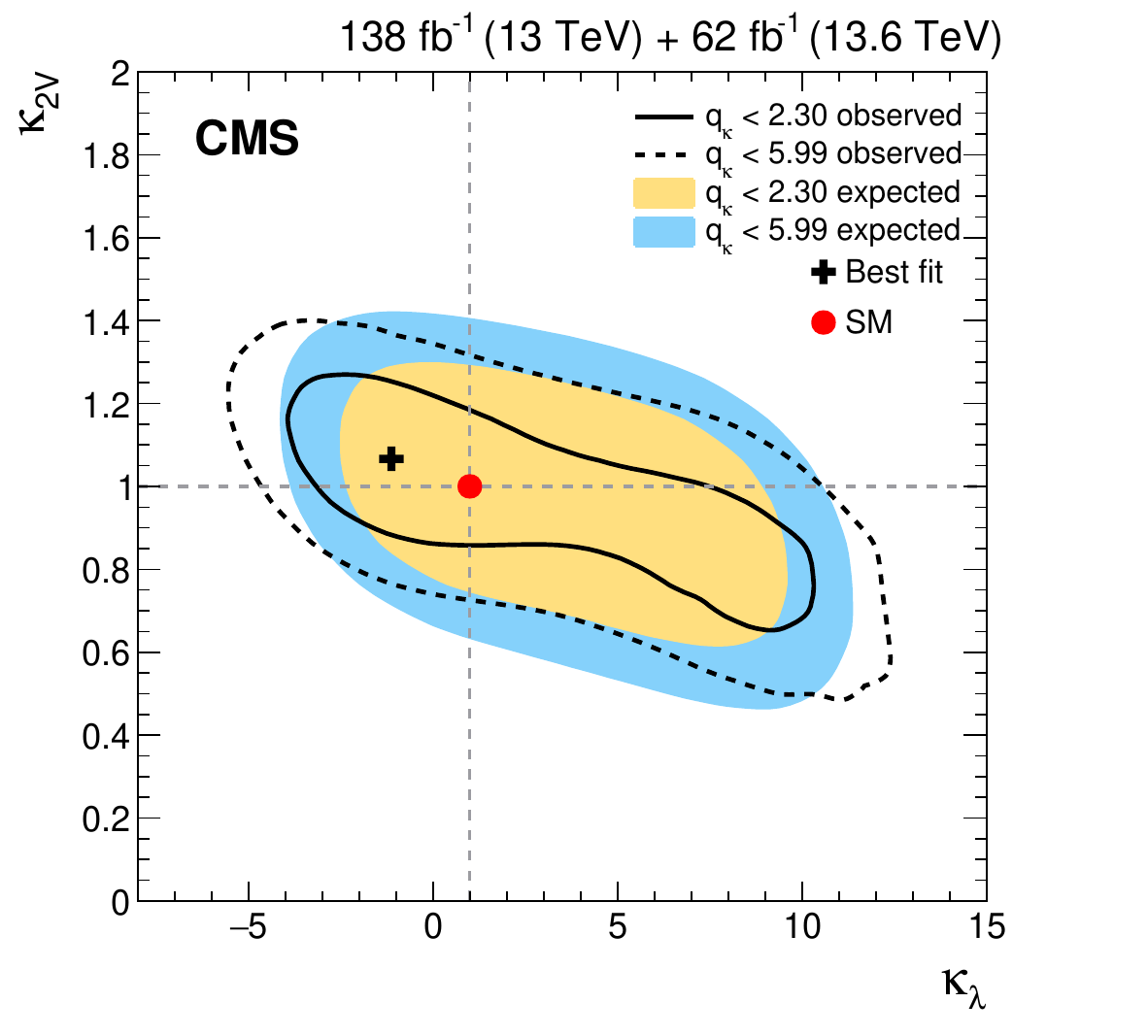}
    \caption{The observed and expected 2D exclusion range for (\mugghh, \muqqhh) (left) and (\kappal,\kappaVV) (right) for the combination of Run~2 and Run~3 \HHbbbb analyses. The black solid and dashed lines show the observed ${q_{\kappa}<2.30}$ and ${q_{\kappa}<5.99}$ exclusion contours, respectively, while the orange and blue shaded regions show the expected exclusion contours. The red circle indicates the SM prediction, while the black cross shows the best fit result. The likelihood function is not well defined in regions with large negative values of \muqqhh, where the total prediction for the sum of signal and backgrounds can become negative in some categories of the input analyses. This region is therefore not shown.}
    \label{fig:combination_run23_scan_kappa2D_results}

\end{figure*}

Figure~\ref{fig:combination_run23_scan_profile_results} shows the observed and expected profiled likelihood ratios as functions of \kappal (left) and \kappaVV (right), while Table~\ref{tab:combination_run23_bestfit} summarizes the observed and expected (for SM) best fit values for the signal strengths, \kappal, and \kappaVV as well as the corresponding ${q_{\kappa}<1}$ intervals. As discussed in Section~\ref{sec:combinations}, intervals for the coupling modifiers are presented using the same bounds on $q_{\kappa}$ considered in the latest \HH combination papers. The observed (expected) allowed range for \kappal with ${q_{\kappa}<3.84}$ is ${[-4.1, 9.6]}\,{([-3.2, 9.8])}$, while for \kappaVV the allowed ${q_{\kappa}<3.84}$ range is ${[0.77, 1.27]}\,{([0.69, 1.35])}$. 
In both figures, all couplings other than the one being varied are fixed to the SM values. Figure~\ref{fig:combination_run23_scan_kappa2D_results} left shows the ${q_{\kappa}<2.3}0$ and ${q_{\kappa}<5.99}$ exclusion contours in the 2D {(\mugghh, \muqqhh)} plane and Fig.~\ref{fig:combination_run23_scan_kappa2D_results} right shows the exclusion contours in the 2D (\kappal, \kappaVV) plane, obtained via a fit with all other couplings fixed to the SM values. The best fit points in the {(\mugghh, \muqqhh)} and {(\kappal, \kappaVV)} planes are observed to be consistent with the SM predictions, within uncertainties. 

\begin{table}[!htb]
  \centering
  \topcaption{Expected and observed best fit values for the signal strengths, \kappal, and \kappaVV from the combined fit of the \HHbbbb Run~2 and Run~3 analyses. The uncertainties given correspond to the ${q_{\kappa}<1}$ intervals. The expected results are calculated with a SM \HHbbbb signal injected. No lower bound is provided for the observed \muqqhh because the likelihood function is not well defined in regions with large negative values of \muqqhh.}
  \renewcommand{\arraystretch}{1.2}
  \begin{scotch}{l c c c c c}
    & \muhh & \mugghh &  \muqqhh & \kappal  & \kappaVV \\
   \hline
   Expected & $1.0^{+1.4}_{-1.2}$ & $1.0^{+1.4}_{-1.3}$ & $1^{+36}_{-31}$ & $1.0^{+7.0}_{-2.5}$ & $1.00^{+0.23}_{-0.20}$ \\
   Observed & $1.7^{+1.6}_{-1.4}$ & $1.8^{+1.6}_{-1.4}$ & $-37^{+34}$ & $-0.7^{+3.2}_{-1.9}$ & $1.02^{+0.13}_{-0.13}$ \\
   \end{scotch}
  \label{tab:combination_run23_bestfit}
\end{table}

\section{Summary} \label{sec:summary}

Measurements of Higgs boson pair (\HH) production in the four bottom quark (4\PQb) final state are presented using proton-proton ($\Pp\Pp$) collision data at ${\sqrt{s}=13.6\TeV}$ collected by the CMS experiment at the CERN LHC during 2022--2023 and corresponding to an integrated luminosity of 62\fbinv. Events in which the Higgs boson decays, \hbb, are separately reconstructed as pairs of small-radius jets (resolved), as well as those where they are reconstructed as single large-radius jets (merged), are studied exclusively. Benefiting from new methods in trigger selection, event selection, and signal extraction, the combination of analyses in the resolved and merged topologies gives an observed (expected) upper limit on the HH signal strength, \muhh, of 4.4~(4.4) at $95\%$ confidence level (\CL). Compared to previously published LHC results in the 4b final state, the expected limit with an equivalent integrated luminosity is improved by more than a factor of two in the resolved topology and is better in the merged topology as well. An updated analysis of the resolved topology using 138\fbinv of 13\TeV $\Pp\Pp$ collision data yields an observed (expected) $95\%$ \CL upper limit on \muhh of 10.0~(5.9), an improvement of about $25\%$ in the expected limit compared to the published results using the same data. Results in the 4\PQb final state with 13 and 13.6\TeV are combined, resulting in an observed (expected) 95\% \CL upper limit on \muhh of 4.7~(2.8). The allowed ranges for the Higgs boson trilinear self-coupling and quartic coupling between two Higgs bosons and two vector bosons, relative to the standard model expectation, in which the test statistic ${q_{\kappa} = -2\Delta \ln \mathcal{L}}$ is less than 3.84 are observed (expected) to be ${[-4.1, 9.6]}\,{([-3.2, 9.8])}$ and ${[0.77, 1.27]}\,{([0.69, 1.35])}$, respectively. These are the most stringent constraints achieved in the 4b final state to date.

\begin{acknowledgments}
We congratulate our colleagues in the CERN accelerator departments for the excellent performance of the LHC and thank the technical and administrative staffs at CERN and at other CMS institutes for their contributions to the success of the CMS effort. In addition, we gratefully acknowledge the computing centers and personnel of the Worldwide LHC Computing Grid and other centers for delivering so effectively the computing infrastructure essential to our analyses. Finally, we acknowledge the enduring support for the construction and operation of the LHC, the CMS detector, and the supporting computing infrastructure provided by the following funding agencies: SC (Armenia), BMBWF and FWF (Austria); FNRS and FWO (Belgium); CNPq, CAPES, FAPERJ, FAPERGS, and FAPESP (Brazil); MES and BNSF (Bulgaria); CERN; CAS, MoST, and NSFC (China); MINCIENCIAS (Colombia); MSES and CSF (Croatia); RIF (Cyprus); SENESCYT (Ecuador); ERC PRG and PSG, TARISTU24-TK10 and MoER TK202 (Estonia); Academy of Finland, MEC, and HIP (Finland); CEA and CNRS/IN2P3 (France); SRNSF (Georgia); BMFTR, DFG, and HGF (Germany); GSRI (Greece); MATE and NKFIH (Hungary); DAE and DST (India); IPM (Iran); SFI (Ireland); INFN (Italy); MSIT and NRF (Republic of Korea); MES (Latvia); LMTLT (Lithuania); MOE and UM (Malaysia); BUAP, CINVESTAV, CONACYT, LNS, SEP, and UASLP-FAI (Mexico); MOS (Montenegro); MBIE (New Zealand); PAEC (Pakistan); MSHE, NSC, and NAWA (Poland); FCT (Portugal); MESTD (Serbia); MICIU/AEI and PCTI (Spain); MOSTR (Sri Lanka); Swiss Funding Agencies (Switzerland); MST (Taipei); MHESI (Thailand); TUBITAK and TENMAK (T\"{u}rkiye); NASU (Ukraine); STFC (United Kingdom); DOE and NSF (USA).

\hyphenation{Rachada-pisek} Individuals have received support from the Marie-Curie program and the European Research Council and Horizon 2020 Grant, contract Nos.\ 675440, 724704, 752730, 758316, 765710, 824093, 101115353, 101002207, 101001205, and COST Action CA16108 (European Union); the Leventis Foundation; the Alfred P.\ Sloan Foundation; the Alexander von Humboldt Foundation; the Science Committee, project no. 22rl-037 (Armenia); the Fonds pour la Formation \`a la Recherche dans l'Industrie et dans l'Agriculture (FRIA) and Fonds voor Wetenschappelijk Onderzoek contract No. 1228724N (Belgium); the Beijing Municipal Science \& Technology Commission, No. Z191100007219010, the Fundamental Research Funds for the Central Universities, the Ministry of Science and Technology of China under Grant No. 2023YFA1605804, the Natural Science Foundation of China under Grant No. 12535004, and USTC Research Funds of the Double First-Class Initiative No.\ YD2030002017 (China); the Ministry of Education, Youth and Sports (MEYS) of the Czech Republic; the Shota Rustaveli National Science Foundation, grant FR-22-985 (Georgia); the Deutsche Forschungsgemeinschaft (DFG), among others, under Germany's Excellence Strategy -- EXC 2121 ``Quantum Universe" -- 390833306, and under project number 400140256 - GRK2497; the Hellenic Foundation for Research and Innovation (HFRI), Project Number 2288 (Greece); the Hungarian Academy of Sciences, the New National Excellence Program - \'UNKP, the NKFIH research grants K 131991, K 138136, K 143460, K 143477, K 147557, K 146913, K 146914, K 147048, TKP2021-NKTA-64, and 2025-1.1.5-NEMZ\_KI-2025-00004, and MATE KKP and KKPCs Research Excellence and Flagship Research Groups grants (Hungary); the Council of Science and Industrial Research, India; ICSC -- National Research Center for High Performance Computing, Big Data and Quantum Computing, FAIR -- Future Artificial Intelligence Research, and CUP I53D23001070006 (Mission 4 Component 1), funded by the NextGenerationEU program, the Italian Ministry of University and Research (MUR) under Bando PRIN 2022 -- CUP I53C24002390006, PRIN PRIMULA 2022RBYK7T (Italy); the Latvian Council of Science; the Ministry of Science and Higher Education, project no. 2022/WK/14, and the National Science Center, contracts Opus 2021/41/B/ST2/01369, 2021/43/B/ST2/01552, 2023/49/B/ST2/03273, and the NAWA contract BPN/PPO/2021/1/00011 (Poland); the Funda\c{c}\~ao para a Ci\^encia e a Tecnologia (Portugal); the National Priorities Research Program by Qatar National Research Fund; MICIU/AEI/10.13039/501100011033, ERDF/EU, ``European Union NextGenerationEU/PRTR", projects PID2022-142604OB-C21, PID2022-139519OB-C21, PID2023-147706NB-I00, PID2023-148896NB-I00, PID2023-146983NB-I00, PID2023-147115NB-I00, PID2023-148418NB-C41, PID2023-148418NB-C42, PID2023-148418NB-C43, PID2023-148418NB-C44, PID2024-158190NB-C22, RYC2021-033305-I, RYC2024-048719-I, CNS2023-144781, CNS2024-154769 and Plan de Ciencia, Tecnolog{\'i}a e Innovaci{\'o}n de Asturias, Spain; the Chulalongkorn Academic into Its 2nd Century Project Advancement Project, the National Science, Research and Innovation Fund program IND\_FF\_68\_369\_2300\_097, and the Program Management Unit for Human Resources \& Institutional Development, Research and Innovation, grant B39G680009 (Thailand); the Eric \& Wendy Schmidt Fund for Strategic Innovation through the CERN Next Generation Triggers project under grant agreement number SIF-2023-004; the Kavli Foundation; the Nvidia Corporation; the SuperMicro Corporation; the Welch Foundation, contract C-1845; and the Weston Havens Foundation (USA).
\end{acknowledgments}\section*{Data availability} Release and preservation of data used by the CMS Collaboration as the basis for publications is guided by the  \href{https://doi.org/10.7483/OPENDATA.CMS.1BNU.8V1W}{CMS data preservation, re-use and open access policy}.
\bibliography{auto_generated}
\cleardoublepage \appendix\section{The CMS Collaboration \label{app:collab}}\begin{sloppypar}\hyphenpenalty=5000\widowpenalty=500\clubpenalty=5000\cmsinstitute{Yerevan Physics Institute, Yerevan, Armenia}
{\tolerance=6000
A.~Hayrapetyan, V.~Makarenko\cmsorcid{0000-0002-8406-8605}, A.~Tumasyan\cmsAuthorMark{1}\cmsorcid{0009-0000-0684-6742}
\par}
\cmsinstitute{Institut f\"{u}r Hochenergiephysik, Vienna, Austria}
{\tolerance=6000
W.~Adam\cmsorcid{0000-0001-9099-4341}, L.~Benato\cmsorcid{0000-0001-5135-7489}, T.~Bergauer\cmsorcid{0000-0002-5786-0293}, M.~Dragicevic\cmsorcid{0000-0003-1967-6783}, P.S.~Hussain\cmsorcid{0000-0002-4825-5278}, M.~Jeitler\cmsAuthorMark{2}\cmsorcid{0000-0002-5141-9560}, N.~Krammer\cmsorcid{0000-0002-0548-0985}, A.~Li\cmsorcid{0000-0002-4547-116X}, D.~Liko\cmsorcid{0000-0002-3380-473X}, M.~Matthewman, J.~Schieck\cmsAuthorMark{2}\cmsorcid{0000-0002-1058-8093}, R.~Sch\"{o}fbeck\cmsAuthorMark{2}\cmsorcid{0000-0002-2332-8784}, M.~Shooshtari\cmsorcid{0009-0004-8882-4887}, M.~Sonawane\cmsorcid{0000-0003-0510-7010}, N.~Van~Den~Bossche\cmsorcid{0000-0003-2973-4991}, W.~Waltenberger\cmsorcid{0000-0002-6215-7228}, C.-E.~Wulz\cmsAuthorMark{2}\cmsorcid{0000-0001-9226-5812}
% ...
\par}
\vskip\cmsinstskip
\dag:~Deceased\\
$^{1}$Also at Yerevan State University, Yerevan, Armenia\\
$^{2}$Also at TU Wien, Vienna, Austria\\
% ...
\end{sloppypar}
\end{document}